\documentclass[12pt]{article}
\usepackage{amsfonts,amssymb,graphicx}
\usepackage{fullpage}
\usepackage[tight]{subfigure}
\usepackage{rotating}
\usepackage{sidecap}
\pdfoutput=1
\usepackage{ifpdf}
\ifpdf
  \usepackage{color}
  \definecolor{darkblue}{rgb}{0.3,0.3,0.6}
    \definecolor{darkgreen}{rgb}{0,0.6,0}
  \usepackage[pdftitle={RigidCycles},pdfauthor={Jill Eckers Gabriele Honecker, Wieland Staessens},pdfsubject={String theory},pdfkeywords={string theory},colorlinks=true,linkcolor=black,urlcolor=darkblue,filecolor=darkblue,citecolor=darkblue,menucolor=darkblue,breaklinks=true,bookmarks=true,anchorcolor=black,unicode=true]{hyperref}
\else
  
\fi
\usepackage[numbers,compress]{natbib}
\usepackage{a4wide}
\usepackage{longtable}
\usepackage{graphicx}
\usepackage{subfigure}
\usepackage[centertags]{amsmath}
\usepackage{multirow}
\usepackage{slashed}
\usepackage{pdflscape}
\usepackage{rotating}
\usepackage{tikz}
\usetikzlibrary{arrows,positioning}
\usepackage{stmaryrd}

\newcommand{\bCentering}{\centering}
\newcommand{\bCaption}{\caption}

\newcommand{\unity}{{\footnotesize\mbox{1\!\!I}}}

\def\muc{\multicolumn}

\def\Z{\mathbb{Z}}

\def\C{\mathbb{C}}
\def\unity{1\!\!{\rm I}}
\def\ov{\overline}
\def\N{\mathbf{N}}
\def\Sym{\mathbf{Sym}}
\def\Anti{\mathbf{Anti}}
\def\Adj{\mathbf{Adj}}
\def\Tr{\text{Tr}}

\def\ov{\overline}
\def\1{{\bf 1}}
\def\2{{\bf 2}}
\def\3{{\bf 3}}
\def\4{{\bf 4}}
\def\6{{\bf 6}}
\def\8{{\bf 8}}
\def\OR{\Omega\mathcal{R}}

\def\pp{\uparrow\uparrow}

\def\targ#1#2{\genfrac{[}{]}{0pt}{}{#1}{#2}}

\def\targ2#1#2{\genfrac{}{}{0pt}{}{#1}{#2}}

\definecolor{blus}{rgb}{0.1,0.1,0.8}
\definecolor{GreenYellow}{cmyk}{0.15,0,0.69,0}
\definecolor{Yellow}{cmyk}{0,0,1,0}
\definecolor{Goldenrod}{cmyk}{0,0.10,0.84,0}
\definecolor{Dandelion}{cmyk}{0,0.29,0.84,0}
\definecolor{Apricot}{cmyk}{0,0.32,0.52,0}
\definecolor{Peach}{cmyk}{0,0.50,0.70,0}
\definecolor{Melon}{cmyk}{0,0.46,0.50,0}
\definecolor{YellowOrange}{cmyk}{0,0.42,1,0}
\definecolor{Orange}{cmyk}{0,0.61,0.87,0}
\definecolor{BurntOrange}{cmyk}{0,0.51,1,0}
\definecolor{Bittersweet}{cmyk}{0,0.75,1,0.24}
\definecolor{RedOrange}{cmyk}{0,0.77,0.87,0}
\definecolor{Mahogany}{cmyk}{0,0.85,0.87,0.35}
\definecolor{Maroon}{cmyk}{0,0.87,0.68,0.32}
\definecolor{BrickRed}{cmyk}{0,0.89,0.94,0.28}
\definecolor{Red}{cmyk}{0,1,1,0}
\definecolor{OrangeRed}{cmyk}{0,1,0.50,0}
\definecolor{RubineRed}{cmyk}{0,1,0.13,0}
\definecolor{WildStrawberry}{cmyk}{0,0.96,0.39,0}
\definecolor{Salmon}{cmyk}{0,0.53,0.38,0}
\definecolor{CarnationPink}{cmyk}{0,0.63,0,0}
\definecolor{Magenta}{cmyk}{0,1,0,0}
\definecolor{VioletRed}{cmyk}{0,0.81,0,0}
\definecolor{Rhodamine}{cmyk}{0,0.82,0,0}
\definecolor{Mulberry}{cmyk}{0.34,0.90,0,0.02}
\definecolor{RedViolet}{cmyk}{0.07,0.90,0,0.34}
\definecolor{Fuchsia}{cmyk}{0.47,0.91,0,0.08}
\definecolor{Lavender}{cmyk}{0,0.48,0,0}
\definecolor{Thistle}{cmyk}{0.12,0.59,0,0}
\definecolor{Orchid}{cmyk}{0.32,0.64,0,0}
\definecolor{DarkOrchid}{cmyk}{0.40,0.80,0.20,0}
\definecolor{Purple}{cmyk}{0.45,0.86,0,0}
\definecolor{Plum}{cmyk}{0.50,1,0,0}
\definecolor{Violet}{cmyk}{0.79,0.88,0,0}
\definecolor{RoyalPurple}{cmyk}{0.75,0.90,0,0}
\definecolor{BlueViolet}{cmyk}{0.86,0.91,0,0.04}
\definecolor{Periwinkle}{cmyk}{0.57,0.55,0,0}
\definecolor{CadetBlue}{cmyk}{0.62,0.57,0.23,0}
\definecolor{CornflowerBlue}{cmyk}{0.65,0.13,0,0}
\definecolor{MidnightBlue}{cmyk}{0.98,0.13,0,0.43}
\definecolor{NavyBlue}{cmyk}{0.94,0.54,0,0}
\definecolor{RoyalBlue}{cmyk}{1,0.50,0,0}
\definecolor{Blue}{cmyk}{1,1,0,0}
\definecolor{Cerulean}{cmyk}{0.94,0.11,0,0}
\definecolor{Cyan}{cmyk}{1,0,0,0}
\definecolor{ProcessBlue}{cmyk}{0.96,0,0,0}
\definecolor{SkyBlue}{cmyk}{0.62,0,0.12,0}
\definecolor{Turquoise}{cmyk}{0.85,0,0.20,0}
\definecolor{TealBlue}{cmyk}{0.86,0,0.34,0.02}
\definecolor{Aquamarine}{cmyk}{0.82,0,0.30,0}
\definecolor{BlueGreen}{cmyk}{0.85,0,0.33,0}
\definecolor{Emerald}{cmyk}{1,0,0.50,0}
\definecolor{JungleGreen}{cmyk}{0.99,0,0.52,0}
\definecolor{SeaGreen}{cmyk}{0.69,0,0.50,0}
\definecolor{Green}{cmyk}{1,0,1,0}
\definecolor{ForestGreen}{cmyk}{0.91,0,0.88,0.12}
\definecolor{PineGreen}{cmyk}{0.92,0,0.59,0.25}
\definecolor{LimeGreen}{cmyk}{0.50,0,1,0}
\definecolor{YellowGreen}{cmyk}{0.44,0,0.74,0}
\definecolor{SpringGreen}{cmyk}{0.26,0,0.76,0}
\definecolor{OliveGreen}{cmyk}{0.64,0,0.95,0.40}
\definecolor{RawSienna}{cmyk}{0,0.72,1,0.45}
\definecolor{Sepia}{cmyk}{0,0.83,1,0.70}
\definecolor{Brown}{cmyk}{0,0.81,1,0.60}
\definecolor{Tan}{cmyk}{0.14,0.42,0.56,0}
\definecolor{Gray}{cmyk}{0,0,0,0.50}
\definecolor{Black}{cmyk}{0,0,0,1}
\definecolor{White}{cmyk}{0,0,0,0}

\definecolor{mygr}{rgb}{0,0.6,0}
\definecolor{mygrey}{rgb}{0,0.1,0.2}
\definecolor{myblue}{rgb}{0,0.5,0.9}
\definecolor{myblue2}{rgb}{0,0.5,0.5}
\definecolor{myorange}{rgb}{1,0.5,0}
\definecolor{mypurple}{rgb}{0.6,0,1}
\definecolor{mygolden}{rgb}{1,0.8,0.2}
\definecolor{mycyan}{rgb}{0,1,1}
\definecolor{mymagenta}{rgb}{1,0,1}

\newcommand{\bCaptionfonts}{\small}
\makeatletter
\long\def\@makecaption#1#2{%
  \vskip\abovecaptionskip
  \sbox\@tempboxa{{\bCaptionfonts #1: #2}}%
  \ifdim \wd\@tempboxa >\hsize
    {\bCaptionfonts #1: #2\par}
  \else
    \hbox to\hsize{\hfil\box\@tempboxa\hfil}%
  \fi
  \vskip\belowcaptionskip}
\makeatother

\makeatletter
\let\ORIGINALlatex@openbib@code=\@openbib@code
\renewcommand{\@openbib@code}{\ORIGINALlatex@openbib@code\setlength{\itemsep}{1ex plus.5ex minus.5ex}\setlength{\parsep}{0pt}}
\makeatother

\def\mathtab#1#2#3{\begin{table}[th]\bCentering$#1$\bCaption{#3}\label{tab:#2}\end{table}}
\def\mathtabfix#1#2#3{\begin{table}[th]\bCentering\resizebox{\linewidth}{!}{$#1$}\bCaption{#3}\label{tab:#2}\end{table}}

\def\mathsidetabfix#1#2#3{\begin{sidewaystable}[H]\bCentering\resizebox{\linewidth}{!}{$#1$}\bCaption{#3}\label{tab:#2}\end{sidewaystable}}

\renewcommand{\arraystretch}{1.3}

\allowdisplaybreaks[4]

\begin{document}
\begin{center}
\begin{flushright}
{\small MITP/14-058\\ 
\today}

\end{flushright}

\vspace{25mm}
{\Large\bf Rigour and Rigidity: Systematics on particle physics D6-brane models on $\Z_2 \times \Z_6$}

\vspace{12mm}
{\large Jill Ecker${}^{\clubsuit}$, Gabriele Honecker${}^{\heartsuit}$ and Wieland Staessens${}^{\spadesuit}$
}

\vspace{8mm}
{
\it PRISMA Cluster of Excellence,\\ Mainz Institute for Theoretical Physics (MITP),\\  \& Institut f\"ur Physik  (WA THEP), \\Johannes-Gutenberg-Universit\"at, D-55099 Mainz, Germany
\;$^{\clubsuit}${\tt eckerji@uni-mainz.de},~$^{\heartsuit}${\tt Gabriele.Honecker@uni-mainz.de} \\ $^{\spadesuit}${\tt wieland.staessens@uni-mainz.de}}

\vspace{15mm}{\bf Abstract}\\[2ex]\parbox{140mm}{
We launch a systematic search for phenomenologically appealing string vacua with intersecting D-branes on the promising $T^6/(\Z_2 \times \Z_6 \times \OR)$ orientifold with discrete torsion.
The number of independent background lattices is reduced from six to two by new symmetries. The occurrence of $USp(2N)$ and $SO(2N)$ versus $U(N)$ gauge groups is classified as 
well as D-branes without matter in the adjoint and/or symmetric representation. Supersymmetric fractional D6-branes allowing for RR tadpole cancellation are fully classified
in terms of all possible values of the one complex structure modulus inherited from the underlying six-torus.
We then systematically investigate the conditions for three particle generations at pairwise intersections of two D6-branes.

Global $SU(5)$ GUT models on $T^6/(\Z_2 \times \Z_6 \times \OR)$ are excluded by demanding three generations and no exotic matter in the {\bf 15} representation. 
Two prototypes of global Pati-Salam models with a mild amount of vector-like exotic matter are found.
}
\end{center}

\thispagestyle{empty}
\clearpage 

\tableofcontents

\setlength{\parskip}{1em plus1ex minus.5ex}
\section{Introduction}\label{S:intro}

String theory is to date the arguably most widely accepted candidate framework for a unified description of quantum field theory and gravity. However, the obvious question of how the Standard Model of particle physics is embedded in string theory still remains open. 

The search for phenomenologically appealing string vacua has over the last years focused on the one hand on supergravity limits on Calabi-Yau manifolds, most notably on scans of large classes of models in the context of the heterotic $E_8 \times E_8$ string with vector bundles~\cite{Bouchard:2005ag,Braun:2005nv,Blumenhagen:2005ga,Blumenhagen:2006ux,Anderson:2011ns,Anderson:2012yf,Anderson:2013xka} 
 and F-theory~\cite{Donagi:2008ca,Beasley:2008dc,Marsano:2009ym,Blumenhagen:2009yv,Weigand:2010wm,Krippendorf:2014xba,Klevers:2014bqa,Lin:2014qga}, and on the other hand on orbifold points with access to the full string spectrum and interactions using the $E_8 \times E_8$ string~\cite{Kim:2007mt,Lebedev:2008un,Blaszczyk:2009in,Pena:2012ki,Nibbelink:2013lua,Blaszczyk:2014qoa} (see e.g. also~\cite{Blaszczyk:2010db} for an orbifold resolution and~\cite{Nilles:2014owa} for a recent review and extended list of references) and Type IIA orientifolds, see e.g. the reviews~\cite{Blumenhagen:2006ci,Cvetic:2011vz,Ibanez:2012zz}.\footnote{
For another large class of exact string models see e.g. the RCFT  and Gepner models~\cite{Dijkstra:2004ym,Dijkstra:2004cc,Anastasopoulos:2006da,Ibanez:2007rs,Kiritsis:2008ry} 
and the free fermionic models~\cite{Assel:2010wj,Faraggi:2014hqa} with MSSM or GUT spectrum. In type IIB string theory, left-right symmetric models, Pati-Salam models and trinification models have been constructed by placing D7- and D3-branes at del Pezzo singularities on compact and non-compact Calabi-Yau three-folds~\cite{Conlon:2008wa,Dolan:2011qu,Cicoli:2012vw,Cicoli:2013mpa}.}

Whithin Type IIA orientifolds, extended scans and proofs of finiteness of the number of solutions to the RR tadpole cancellation and supersymmetry conditions have been performed for various orbifolds~\cite{Blumenhagen:2004xx,Denef:2006ad,Douglas:2006xy,Gmeiner:2008xq} with only a tiny fraction ${\cal O}(10^{-9} -10^{-8})$ having Standard Model-like features~\cite{Gmeiner:2005vz,Gmeiner:2007zz}.\footnote{ 
See also~\cite{Cvetic:2014gia} for the finiteness of the number of solutions on smooth Calabi-Yau manifolds and~\cite{Denef:2004ze,Denef:2004cf,MartinezPedrera:2012rs} for inclusions of closed string fluxes.} 
Fractional D6-branes on toroidal orbifolds have turned out to be of particular interest for model building since unwanted matter in the adjoint representation can in principle be projected out by construction in the presence of some $\Z_2$ symmetry~\cite{Blumenhagen:2002wn,Blumenhagen:2002gw,Honecker:2004kb,Honecker:2004np,Blumenhagen:2005tn,Gmeiner:2006vb,Bailin:2006zf,Gmeiner:2007we,Gmeiner:2007zz,Bailin:2007va,Gmeiner:2008xq,Bailin:2008xx,Forste:2008ex,Gmeiner:2009fb,Forste:2010gw,Bailin:2011am,Honecker:2012qr,Honecker:2013kda,Bailin:2013sya}.  Most notably, on orbifolds with discrete torsion due to some $\Z_2 \times \Z_2$ subgroup, completely rigid D-brane occur~\cite{Blumenhagen:2005tn,Forste:2008ex,Honecker:2012qr,Honecker:2013kda} which do not contain any matter in the adjoint representation of the Standard Model or some GUT gauge group.

We expect the last remaining orientifold with factorisable tori and orbifold group acting only by rotations, $T^6/(\Z_2 \times \Z_6 \times \OR)$ with discrete torsion, to provide a very fertile patch in the string landscape.
This conjecture is based on the one hand on the observation that this particular orientifold contains subsectors related to $T^6/(\Z_6' \times \OR)$~\cite{Forste:2010gw}, which provided a plethora of phenomenologically appealing spectra with fractional, but non-rigid D-branes~\cite{Bailin:2006zf,Gmeiner:2007zz,Bailin:2007va,Gmeiner:2008xq,Bailin:2008xx,Gmeiner:2009fb,Bailin:2011am}.\footnote{For field theoretical investigations on $T^6/(\Z_6' \times \OR)$ see also~\cite{Honecker:2011sm,Honecker:2011hm,Honecker:2012jd,Honecker:2012fn,Honecker:2013hda}.}
On the other hand, on $T^6/(\Z_2 \times \Z_6' \times \OR)$ with discrete torsion and a different $\Z_6'$ action global Pati-Salam models on rigid D-branes could be constructed~\cite{Honecker:2012qr,Honecker:2013kda}, but the $\Z_2 \times \Z_6'$ geometry {\it a priori} constrained model building, e.g. by the fact that three particle generations in the antisymmetric representation would necessarily be accompanied by three exotic matter states in the symmetric representation of the same non-Abelian gauge group. 
The present  $T^6/(\Z_2 \times \Z_6 \times \OR)$ orientifold not only possesses a wider model building freedom due to the one complex structure modulus inherited from the underlying six-torus,  but also the number of chiral states in the antisymmetric and symmetric representation are {\it a priori} different.

This article is organized as follows: in section~\ref{S:Geometry}, we review the geometry of the $T^6/(\Z_2 \times \Z_6 \times \OR)$ orientifold with rigid and fractional three-cycles. We unveil previously unknown maps between model building ingredients on the {\it a priori} six different background lattice orientations, leaving only two distinct lattices to be explored in view of phenomenologically appealing spectra.
In section~\ref{S:StepsInD6BraneMB}, we perform first steps in the search for three generation models without exotic matter by classifying when gauge group enhancements $U(N) \hookrightarrow USp(2N)$ or $SO(2N)$ occur, which bulk three-cycles can be used to construct completely rigid fractional three-cycles, which three-cycles do not support chiral matter in the symmetric representation and which pairwise intersection numbers occur among such favourable cycles. In section~\ref{S:GlobalModels}, we assemble the ingredients to search for global $SU(5)$ and Pati-Salam models.
Our conclusions and outlook are given in section~\ref{S:Conclusions}. 
Appendix~\ref{A:ClassBulkThreeCycles} contains the full classification of supersymmetric three-cycles not overshooting the bulk RR tadpole cancellation conditions on the two distinct background lattices.

\section{Geometric Considerations }\label{S:Geometry}

A first look into the geometric characteristics of Type IIA string theory on the factorisable toroidal orbifold $T^6/(\Z_2 \times \Z_6\times \OR)$ with discrete torsion was performed in~\cite{Forste:2010gw}, including both the mathematical description of rigid {\it special Lagrangian} three-cycles as well as global consistency conditions, i.e.~RR~tadpole cancellation conditions and K-theory constraints. These aspects will be briefly reviewed in section~\ref{Ss:IIAonZ2Z6}, combined with a new discussion in which the six a priori independent background lattices are explicitly related to each other three by three, leaving only two physically inequivalent options. The philosophy behind the identification mimics the one presented in~\cite{Honecker:2012qr} for $T^6(\Z_2 \times \Z_6' \times \OR)$ with $\Z_6' \neq \Z_6$, where the equivalence between two seemingly different lattices has been fully proven for the first time on the level of the massless spectra, global consistency conditions and CFT results for one-loop vacuum amplitudes.  

In section~\ref{Ss:Reduction_Z6p}, we briefly comment on how the relations among lattices can be truncated to the $T^6/(\Z_6' \times \OR)$ case. This completes the proof of equivalent lattices initiated in appendix~D of~\cite{Gmeiner:2008xq}.

\subsection{Type IIA on $T^6/(\Z_2 \times \Z_6\times \OR)$ with discrete torsion}\label{Ss:IIAonZ2Z6}

\begin{figure}[ht]
\begin{center}
\begin{tabular}{c@{\qquad}c@{\hspace{-0.2in}}c}
\vspace{-0.7in}\includegraphics[width=4cm]{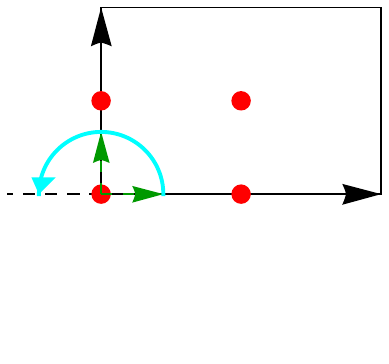} \begin{picture}(0,0) \put(-100,55){\bf \color{mygr} a} \put(-50,110){$R_1$} \put(0,75){$R_2$}  \put(-95,35){\bf \color{red} \bf 1} \put(-100,75){\bf \color{red} \bf 4} \put(-45,35){\bf \color{red} \bf 2} \put(-42,75){\bf \color{red} \bf 3}
\put(-15,35){$\pi_{1}$} \put(-105,105){$\pi_2$} \put(-70,65){\bf \color{mycyan}$\theta$}   \end{picture} &&\\
\vspace{-1.2in}  & \includegraphics[width=6cm]{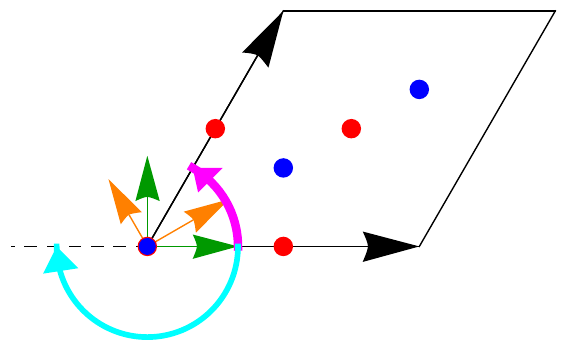} \begin{picture}(0,0) \put(-60,110){$r_2$} \put(-135,65){\color{mygr} {\bf A}} \put(-155,55){\color{myorange} {\bf B}} \put(-135,15){\color{red} \bf 1}  \put(-90,15){\color{red} \bf 4}  \put(-120,65){\color{red} \bf 5}  \put(-62,62){\color{red} \bf 6}  \put(-85,50){\color{blue} \bf 2}  \put(-42,75){\color{blue} \bf 3} \put(-60,15){$\pi_{3}$} \put(-105,105){$\pi_4$}  \put(-105,0){\bf \color{mycyan}$\theta$}   \put(-100,40){\bf \color{mymagenta}$\omega$}  \end{picture}  
&\includegraphics[width=6cm]{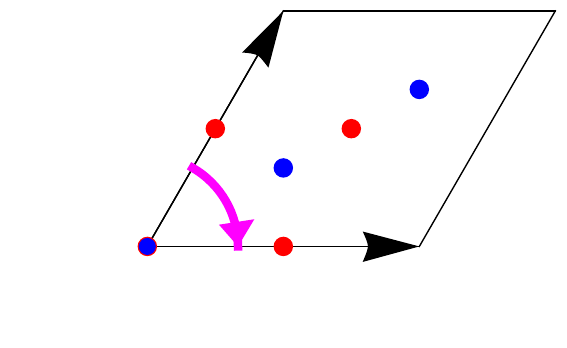} \begin{picture}(0,0) \put(-60,110){$r_3$}  \put(-135,15){\color{red} \bf 1}  \put(-90,15){\color{red} \bf 4}  \put(-120,65){\color{red} \bf 5}  \put(-62,62){\color{red} \bf 6}  \put(-85,50){\color{blue} \bf 2}  \put(-42,75){\color{blue} \bf 3} \put(-60,15){$\pi_{5}$} \put(-105,105){$\pi_6$}  \put(-107,52){\bf \color{mymagenta}$\omega$}  \end{picture} \\
\includegraphics[width=3.cm]{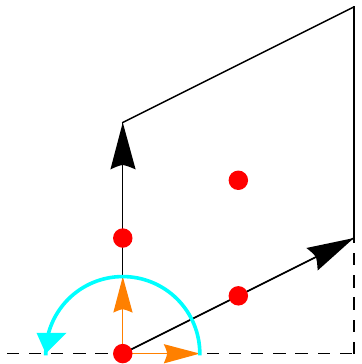} \begin{picture}(0,0) \put(-72,8){\bf \color{myorange} b}  \put(-40,-10){$R_1$} \put(0,55){$R_2$} \put(-68,-10){\bf \color{red} \bf 1} \put(-72,28){\bf \color{red} \bf 4} \put(-28,6){\bf \color{red} \bf 2} \put(-27,40){\bf \color{red} \bf 3}  \put(0,25){$\pi_{1}$} \put(-75,58){$\pi_2$}  \put(-50,22){\bf \color{mycyan}$\theta$}  \end{picture}  & &\\
\end{tabular}
\caption{The $SU(2)^2 \times SU(3) \times SU(3)$ compactification lattice for the $T^6/(\Z_2 \times \Z_6 \times \OR)$ orientifold defined by the action in equation~(\ref{Eq:Z2Z6action}) with complex structure modulus $\varrho \equiv \sqrt{3}{R_2/R_1}$ on $T^2_{(1)}$. The $\Z_2$ fixed points are labeled by the red points $(1, 2, 3, 4)$ on the first two-torus and $(1, 4, 5, 6)$ on the second and third two-torus. The $\Z_6$ action is trivial on the first torus and cyclically permutes three $\Z_2$ fixed points on the second and third torus: $1\stackrel{e^{\pi i/3}}{\circlearrowleft},4\stackrel{e^{\pi i/3}}{\to}5\stackrel{e^{\pi i/3}}{\to}6\stackrel{e^{\pi i/3}}{\to}4$.
The blue points denote the $\Z_3$ fixed points with $2\stackrel{e^{\pi i/3}}{\leftrightarrow} 3$. Invariance under the anti-holomorphic involution~(\ref{Eq:AntiHoloInvolution}) permits an untilted {\color{mygr}{\bf a}}-type (with $2  \stackrel{\cal R}{\circlearrowleft}, 3  \stackrel{\cal R}{\circlearrowleft}, 4 \stackrel{\cal R}{\circlearrowleft}$) or a tilted {\color{myorange}{\bf b}}-type lattice for $T_{(1)}^2$ (with $4 \stackrel{\cal R}{\circlearrowleft}, 2 \stackrel{\cal R}{\leftrightarrow} 3$), and two orientation choices for the other two-tori as well: {\color{mygr}{\bf A}}-type (with $4 \stackrel{\cal R}{\circlearrowleft}, 5 \stackrel{\cal R}{\leftrightarrow} 6$) and {\color{myorange}{\bf B}}-type (with $4 \stackrel{\cal R}{\leftrightarrow} 5, 6 \stackrel{\cal R}{\circlearrowleft}$). 
\label{Fig:LatticesZ2Z6}}
\end{center}
\end{figure}

The point group $\Z_2 \times \Z_6$ acts on the factorisable six-torus $T^2_{(1)} \times T^2_{(2)} \times T^2_{(3)}$ by rotating the complex coordinate $z^k$ per two-torus $T^2_{(k)}$ with $k \in \{1,2,3 \}$ as follows: 
\begin{equation}\label{Eq:Z2Z6action}
\theta^m \omega^n : z^k \rightarrow e^{2 \pi i (m v_k + n w_k)} z^k, \qquad \text{ with } \quad \vec{v} = \frac{1}{2} (1,-1,0), \quad \vec{w} = \frac{1}{6} (0,1,-1).
\end{equation}
Here, $\theta^m \omega^n$ represents a generic element of the point group, with $\theta$ generating the $\Z_2$ part of the orbifold group, while the generator $\omega$ only acts on $T^2_{(2)} \times T^2_{(3)}$ by a $\Z_6$ rotation.\footnote{Note the contrast to the $\Z_2 \times \Z_6'$ orbifold with $\vec{w}^{\, \prime} = \frac{1}{6} (-2,1,1)$, for which D6-brane models were explored in~\cite{Honecker:2012qr,Honecker:2013kda}.} 
The shift vectors $\vec{v} + \vec{w}= \frac{1}{6}(3,-2,-1)$ and $\vec{v} + 2 \vec{w}= \frac{1}{6}(3,-1,-2)$ generate two different $\Z_6'$ subsectors, indicating that the bulk three-cycles of the orbifold $T^6/(\Z_2 \times  \Z_6)$ are identical (up to normalization) to those on the $T^6/\Z_6'$ orbifold~\cite{Bailin:2006zf,Gmeiner:2007we,Gmeiner:2007zz,Bailin:2007va,Gmeiner:2008xq,Bailin:2008xx,Bailin:2011am}, as we will see later on.
The crystallographic action of the $\Z_6$ generator $\omega$ constrains the complex structures of the factorisable four-torus $T^2_{(2)} \times T^2_{(3)}$, such that their lattices take the shape of $SU(3)$ root lattices, as depicted in figure~\ref{Fig:LatticesZ2Z6}. Due to the trivial $\Z_6$ action on $T^2_{(1)}$, the lattice corresponds to the $SU(2)^2$ root lattice with an a priori unconstrained complex structure parameter. Nonetheless, the complex structure is restricted by the anti-holomorphic involution ${\cal R}$,   
\begin{equation}\label{Eq:AntiHoloInvolution}
{\cal R}: z^k \rightarrow \ov{z}^k,
\end{equation}
accompanying the worldsheet parity $\Omega$ when defining the Type IIA orientifold on $T^6/(\Z_2~\times~\Z_6~\times~\OR)$. 
As a result, the complex structure of $T^2_{(1)}$ contains only one real free parameter, the  ratio $\varrho  \equiv \sqrt{3} R_2 /R_1$, and
the shape of the first two-torus is either untilted (rectangular) or tilted, parametrised by  $b=0, 1/2$ respectively as depicted in figure~\ref{Fig:LatticesZ2Z6}.  Invariance under the orientifold projection also limits the possible orientations of the two-torus lattices for $T_{(l=2,3)}^2$ w.r.t.~the orientifold invariant direction: the {\bf A} orientation has the one-cycle $\pi_{2l-1}$ along the $\OR$-invariant plane, whereas the {\bf B} orientation corresponds to the configuration with the one-cycle $\pi_{2l-1} + \pi_{2l}$ along the $\OR$-invariant direction.
%

For orbifold groups of the type $\Z_N\times \Z_M$, each generator of $\Z_N$ can act on the $\Z_M$ twisted sector with a phase factor $\eta = e^{2 \pi n\,  i/ \text{gcd}(N,M)}$ (with $n \in \Z$) and vice versa~\cite{Vafa:1986wx,Font:1988mk}, which for the toroidal orbifold $T^6/(\Z_2 \times  \Z_6)$ at hand boils down to the choice of a sign factor,
\begin{equation}\label{Def:torsion}
\eta = \left\{\begin{array}{cc} 1 & \text{without}
\\ -1 & \text{with}\end{array}  \right. 
\text{ discrete torsion}
.
\end{equation}
The impact of the absence or presence of discrete torsion is first of all reflected in the Hodge numbers counting the number of two- and three-cycles in the twisted sectors of the toroidal orbifold:
\begin{equation}\label{Eq:HodgeNumbersZ2Z6}
\mbox{\resizebox{0.94\textwidth}{!}{%
$
\hspace{-4mm}
\left(\begin{array}{c} h_{11} \\ h_{21} \end{array}\right)
=
\left(\begin{array}{c} h_{11}^U + h_{11}^{\Z_6} + h_{11}^{\Z_3} + h_{11}^{\Z_2}  \\  h_{21}^U + h_{21}^{\Z_6} + h_{21}^{\Z_3} + h_{21}^{\Z_2}   
\end{array}\right)
=\left\{
\begin{array}{cc}
\left(\begin{array}{c} 3+ (2+ 2 \times 8) + 8+ ( 6 + 2 \times 8) =51\\  1+ 0+2+0 =3\end{array}\right) & \eta=1
\\
\left(\begin{array}{c} 3+ (0 +2 \times 4) +8+0 =19\\ 1+ 2+2+ (6 + 2 \times 4) =19 \end{array}\right) & \eta=-1
\end{array}\right.
.
$}}
\end{equation}
In order to determine the Hodge numbers for the twisted sectors, one counts the number of $\Z_6$ invariant orbits of fixed points per twisted sector and takes into account that $\Z_N$ fixed points along $T^6$ 
(generators of $\Z_6'$ here: $\theta\omega$, $\theta\omega^2$) or some $T^4$ (generators of $\Z_N$ here: $\omega_{(N=6)}$, $\omega^2_{(N=3)}$, $\{\theta,\omega^3,\theta\omega^3\}_{(N=2)}$)
support two-forms and their dual four-forms, while $\Z_N$ fixed lines support three-forms dual to three-cycles consisting of exceptional divisors at fixed points along some $T^4$ tensored with a one-cycle on the remaining two-torus. Finally, taking into account the action of one $\Z_N$ onto the $\Z_M$ twisted sector through the phase factor $\eta$ results in the Hodge numbers in equation~(\ref{Eq:HodgeNumbersZ2Z6}). 

In this article, we focus on model building with rigid D6-branes constructed from exceptional three-cycles stuck at $\Z_2$ fixed points. The Hodge numbers above clearly indicate that such rigid D6-branes only exist in the presence of discrete torsion $(\eta=-1$), and in the remainder of the article we restrict to this case.       

Combining the orbifold group with an orientifold projection $\OR$ allows for compactifications of Type IIA string theory on $T^6/(\Z_2\times\Z_6\times\OR)$ preserving ${\cal N} = 1$ supersymmetry. The respective O6-planes can be grouped~\cite{Forste:2010gw} into four inequivalent orbits under the $\Z_6$-action, each containing one of the $\OR$- and $\OR\Z_2^{(k=1,2,3)}$-invariant planes. Invariance under the orientifold projection can be achieved for a priori six different lattice configurations: {\bf aAA}, {\bf aAB}, {\bf aBB}, {\bf bAA}, {\bf bAB} and {\bf bBB},  see figure~\ref{Fig:LatticesZ2Z6} for details.  

Denoting the sign of RR charges of the various (orbits of) O6-planes by $\eta_{\OR}$ and \mbox{$\eta_{\OR\Z_2^{(k)}} \in \{\pm 1\}$}, it was shown~\cite{Blumenhagen:2005tn,Forste:2010gw} that
worldsheet consistency of the Klein bottle amplitude enforces the following relation:
\begin{equation}\label{Eq:eta-for-ORZ2s}
\eta = \eta_{\OR} \prod_{k=1}^3 \eta_{\OR\Z_2^{(k)}}.
\end{equation} 
This implies that in the presence of discrete torsion, an odd number of exotic O6-planes with positive RR charges ($\eta_{\OR(\Z_2^{(k)})} = -1$)
occurs. We will see below that the choice of three exotic O6-planes is not compatible with supersymmetric D6-model building.

Under the orientifold projection, the two- and three-cycles counted by the Hodge numbers in (\ref{Eq:HodgeNumbersZ2Z6}) decompose into $\OR$-even and $\OR$-odd cycles, 
and the action of $\OR$ on the $\Z_N$ twisted sectors  is correlated with the presence of discrete torsion. For $\Z_2^{(k)}$ twisted sectors, we find~\cite{Blumenhagen:2005tn,Forste:2010gw}
\begin{equation}\label{Eq:eta-for-Z2s}
\eta_{(k)} \equiv \eta_{\OR} \eta_{\OR \Z_2^{(k)}} \qquad \text{with} \quad \eta = \prod_{k=1}^3 \eta_{(k)}.
\end{equation}
The massless closed string spectrum for Type IIA string theory on $T^6/(\Z_2\times \Z_6\times \OR)$ with discrete torsion is summarised in table~\ref{tab:ClosedSpectrum_Z2Z6}, with the Hodge numbers $h_{11}^+$ counting the number of ${\cal N}=1$ vector multiplets, $h_{11}^-$ the number of  chiral multiplets with K\"ahler moduli as real scalar components and $h_{21}$ the number of chiral multiplets containing complex structure moduli.

\mathtabfix{
\begin{array}{|c|c||c|c|c|}\hline
\muc{5}{|c|}{\text{\bf Closed string spectrum on } T^6/(\Z_2 \times \Z_6 \times \OR)  \; \text{\bf with discrete torsion } \eta=-1}
\\\hline\hline
{\cal N}=1 \text{ multiplet} & \# & {\bf a/bAA} & {\bf a/bAB} & {\bf a/bBB} 
\\\hline\hline
\text{gravity} 
&  \muc{4}{|c|}{ 1 }
\\\hline
\text{dilaton - axion} 
&  \muc{4}{|c|}{ 1 }
\\\hline
\text{vector} & h_{11}^+= & 4 + 2(1-b) (\eta_{(2)}+\eta_{(3)}) & 4 + 2(1-b) (\eta_{(2)}-\eta_{(3)}) & 4 - 2(1-b) (\eta_{(2)}+\eta_{(3)}) \\
\hline
\text{K\"ahler moduli} & h_{11}^-= &  15- 2(1-b) (\eta_{(2)}+\eta_{(3)})  &   15- 2(1-b) (\eta_{(2)}-\eta_{(3)})&  15+ 2(1-b) (\eta_{(2)}+\eta_{(3)})\\\hline
\text{complex structure} & h_{21}= & \muc{3}{|c|}{ 19} 
\\\hline
\end{array}
}{ClosedSpectrum_Z2Z6}{${\cal N}=1$ supersymmetric closed string spectrum of Type IIA string theory on $T^6/(\Z_2 \times \Z_6 \times \OR)$ with discrete torsion,  as first computed in~\cite{Forste:2010gw}.
Counting $(h_{11}^+,h_{11}^-)$ gives the first indication that for fixed $b \in \{0,\frac{1}{2}\}$, lattices can be related by permuting the exotic O6-plane label.}

\subsubsection{Bulk three-cycles}\label{Sss:BulkCycles}

In the lattice of three-cycles  on the orbifold $T^6/(\Z_2 \times \Z_6)$ with discrete torsion, \mbox{$b_3^{\text{bulk}} = 2 + 2 h_{21}^U = 4$} three-cycles can be identified as {\it bulk cycles} inherited from the underlying factorable six-torus. A basis is given by the following four three-cycles:
\begin{equation}
\begin{array}{l}
\rho_1 \equiv 4 (1 + \omega + \omega^2) \pi_{135} = 4( \pi_{135} + \pi_{145} - 2 \pi_{146} + \pi_{136} ),\\
\rho_2 \equiv 4 (1 + \omega + \omega^2) \pi_{136} = 4( 2\pi_{136} + 2\pi_{145} -  \pi_{146} - \pi_{135} ),\\
\rho_3 \equiv 4 (1 + \omega + \omega^2) \pi_{235} = 4( \pi_{235} + \pi_{245} - 2 \pi_{246} + \pi_{236} ),\\
\rho_4 \equiv 4 (1 + \omega + \omega^2) \pi_{236} = 4( 2\pi_{236} +2 \pi_{245} -  \pi_{246} - \pi_{235} ),
\end{array}
\end{equation}
for which the only non-vanishing intersection numbers are given by,
\begin{equation}\label{Eq:BulkIntersectionNumbers}
\begin{array}{l}
\rho_1 \circ \rho_3 = \rho_2 \circ \rho_4 = 8,\\
\rho_1 \circ \rho_4 = \rho_2 \circ \rho_3 = 4.
\end{array}
\end{equation}
One can then express a generic bulk three-cycle in terms of this basis as:
\begin{equation}
\Pi^{\text{bulk}}_a = P_a \, \rho_1 + Q_a \, \rho_2 + U_a \, \rho_3 + V_a \, \rho_4,
\end{equation}
by introducing the bulk wrapping numbers, 
\begin{equation}\label{Eq:BulkWrappingNumbers}
\begin{array}{l}
P_a \equiv n^1_a X_a, \qquad Q_a \equiv n^1_a Y_a, \qquad U_a \equiv m^1_a X_a, \qquad V_a \equiv m^1_a Y_a,\\
 \text{with}\quad  X_a \equiv n^2_a n^3_a - m^2_a m^3_a, \qquad Y_a \equiv n^2_a m^3_a + m^2_a n^3_a + m^2_a m^3_a.
\end{array}
\end{equation}
The intersection number of two generic bulk three-cycles follows directly from equation~(\ref{Eq:BulkIntersectionNumbers}):
\begin{equation}
\Pi_a^{\text{bulk}} \circ \Pi_b^{\text{bulk}} = 8 \left( P_a U_b - P_b U_a + Q_a V_b - Q_b V_a \right) + 4 \left(  P_a V_b - P_b V_a +  Q_a U_b -  Q_b U_a \right).
\end{equation}
The above basis of bulk cycles is - up to normalisation and permutation of two-torus indices - identical to the one for $T^6/\Z_6'$ introduced in~\cite{Gmeiner:2007zz}.

The bulk wrapping numbers $(P_a,Q_a,U_a,V_a)$ are orbifold-invariant combinations of the torus wrapping numbers $(n_a^i, m_a^i)_{i=1,2,3}$ which transform non-trivially under the $\Z_6$ action as:
\begin{equation}\label{Eq:1-cycle-orbits}
\left(\begin{array}{cc}
n^1_a & m^1_a \\ n^2_a & m^2_a \\ n^3_a & m^3_a 
\end{array}\right) \stackrel{\omega}{\rightarrow}
\left(\begin{array}{cc}
n^1_a & m^1_a \\ m^2_a & -(n^2_a +m^2_a) \\ -(n^3_a+m^3_a) & n^3_a 
\end{array}\right) \stackrel{\omega}{\rightarrow}
\left(\begin{array}{cc}
n_a^1& m^1_a\\ -(n^2_a+m^2_a) & n^2_a \\ m^3_a & -(n^3_a+m^3_a) 
\end{array}\right) 
,
\end{equation}
where an overall sign-flip along $T^2_{(2)}\times T^2_{(3)}$ has been taken into account in comparison to~\cite{Forste:2010gw}. With the sign-flip, not only are the bulk wrapping numbers independent of the choice of the representant of a given $\Z_6$ orbit, i.e.~$a$ or $(\omega a)$ or $(\omega^2 a)$, but also the fractional three-cycle - consisting of linear combinations of bulk and exceptional three-cycles as detailed below in section~\ref{Sss:FractionalCycles} - will be independent of the choice of orbifold representant, analogously to the considerations in~\cite{Honecker:2012qr} for $\Z_2 \times \Z_6'$.

The bulk wrapping numbers $(P_a,Q_a,U_a,V_a)$  transform under the orientifold projection, as can be derived from the behaviour of the basic bulk three-cycles under $\OR$  in table~\ref{Tab:Z2Z6bulk-Orient}, 
\begin{table}[h!]
\renewcommand{\arraystretch}{1.3}
  \begin{center}
\begin{equation*}
\begin{array}{|c||c|c|c|c|}\hline
 \multicolumn{5}{|c|}{\text{\bf Orientifold images of bulk 3-cycles on } T^6/(\Z_2 \times \Z_6 \times \OR)}
\\\hline\hline
{\rm 3-cycle} & \rho_1 & \rho_2 & \rho_3 & \rho_4
\\\hline\hline
{\bf a/bAA} & \rho_1-(2b)\rho_3 & \rho_1 - \rho_2-(2b)[\rho_3-\rho_4] & -\rho_3 & \rho_4 - \rho_3 
\\\hline
{\bf a/bAB} & \rho_2-(2b)\rho_4 & \rho_1-(2b)\rho_3 & -\rho_4 & -\rho_3
\\\hline
{\bf a/bBB} & \rho_2 -\rho_1 - (2b)[\rho_4-\rho_3] & \rho_2 - (2b) \rho_4 & \rho_3 - \rho_4 & -\rho_4
\\\hline
\end{array}
\end{equation*}
\end{center}
\caption{The orientifold projection on bulk three-cycles on  $T^6/(\Z_2 \times \Z_6 \times \OR)$ without and with discrete torsion, depending on the choice of background lattice orientation.}
\label{Tab:Z2Z6bulk-Orient}
\end{table}
or by using the transformation properties of the toroidal one-cycle wrapping numbers under $\OR$, depending on the two-torus lattice orientation:
\begin{equation}\label{Eq:OR_on_n+m}
(n^1_{a'},m^1_{a'}) = \left\{\begin{array}{cc}
(n^1_a  , \, - m^1_a) & ({\bf a}) \\
(n^1_a , -n^1_a-m_a^1) & ({\bf b})
\end{array}\right. 
,
\qquad
(n^i_{a'},m^i_{a'})_{i=2,3} 
= \left\{\begin{array}{cc}
(n^i_a+m^i_a \, , \, - m^i_a) & ({\bf A}) \\
(m^i_a,n^i_a) & ({\bf B})
\end{array}\right.
.
\end{equation}
Only bulk cycles parallel to one of the (orbits of) orientifold invariant planes  have $\OR$-invariant bulk wrapping numbers, see the O6-plane orbits in table~\ref{tab:Z2Z6-Oplanes-torus}. Notice, however, that the geometry allows for orientifold invariant bulk three-cycles displaced from the O6-planes by one-half of a lattice vector
and that in case of a tilted torus, there is no O6-plane along the displaced position.
\mathtabfix{
\begin{array}{|c|c||c|c||c|c||c|c|}\hline
\multicolumn{8}{|c|}{\text{\bf Torus and bulk wrapping numbers for the four O6-plane orbits on  } T^6/(\Z_2 \times \Z_6 \times \OR)}
\\\hline\hline
\text{O6-plane} & \frac{\rm angle}{\pi} & \muc{2}{|c|}{\bf a/bAA} &  \muc{2}{|c|}{\bf a/bAB} &  \muc{2}{|c|}{\bf a/bBB}
\\\hline
& & (n^i,m^i) & (P,Q,U,V)  & (n^i,m^i) & (P,Q,U,V)  & (n^i,m^i) & (P,Q,U,V) 
\\\hline\hline
\OR & (0,0,0) & (\frac{1}{1-b},\frac{-b}{1-b};1,0;1,0) & \frac{1}{1-b}(1,0,-b,0)
&  (\frac{1}{1-b},\frac{-b}{1-b};1,0;1,1) & \frac{1}{1-b}(1,1,-b,-b)
&  (\frac{1}{1-b},\frac{-b}{1-b};1,1;1,1) & \frac{3}{1-b}(0,1,0,-b)
\\
\OR\Z_2^{(1)} & (0,\frac{1}{2},\frac{-1}{2}) &  (\frac{1}{1-b},\frac{-b}{1-b};-1,2;1,-2) & \frac{3}{1-b}(1,0,-b,0)
&   (\frac{1}{1-b},\frac{-b}{1-b};-1,2;1,-1) &  \frac{1}{1-b}(1,1,-b,-b)
&   (\frac{1}{1-b},\frac{-b}{1-b};-1,1;1,-1) & \frac{1}{1-b}(0,1,0,-b)
\\
\OR\Z_2^{(3)} & (\frac{1}{2},\frac{-1}{2},0) & (0,1;1,-2;1,0) & (0,0,1,-2)
& (0,1;1,-2;1,1) & (0,0,3,-3)
& (0,1;1,-1;1,1) & (0,0,2,-1)
\\
\OR\Z_2^{(2)}  &  (\frac{1}{2},0,\frac{-1}{2}) &  (0,1;1,0;1,-2) & (0,0,1,-2)
&  (0,1;1,0;1,-1) & (0,0,1,-1)
& (0,1;1,1;1,-1) & (0,0,2,-1)
\\\hline
\end{array}
}{Z2Z6-Oplanes-torus}{The torus wrapping numbers $(n^i,m^i)_{i\in \{1,2,3\}}$ are given for one representant of each O6-plane orbit on $T^6/(\Z_2 \times \Z_6 \times \OR)$. 
The angle w.r.t. the $\OR$-invariant  plane is listed in the second column. The bulk wrapping numbers $(P,Q,U,V)$ are independent of the choice of representant. 
The number of identical O6-planes is $N_{O6}=2(1-b)$ with $b=0,1/2$ for the {\bf a}- and {\bf b}-type torus $T^2_{(1)}$, respectively.
}

Combining all the previous information allows us to write down the bulk RR tadpole cancellation conditions and bulk supersymmetry conditions for D6-branes on $T^6/(\Z_2 \times \Z_6 \times \OR)$ with discrete torsion as displayed in table~\ref{tab:Bulk-RR+SUSY-Z2Z6}. Note that the necessary and sufficient conditions for supersymmetric D6-branes are identical to those on $T^6/\Z_6'$ given in~\cite{Gmeiner:2007zz}. Both supersymmetry conditions depend on the complex structure modulus $\varrho$ of the first two-torus $T^2_{(1)}$, implying that a classification of supersymmetric three-cycles on $T^6/(\Z_2 \times \Z_6 \times \OR)$ has to be done for separate values of $\varrho$. It is obvious that for irrational values of $\varrho$, the only supersymmetric three-cycles are those cycles whose bulk part is parallel to one of the orientifold invariant planes along $T^2_{(1)}$, i.e. $(n^1,\tilde{m}^1) \in \{(\frac{1}{1-b},0),(0,1)\}$ as listed in tables~\ref{tab:Z2Z6-Oplanes-torus}
and~\ref{tab:RhoIndependentaAA} of appendix~\ref{A:ClassBulkThreeCycles}. We will come back to the counting of supersymmetric bulk three-cycles for rational values of $\varrho$ 
at the end of this section and provide a full classification in appendix~\ref{A:ClassBulkThreeCycles}.

\mathtabfix{
\begin{array}{|c||c|c|}\hline
\multicolumn{3}{|c|}{\text{\bf Global bulk consistency conditions on } \; T^6/(\Z_2 \times \Z_6 \times \OR) \; \text{\bf with discrete torsion}\, (\eta = -1)}
\\\hline\hline
\text{\bf lattice} & \text{\bf Bulk RR tadpole cancellation} &  \text{\bf SUSY conditions}  
\\\hline\hline
{\bf a/bAA}& \begin{array}{l} 
\sum_a N_a \left( 2 \, P_a + Q_a \right) =   8 \, \left(\eta_{\OR} + 3 \,     \eta_{\OR\Z_2^{(1)}}\right)  \\
     - \sum_a N_a \frac{V_a + b \, Q_a}{1-b} =  8  \, \left(\eta_{\OR\Z_2^{(2)}} + \eta_{\OR\Z_2^{(3)}}\right)
    \end{array}
    & 
   \begin{array}{ll} 
   \text{\bf necessary:}& 3 Q_a + \varrho \, [ 2 U_a + V_a + b (2 P_a + Q_a)  ] =0 \\
    \text{\bf sufficient:}& 2 P_a + Q_a - \varrho \, [V_a + b Q_a] > 0 
    \end{array} \\
\hline 
{\bf a/bAB}& 
\begin{array}{l} 
 \sum_a N_a \left(P_a + Q_a  \right) =  8  \left(\eta_{\OR} + \eta_{\OR\Z_2^{(1)}}\right) \\
     \sum_a N_a \frac{U_a - V_a +b\, (P_a - Q_a)}{1-b}= 8 \,  \left( \eta_{\OR\Z_2^{(2)}} + 3     \, \eta_{\OR\Z_2^{(3)}}\right)
    \end{array}
& 
\begin{array}{ll}
\text{\bf necessary:}& Q_a - P_a + \varrho \, [ U_a + V_a + b(P_a + Q_a)] = 0   \\
  \text{\bf sufficient:}& 3 (P_a + Q_a ) + \varrho \, [U_a - V_a + b (P_a - Q_a)] >0
\end{array}
\\
\hline
{\bf a/bBB}& 
\begin{array}{l}
\sum_a N_a \left(P_a + 2 \, Q_a  \right) = 8 \left( 3\, \eta_{\OR} +    \eta_{\OR\Z_2^{(1)}} \right)\\
 \sum_a N_a \frac{ U_a + b \, P_a}{1-b} =8 \,  \left(\eta_{\OR\Z_2^{(2)}} + \eta_{\OR\Z_2^{(3)}}\right) 
\end{array}
&
\begin{array}{ll}
  \text{\bf necessary:}& -3 P_a + \varrho \, [U_a + 2 V_a + b (P_a+2 Q_a)] = 0  \\
   \text{\bf sufficient:} & P_a + 2 Q_a + \varrho \, (U_a + b P_a) >0 
 \end{array}\\
\hline
\end{array}
}{Bulk-RR+SUSY-Z2Z6}{Model building constraints for the bulk part of fractional D6-branes defined in equation~(\protect\ref{Eq:Z2Z6FractCycles}), as first derived in~\cite{Forste:2010gw}.
The structure of bulk RR tadpole and supersymmetry conditions supports the conjectured equivalence of lattices in table~\protect\ref{tab:ClosedSpectrum_Z2Z6} by
the identifications in equation~(\protect\ref{Eq:bulk-identifications}).
}

By combining the bulk supersymmetry conditions with the bulk RR tadpole cancellation conditions, we can on the one hand immediately exclude some choices of exotic O6-planes for supersymmetric D6-brane model building, such as any choice of three exotic O6-planes or the particular choice $\eta_{\OR\Z_2^{(1)}}=-1$ of one exotic O6-plane on {\bf a/bAA}.
On the other hand, we find identifications between the bulk wrapping numbers and complex structure parameters $\varrho$ on the {\bf a/bAA} and {\bf a/bAB} and {\bf a/bBB} lattices, which relate physically identical vacua,
\begin{equation}\label{Eq:bulk-identifications}
\left(\begin{array}{c}  \eta_{\OR\Z_2^{(1)}} \\ \eta_{\OR} \\\hline \varrho  \\\hline 2 \, P_a + Q_a \\ -\frac{\tilde{V}_a}{1-b} \\ Q_a \\ \frac{2\, \tilde{U}_a + \tilde{V}_a}{1-b} \end{array}\right)_{\bf a/bAA}
\longleftrightarrow \quad
\left(\begin{array}{c}  \eta_{\OR\Z_2^{(3)}}  \\ \eta_{\OR\Z_2^{(2)}}\\\hline \frac{3}{\varrho \, (1-b)^2} \\\hline \frac{\tilde{U}_a - \tilde{V}_a }{1-b} \\ P_a + Q_a \\ \frac{\tilde{U}_a + \tilde{V}_a}{1-b} \\ Q_a - P_a \end{array}\right)_{\bf a/bAB}
\longleftrightarrow \quad 
\left(\begin{array}{c} \eta_{\OR} \\ \eta_{\OR\Z_2^{(1)}} \\\hline \varrho  \\\hline P_a + 2 \, Q_a\\ \frac{\tilde{U}_a}{1-b} \\ -P_a \\ \frac{\tilde{U}_a + 2\, \tilde{V}_a}{1-b} \end{array}\right)_{\bf a/bBB}
,
\end{equation}
where we used the standard abbreviation $\tilde{m}^1_a \equiv m^1_a + b \, n^1_a$ and analogously for $\tilde{U}_a$, $\tilde{V}_a$ with the definitions~(\ref{Eq:BulkWrappingNumbers}) inserted. 
Observe that the identification of the sufficient supersymmetry conditions in table~\ref{tab:Bulk-RR+SUSY-Z2Z6} implies that the normalized three-cycle volume and thus tree-level value of the gauge coupling remains unchanged,
\begin{equation}\label{Eq:g_at_tree}
\frac{4\pi}{g^2_a} \propto \frac{\text{Vol}_a}{\sqrt{\text{Vol}_6}} = \sqrt{\frac{4}{\sqrt{3} \varrho} \Bigl( [P_a^2 + P_aQ_a + Q_a^2] + \frac{\varrho}{3} [\tilde{U}_a^2 +\tilde{U}_a\tilde{V}_a + \tilde{V}_a^2] \Bigr) }
\stackrel{\text{SUSY}}{=} \left\{\begin{array}{cc}
\frac{2 P_a + Q_a - \varrho \, \tilde{V}_a}{\sqrt{\sqrt{3}  \varrho}} & {\bf a/bAA} \\
\frac{3 (P_a + Q_a ) + \varrho \, [ \tilde{U}_a - \tilde{V}_a ]}{\sqrt{\sqrt{3}  \varrho}}
& {\bf a/bAB}\\
\frac{P_a + 2 Q_a + \varrho \, \tilde{U}_a}{\sqrt{\sqrt{3}  \varrho}} & {\bf a/bBB} 
\end{array}\right.
,
\end{equation}
where the relations in appendix~B.2 of~\cite{Honecker:2011sm} have been used.

To obtain a unique identification at the level of toroidal one-cycle wrapping numbers $(n^i,m^i)$ as well as the correct permutation of the remaining two O6-plane orbits, the bulk constraints need to be confronted with the exceptional RR tadpole cancellation conditions. This will be done in section~\ref{Sss:FractionalCycles}.

The {\it naive} maximally allowed rank of the gauge group for any {\it globally} defined supersymmetric D6-brane model can now be obtained as follows: using the definitions~(\ref{Eq:BulkWrappingNumbers}),
we can manipulate the bulk supersymmetry conditions in table~\ref{tab:Bulk-RR+SUSY-Z2Z6} and obtain e.g. for the {\bf a/bAA} torus $2P_a + Q_a \geqslant 0$ and $- (V_a + b Q_a) \geqslant 0$. As a consequence, we can simply add the right hand sides of 
the bulk RR tadpole cancellation conditions in table~\ref{tab:Bulk-RR+SUSY-Z2Z6} for {\it one} exotic O6-plane,
\begin{equation}
\text{max. SUSY rank} \leqslant 32 \text{ for }  \left\{ \begin{array}{cc} {\bf a/bAA}: & \eta_{\OR\Z_2^{(1)}}= 1 \\ {\bf a/bAB}: &  \eta_{\OR\Z_2^{(3)}}= 1\\ {\bf a/bBB}: &   \eta_{\OR}= 1
 \end{array}\right.
 ,
 \text{ and no SUSY solution otherwise.}
\end{equation}
In particular, also {\it three} exotic O6-planes are excluded by bulk supersymmetry of D6-branes.

When classifying supersymmetric bulk three-cycles for D6-brane model building, similar considerations can be used to impose a strict upper bound on the bulk wrapping numbers and by the definitions~(\ref{Eq:BulkWrappingNumbers}) also on the toroidal ones.  For example, the bulk RR tadpole cancellation conditions on the {\bf a/bAA} lattice imply
$0 \leqslant \mathfrak{X}_a \leqslant 16$ for $\mathfrak{X}_a \in \{n^1_a, \frac{\tilde{m}^1_a}{1-b}, (2\, X_a + Y_a), -Y_a   \}$ for $\eta_{\OR}=-1$. In any search for the Standard Model of particle physics, at least the minimal gauge group $U(3)_a \times USp(2)_b \times U(1)_c$ has to occur, which reduces the upper bound  for $\eta_{\OR}=-1$ to $0 \leqslant \mathfrak{X}_a \leqslant 4$, 
$0 \leqslant \mathfrak{X}_{b,c} \leqslant 12$.
The other allowed choices $(\eta_{\OR\Z_2^{(2)}},\eta_{\OR\Z_2^{(3)}}) \in \{(1,-1),(-1,1)\}$ of the exotic O6-plane on the {\bf a/bAA} lattice lead to the constraints $0 \leqslant \mathfrak{Y}_a \leqslant 32$ for $ \mathfrak{Y}_a \in \{n^1_a,(2\, X_a + Y_a)\}$ and $\tilde{m}^1_a=0$ or $Y_a=0$, which is again further limited by imposing the embedding of the Standard Model gauge group, e.g. $0 \leqslant \mathfrak{Y}_a \leqslant 10$ for the {\it QCD} stack.

Before turning to the contributions from exceptional three-cycles, let us  briefly summarise how to avoid multiple counting of orbifold  or orientifold images of the toroidal (pairwise co-prime) wrapping numbers $(n^i_a,m^i_a)$
for rational values of the  complex structure modulus $\varrho$:
 \begin{itemize}
\item[1.] $(n^3_a, m^3_a) = $ (odd, odd) selects one orbifold image $(\omega^k \,  a)_{k \in \{0,1,2\}}$ under the $\omega$ action (cf. equation~\eqref{Eq:1-cycle-orbits}), and $n^3_a >0$ fixes the orientation of the one-cycle along the two-torus $T^2_{(3)}$, excluding simultaneous orientation flips by angles $\pi$ on $T_{(2)}^2 \times T_{(3)}^2$. 
\item[2.] Choosing the angle $ 0\leq \pi \phi_a^{(1)} \leq \frac{\pi}{2}$ on $T^2_{(1)}$ singles out a D6-brane compared to its orientifold image. This condition amounts to ensuring that the torus wrapping numbers satisfy the condition $(n^1_a, \tilde m^1_a ) \in \{ (\frac{1}{1-b},0), (0,1), (n^1_a>0, \tilde m^1_a >0) \}$.
\end{itemize} 
By these two conditions, the toroidal wrapping numbers $(n^2_a,m^2_a)$ are  uniquely fixed when the bulk supersymmetry conditions of table~\ref{tab:Bulk-RR+SUSY-Z2Z6} are imposed.

Take for example the {\bf a/bAA} lattice configuration:
\begin{itemize}
\item[(i)] $(n^1_a, \tilde m^1_a ) = (\frac{1}{1-b},0)$  implies the SUSY conditions: $Y_a = 0$ and $X_a >0$, 
\item[(ii)] $(n^1_a, \tilde m^1_a ) = (0,1)$ implies the SUSY conditions: $-Y_a = 2 X_a >0$,
\item[(iii)] $n^1_a > 0, \tilde m^1_a >0$ implies the SUSY conditions: $Y_a = -\frac{\varrho}{3} \frac{\tilde m_a^1}{n^1_a} [2 X_a + Y_a]$ and $2X_a + Y_a >0$.  
\end{itemize}
The conditions on $X_a$ and $Y_a$ then allow to classify all torus wrapping numbers $(n_a^2, m_a^2; n_a^3, m_a^3)$ satisfying these constraints. 
 In appendix~\ref{A:ClassBulkThreeCycles}, the full classification is performed for the {\bf aAA} and {\bf bAA} lattices.

\subsubsection{Exceptional three-cycles}\label{Sss:Ex-3-cycles}

As encoded in the  Hodge numbers  of the orbifold $T^6/(\Z_2 \times \Z_6)$ with discrete torsion in equation~(\ref{Eq:HodgeNumbersZ2Z6}), besides the four basic bulk three-cycles there exist  $b_3^{\Z_2} = 2 h_{21}^{\Z_2} = 2 (6 + 2 \times 4) = 28$
exceptional three-cycles, which arise as tensor products of exceptional divisors at  $\Z_2^{(k)}$ singularities along $T^4_{(k)} \equiv T^2_{(i)} \times T^2_{(j)}$ with some one-cycle along $T^2_{(k)}$ inherited from the underlying two-torus. In addition, there exist $b_3^{\Z_6 + \Z_3}= 2 \cdot (2+2)$ exceptional three-cycles located at fixed points of $\omega$ and $\omega^2$ along $T^4_{(1)}$.
Their existence must be taken into account when determining the uni-modular basis of the full lattice of three-cycles, but for model building purposes only bulk three-cycles and exceptional three-cycles from $\Z_2^{(k)}$ twisted sectors as well as fractional combinations thereof will be used. The reason is that exceptional three-cycles from $\Z_{N \neq 2}$ twisted sectors cannot be described by the standard Conformal Field Theory tools for Type II/$\OR$ orientifolds developed in~\cite{Blumenhagen:1999md,Blumenhagen:1999ev,Forste:2000hx}, since only $\unity$ and $\Z_2$ projector insertions in the open-string loop-channel annulus amplitude produce non-vanishing contributions~\cite{Blumenhagen:2002wn,Blumenhagen:2006ci}.

For each $\Z_2^{(k)}$ twisted sector, the orbifold singularities correspond locally to $\C^2_{(k)}/\Z_2^{(k)}$-type singularities ($k \in \{1,2,3 \}$), which support exceptional divisors $e^{(k)}_{\kappa_1 \lambda_1}$ with label $\kappa_i, \lambda_i$ denoting the fixed points along the four-torus $T^4_{(k)}$, more explicitly $\kappa_i, \lambda_i \in \{1,4,5,6 \}$ for the $\Z_2^{(1)}$ sector, $\kappa_i \in \{ 1,2,3,4 \}$ and $\lambda_i \in \{1,4,5,6 \}$ for the $\Z_2^{(2)}$ and $\Z_2^{(3)}$ sectors. The intersection numbers are given by $e^{(k)}_{\kappa_1 \lambda_1} \circ e^{(l)}_{\kappa_2 \lambda_2} = - 2\, \delta^{kl} \delta_{\kappa_1 \kappa_2} \delta_{\lambda_1 \lambda_2}$. A basis of exceptional three-cycles at $\Z_2$ fixed points on $T^6/(\Z_2 \times \Z_6)$ can be constructed by taking the $\Z_6$-invariant orbits of such exceptional divisors tensored with toroidal one-cycles $\pi_{2k-1}$ or $\pi_{2k}$ on the $\Z_2^{(k)}$-invariant two-torus $T^2_{(k)}$. 
At this level, the differences between the $\Z_2^{(1)}$ sector on the one hand and the $\Z_2^{(2)}$ and $\Z_2^{(3)}$ sectors on the other hand become best apparent through two explicit examples. In the $\Z_2^{(1)}$ twisted sector, one of the six basic exceptional three-cycles has the form
\begin{equation}
\sum_{k=0}^1 \sum_{l=0}^2 \theta^k \omega^l \left( e^{(1)}_{44} \otimes \pi_1  \right) = 2 \left[ e^{(1)}_{44} + e^{(1)}_{56}  + e^{(1)}_{65}    \right] \otimes \pi_1,
\end{equation}
while any basic exceptional three-cycle in the $\Z_2^{(2)}$ twisted sector takes the form for $\kappa~\in~\{1,2,3,4\}$:
\begin{equation}
\sum_{k=0}^1 \sum_{l=0}^2 \theta^k\omega^l \left( e^{(2)}_{\kappa 4} \otimes \pi_3 \right) = e^{(2)}_{\kappa 4} \otimes \pi_3 + e^{(2)}_{\kappa 6} \otimes (- \pi_4) + e^{(2)}_{\kappa 5} \otimes (\pi_4 - \pi_3).
\end{equation}
The distinction between the $\Z_2^{(1)}$ twisted sector and the other two twisted sectors can be traced back to the fact the $\Z_2^{(1)}$ factor forms a subsector of the $\Z_6$ symmetry, while the other two $\Z_2$ twisted sectors 
feel the $\Z_6$ symmetry along one two-torus as permutation of $\Z_2$ fixed points and along the other as permutation of toroidal one-cycles.  
In~\cite{Forste:2010gw} the full basis of $\Z_2^{(k)}$ exceptional three-cycles was constructed in detail with intersection form:
\begin{equation}
\begin{array}{lll}
\varepsilon^{(1)}_0 \circ \tilde \varepsilon^{(1)}_0 = -12, & \varepsilon^{(1)}_\alpha \circ \tilde \varepsilon^{(1)}_\beta= - 4\, \delta_{\alpha \beta}, & \alpha,\beta \in \{1,2,3,4,5\},\\ 
\varepsilon^{(l)}_\alpha \circ \tilde \varepsilon^{(l)}_\beta =  - 4\, \delta_{\alpha \beta} & \text{with }\,  l  = 2,3 & \alpha, \beta \in \{1,2,3,4\} .
\end{array}
\end{equation}
The full basis of exceptional three-cycles is summarised in the form of `fixed point orbits' for all three $\Z_2$ twisted sectors in table~\ref{tab:Z2Z6ExceptionalCycles}.
\mathtabfix{
\begin{array}{|c|c||c|c|}
\hline \multicolumn{4}{|c|}{\Z_2^{(k)} \;  \text{\bf fixed points and exceptional 3-cycles on}\, T^6/(\Z_2 \times \Z_6)\, \text{\bf with discrete torsion }\, (\eta = -1)}\\
\hline
\hline
\multicolumn{2}{|c||}{\Z_2^{(1)}\, \text{\bf twisted sector}} & \multicolumn{2}{|c|}{\Z_2^{(l)}\, \text{\bf twisted sector with }\; l=2,3} \\
\hline
\hline
{\rm f.p.}^{(1)} \otimes (n^1 \pi_{1} + m^1 \pi_{2})\!\! & {\rm orbit} & {\rm f.p.}^{(l)} \otimes (n^{l} \pi_{2l-1} + m^{l} \pi_{2l})\!\! & {\rm orbit} \\
\hline
\hline
11 & n^1 \varepsilon^{(1)}_0 + m^1 \tilde{\varepsilon}^{(1)}_0 & \kappa 1 & - \\
\hline
41, \; 51, \; 61 
 & n^1 \varepsilon^{(1)}_1 + m^1 \tilde{\varepsilon}^{(1)}_1 & \kappa 4 & n^{l}  \varepsilon^{(l)}_{\kappa} + m^{l} \tilde{\varepsilon}^{(l)}_{\kappa} \\
 \hline
 14, \; 15, \; 16  & n^1 \varepsilon^{(1)}_2 + m^1 \tilde{\varepsilon}^{(1)}_2 & \kappa 5 & m^{l}  \varepsilon^{(l)}_{\kappa} -(n^{l} + m^{l}) \tilde{\varepsilon}^{(l)}_{\kappa} \\
 \hline
 44, \; 56, \; 65 &  n^1 \varepsilon^{(1)}_3 + m^1 \tilde{\varepsilon}^{(1)}_3 & \kappa 6 &  -(n^{l} +m^{l}) \varepsilon^{(l)}_{\kappa}  + n^{l} \tilde{\varepsilon}^{(l)}_{\kappa} \\
 \hline
 45, \; 54, \; 66 &  n^1 \varepsilon^{(1)}_4 + m^1 \tilde{\varepsilon}^{(1)}_4 & \muc{2}{|c|}{} \\
 \cline{1-2}
46, \; 55,  \; 64 &  n^1 \varepsilon^{(1)}_5 + m^1 \tilde{\varepsilon}^{(1)}_5 & \muc{2}{|c|}{}  \\
\hline
\end{array}
}{Z2Z6ExceptionalCycles}{Complete list of $\Z_6$-invariant orbits for exceptional divisors $e_{\kappa\lambda}^{(k)}$ at $\Z_2^{(k)}$ fixed points tensored with one-cycles $n^{k} \pi_{2k-1} + m^{k} \pi_{2k}$ along the $\Z_2^{(k)}$-invariant two-torus $T^2_{(k)}$, as first computed in~\cite{Forste:2010gw}.}

Also the exceptional three-cycles transform non-trivially under the $\OR$-projection, as $\OR$ acts in the $\Z_2^{(k)}$ sector on the one-cycles along $T^2_{(k)}$ by permutations and on the exceptional divisors at the $\C^2_{(k)}/\Z_2^{(k)}$ singularities according to, 
\begin{equation}
\OR: e_{\kappa\lambda}^{(k)} \rightarrow - \eta_{(k)} \; e_{\kappa' \lambda'}^{(k)},
\end{equation}
where $\kappa', \lambda'$ represent the ${\cal R}$-images of the $\Z_2^{(k)}$ fixed points $\kappa, \lambda$ as explained in the caption of figure~\ref{Fig:LatticesZ2Z6}. The orientifold projection depends on the choice of the exotic O6-plane through the sign-factor $\eta_{(k)}$ defined in equation~(\ref{Eq:eta-for-Z2s}). See table~\ref{tab:Z2Z6OrientifoldExceptionalCycles} for the complete overview of the orientifold action on the basis of exceptional three-cycles for all $\Z_2$ sectors and any choice of lattice orientation.

\mathtabfix{
\begin{array}{|c||c|c|c|c||c|c|c|c|}
\hline
\multicolumn{9}{|c|}{\text{\bf Orientifold images of exceptional three-cycles on}\, T^6/(\Z_2 \times \Z_6\times \OR) \, \text{\bf with discrete torsion } \, (\eta = -1)  }\\
\hline
\hline
& \multicolumn{4}{|c||}{ \Z_2^{(1)} \text{\bf twisted sector}} &  \multicolumn{4}{|c|}{ \Z_2^{(k)} \text{\bf twisted sector with}\; l = 2,3} \\
\hline
\hline
\text{3-cycle} & \OR(\varepsilon^{(1)}_{\alpha}) & \OR(\tilde{\varepsilon}^{(1)}_{\alpha}) & \alpha=\alpha' & \alpha \leftrightarrow \alpha' &  \OR(\varepsilon^{(l)}_{\alpha}) & \OR(\tilde{\varepsilon}^{(l)}_{\alpha}) & \alpha=\alpha' & \alpha \leftrightarrow \alpha' \\
\hline
\hline
\begin{array}{c}
{\bf a/bAA} \\\hline {\bf a/bAB} \\\hline {\bf a/bBB} 
\end{array} 
&  \eta_{(1)} \left( -\varepsilon^{(1)}_{\alpha'} + (2b) \tilde{\varepsilon}^{(1)}_{\alpha'} \right)
& \eta_{(1)} \, \tilde{\varepsilon}^{(1)}_{\alpha'}
&\begin{array}{c}
0,1,2,3 \\\hline 0,1,2,5 \\\hline 0,1,2,4  
\end{array} 
&\begin{array}{c}
4,5 \\\hline 3,4  \\\hline  3,5 
\end{array} 
&
\begin{array}{c} - \eta_{(l)} \, \varepsilon^{(l)}_{\alpha'} \\\hline
(-)^l \, \eta_{(l)} \,  \tilde{\varepsilon}^{(l)}_{\alpha'} \\\hline
\eta_{(l)} \left( \tilde{\varepsilon}^{(l)}_{\alpha'} -\varepsilon^{(l)}_{\alpha'} \right)
\end{array}
&\begin{array}{c} \eta_{(l)} \left( \tilde{\varepsilon}^{(l)}_{\alpha'} -\varepsilon^{(l)}_{\alpha'} \right)\\\hline
(-)^l\,  \eta_{(l)} \, \varepsilon^{(l)}_{\alpha'} \\\hline
\eta_{(l)} \,  \tilde{\varepsilon}^{(l)}_{\alpha'}
\\\end{array}
& 1,4
& 2+2b, 3-2b\\
\hline
\end{array}
}{Z2Z6OrientifoldExceptionalCycles}{Orientifold images of the $\Z_2^{(k)}$ exceptional three-cycles, depending on the background lattice and the choice of the exotic O6-plane orbit with sign factor $\eta_{(k)} \equiv \eta_{\OR} \eta_{\OR\Z_2^{(k)}}$.}

\subsubsection{Fractional three-cycles}\label{Sss:FractionalCycles}

Supersymmetric fractional three-cycles can now be constructed as linear combinations of bulk and exceptional three-cycles,
\begin{equation}\label{Eq:Z2Z6FractCycles}
\mbox{\resizebox{0.92\textwidth}{!}{
$
\begin{aligned}\hspace{-4mm}
\Pi^{\rm frac}_a =&  \frac{1}{4} \Pi^{\rm bulk}_a  + \frac{1}{4} \sum_{i=1}^3 \Pi^{\Z_2^{(i)}}_a\\
=& \frac{1}{4} \left( P_a \rho_1 + Q_a  \rho_2 + U_a \rho_3 + V_a \rho_4 \right) + \frac{1}{4} \sum_{\alpha=0}^5 \left( x^{(1)}_{\alpha, a} \,
    \varepsilon_\alpha^{(1)} + y ^{(1)}_{\alpha, a} \, \tilde{\varepsilon}_\alpha^{(1)}
  \right) 
+ \frac{1}{4} \sum_{l=2,3} \sum_{\alpha=1}^4 \left( x^{(l)}_{\alpha, a} \, \varepsilon_\alpha^{(l)} + y ^{(l)}_{\alpha, a} \, \tilde{\varepsilon}_\alpha^{(l)} \right),
\end{aligned}
$}}
\end{equation}
with the bulk wrapping numbers defined in equation~(\ref{Eq:BulkWrappingNumbers}) and the exceptional wrapping numbers $(x^{(k)}_{\alpha, a}, y^{(k)}_{\alpha, a})$ defined as linear combinations of the torus wrapping numbers
$(n^k_a,m^k_a)$ of the one-cycle along the $\Z_2^{(k)}$-invariant two-torus $T^2_{(k)}$ dressed with additional discrete parameters:
\begin{itemize}
\item \vspace{-0.1in} three discrete displacement parameters $\vec{\sigma}_a$ parameterising whether the bulk cycle passes through the origin ($\sigma_a^i = 0$) or is shifted by one half of a lattice vector ($\sigma_a^i = 1$) on the two-torus $T^2_{(i)}$, i.e. for which $\alpha$ one has $(x^{(l)}_{\alpha,a} \, , \, y^{(l)}_{\alpha,a} ) \neq (0,0)$,
\item \vspace{-0.2in}  three $\Z_2^{(k)}$ eigenvalues $(-)^{\tau_a^{\Z_2^{(k)}}}$, which can naively be viewed as 
 the exceptional part encircling a reference fixed point on the four-torus  $T^{4}_{(k)}$ `clockwise' ($\tau_a^{\Z_2^{(k)}}=0$) or `counter-clockwise' ($\tau_a^{\Z_2^{(k)}}=1$). Due to the relation $(-)^{\tau_a^{\Z_2^{(3)}}} = (-)^{\tau_a^{\Z_2^{(1)}}+ \tau_a^{\Z_2^{(2)}}}$, only two $\Z_2^{(k)}$ eigenvalues are independent,
\item \vspace{-0.1in} three discrete Wilson lines $\vec{\tau}_a$, which can naively be viewed as  
encoding how the exceptional cycle encircles another fixed point on $T^4_{(k)}$ with respect to the reference point: same orientation ($\tau_a^i = 0$) or opposite orientation ($\tau_a^i = 1$) around the $\Z_2^{(k)}$ fixed point per two-torus $T_{(i)}^2$.
\end{itemize}
A schematic overview of the form of exceptional wrapping numbers $(x^{(k)}_{\alpha, a}, y^{(k)}_{\alpha, a})$ is given in table~\ref{tab:Z2Z6ExceptionalWrappingNumbers}, taking into account the amendments introduced in section 2.1.2 of~\cite{Honecker:2012qr} compared to~\cite{Forste:2010gw}, such that the assignment of pre-factors follows the conventions presented in table~\ref{tab:Z2Z6SignAssignment}. The conventions for the two-tori on which the $\Z_6$ symmetry acts, i.e.~$T^2_{(2)}$ and $T^2_{(3)}$, are the same as in~\cite{Honecker:2012qr}. The invariance of the first two-torus under $\Z_6$ leaves an ambiguity when fixing the order of the fixed points on $T_{(1)}^2$, based on the requirement that the $\Z_6$-invariant orbit $\Pi_a^{\Z_2^{(1)}}$ is the same for all orbifold images $(\omega^k \,  a)_{k\in \{0,1,2\}}$. The only consistency check is provided by ensuring that the $\OR$-image of an exceptional cycle $\Pi_a^{\Z_2^{(1)}}$ upon using table~\ref{tab:Z2Z6OrientifoldExceptionalCycles} is equivalent to the exceptional cycle $\Pi_{a'}^{\Z_2^{(1)}}$ on the {\bf b}-type lattice, where the orbit $a'$ corresponds to the orientifold image of the bulk orbit $a$.

\mathtabfix{
\begin{array}{|c|c||c|c|}
\hline \multicolumn{4}{|c|}{\text{\bf Exceptional wrapping numbers } (x^{(k)}_{\alpha, a}, y^{(k)}_{\alpha, a})\, \text{\bf on $T^6/(\Z_2 \times \Z_6 \times \OR)$ in terms of torus wrapping numbers } (n_a^i, m_a^i) }\\
\hline
\hline
\multicolumn{2}{|c||}{\Z_2^{(1)} \text{\bf twisted sector}} & \multicolumn{2}{|c|}{\Z_2^{(l)} \text{\bf twisted sector with } l=2,3 }\\
\hline
\hline
{\rm I.} & {\rm II.} & {\rm I.} & {\rm II.}\\
\hline
\hline
 ( z_{\alpha, a}^{(1)}\, n^1_a, z_{\alpha, a}^{(1)}\, m^1_a) &   ( \hat z_{\alpha, a}^{(1)}\, n^1_a, \hat z_{\alpha, a}^{(1)}\, m^1_a) & 
 \begin{array}{c}
 (\zeta^{(l)}_{\alpha,a} \; n^l_a \;, \; \zeta^{(l)}_{\alpha,a} \, m^l_a) \\
 (\zeta^{(l)}_{\alpha,a} \; m^l_a \;, \; -\zeta^{(l)}_{\alpha,a} \; (n^l_a+m^l_a) )\\
  (-\zeta^{(l)}_{\alpha,a} \; (n^l_a+m^l_a)  \; , \; \zeta^{(l)}_{\alpha,a} \; n^l_a)
 \end{array}
 &
  \begin{array}{c}
  \left(-\zeta^{(l)}_{\alpha,a} \;  n^l_a+ (\hat \zeta^{(l)}_{\alpha,a}-\zeta^{(l)}_{\alpha,a}) \, m^l_a \; , \, (\zeta^{(l)}_{\alpha,a}-\hat \zeta^{(l)}_{\alpha,a}) \; n^l_a -\hat \zeta^{(l)}_{\alpha,a} \, m^l_a \right)\\
 \left( (\zeta^{(l)}_{\alpha,a}-\hat \zeta^{(l)}_{\alpha,a}) \; n^l_a -\hat \zeta^{(l)}_{\alpha,a} \; m^l \; , \;  \zeta^{(l)}_{\alpha,a}\; m^l_a+ \hat\zeta^{(l)}_{\alpha,a} \; n^l_a \right)
 \\
 \left( \hat\zeta^{(l)}_{\alpha,a} \; n^l_a+\zeta^{(l)}_{\alpha,a} \; m^l_a \; , \; -\zeta^{(l)}_{\alpha,a}\; n^l_a +(\hat \zeta^{(l)}_{\alpha,a} -\zeta^{(l)}_{\alpha,a} ) \; m^l_a \right)
 \end{array}
 \\
 \hline
\end{array}
}{Z2Z6ExceptionalWrappingNumbers}{The exceptional wrapping numbers of type I stem from a single contribution of an orbit in table~\ref{tab:Z2Z6ExceptionalCycles}, while those of type II result from an orbit contributing twice
due to two different $\Z_2$ fixed points on $T^2_{(2)} \times T^2_{(3)}$. Clarifications about the form of the pre-factors $z_{\alpha, a}^{(1)}, \zeta^{(l)}_{\alpha,a},  \hat \zeta^{(l)}_{\alpha,a} \in \{\pm 1\}$ and $\hat z_{\alpha, a}^{(1)} \in \{0,\pm 2\}$,  are presented in the main text.}

\mathtabfix{
\begin{array}{|c||ccc||ccc|ccc|}
\hline \multicolumn{10}{|c|}{\text{\bf Assignment of prefactors $(-1)^{\tau^{\Z_2^{(i)}}_a}$ \!\!\!\!\! or $(-1)^{\tau^{\Z_2^{(i)}}_a \!\!\!+ \tau^i_a}$ }}\\
\hline
\hline
&\multicolumn{3}{|c||}{\text{\bf Assignment on }\, T_{(1)}^2} & \multicolumn{3}{|c|}{\text{\bf Assignment on }\, T_{(2)}^2 }& \multicolumn{3}{|c|}{\text{\bf Assignment on }\, T_{(3)}^2}\\
\hline
\hline
(n^i_a,m^i_a) & \text{(odd,odd)}  & \text{(odd,even)} & \text{(even,odd)} & \text{(odd,odd)} \stackrel{\omega}{\to} & \!\!\!\!\text{(odd,even)} \stackrel{\omega}{\to} & \!\!\!\!\text{(even,odd)}
& \text{(odd,odd)} \stackrel{\omega}{\to} & \!\!\!\!\text{(even,odd)} \stackrel{\omega}{\to} & \!\!\!\!\text{(odd,even)}\\
\hline
\hline
\sigma^i_a=0&\left(\begin{array}{c} 1 \\  3\end{array}\right) \stackrel{\OR_{\bf b}}{\to} &\left(\begin{array}{c} 1 \\  2\end{array}\right) & \left(\begin{array}{c} 1 \\  4\end{array}\right) 
\stackrel{\OR_{\bf b}}{\circlearrowright} 
& 
\left(\begin{array}{c} 1 \\  6\end{array}\right) \to 
& \left(\begin{array}{c} 1 \\ 4 \end{array}\right) \to & \left(\begin{array}{c} 1 \\ 5 \end{array}\right)
&
\left(\begin{array}{c} 1 \\  6\end{array}\right) \to 
& \left(\begin{array}{c} 1 \\ 5 \end{array}\right) \to & \left(\begin{array}{c} 1 \\ 4 \end{array}\right)
\\
\hline
\sigma^i_a=1 & \left(\begin{array}{c} 2 \\ 4\end{array}\right) \stackrel{\OR_{\bf b}}{\to} & \left(\begin{array}{c} 3 \\  4\end{array}\right) & \left(\begin{array}{c} 2 \\  3\end{array}\right)  \updownarrow \OR_{\bf b} & 
\left(\begin{array}{c}  4 \\ 5 \end{array}\right) \to & \left(\begin{array}{c} 5 \\ 6 \end{array}\right)\to  
& \left(\begin{array}{c} 6 \\ 4 \end{array}\right)  
&
\left(\begin{array}{c}  4 \\ 5 \end{array}\right) \to & \left(\begin{array}{c} 6 \\ 4 \end{array}\right)\to  
& \left(\begin{array}{c} 5 \\ 6 \end{array}\right)  
\\
\hline
\end{array}
}{Z2Z6SignAssignment}{Consistent assignment of the reference point (upper entry)  and the second $\Z_2^{(i)}$ fixed point (lower entry) contributing with  sign factor  $(-1)^{\tau^{\Z_2^{(i)}}_a}$ and $(-1)^{\tau^{\Z_2^{(i)}}_a+ \tau^i_a}$, respectively, to $\Pi^{\Z_2^{(j),j\neq i}}_a$. The assignment on the two-tori $T^2_{(l=2,3)}$ is set by ensuring that the $\Z_6$-invariant orbit is independent of the representant $(\omega^k \,  a)_{k\in \{0,1,2\}}$. For the two-torus $T^2_{(1)}$, the only constraint on the fixed point ordering arises from requiring consistency under the $\OR$-projection on the {\bf b}-type lattice.}

In the $\Z_2^{(1)}$ twisted sector, a generic three-cycle $\Pi_a^{\Z_2^{(1)}}$ can receive two different kinds of contributions:
\begin{enumerate}
\item
For $(\sigma^2_a,\sigma^3_a) =(0,0)$ and using the abbreviation ${\cal E}^{(n^1,m^1)}_{\alpha} \equiv n^1_a \varepsilon^{(1)}_{\alpha} + m^1_a \tilde{\varepsilon}^{(1)}_{\alpha}$, the cycle can be written as
\begin{equation}
\Pi_a^{\Z_2^{(1)}} = (-1)^{\tau^{\Z_2^{(1)}}_a} \hspace{-2mm} \left({\cal E}^{(n^1_a,m^1_a)}_0 
+ (-1)^{\tau_a^2} \, {\cal E}^{(n^1_a,m^1_a)}_1
+ (-1)^{\tau_a^3} \, {\cal E}^{(n^1_a,m^1_a)}_2
+ (-1)^{\tau_a^2 + \tau_a^3} \, {\cal E}^{(n^1_a,m^1_a)}_{\beta}
\right)
\end{equation}
for one value $\beta \in \{3,4,5\}$, which depends on the even-/oddness of the torus wrapping numbers $(n^2_a, m^2_a)$ for fixed choice $(n^3_a, m^3_a)=(\text{odd},\text{odd})$. Each fixed point contribution belongs to class I, and the prefactors in table~\ref{tab:Z2Z6ExceptionalWrappingNumbers} are 
\linebreak
 $z^{(1)}_{\alpha, a} =(-1)^{\tau^{\Z_2^{(1)}}_a} \hspace{-2mm}, (-1)^{\tau^{\Z_2^{(1)}}_a  \hspace{-1mm}+ \tau_a^2 },  (-1)^{\tau^{\Z_2^{(1)}}_a  \hspace{-1mm}+ \tau_a^3 }, (-1)^{\tau^{\Z_2^{(1)}}_a  \hspace{-1mm}+ \tau_a^2 + \tau_a^3}$ for $\alpha=0,1,2,\beta$, respectively. 
\item
For $(\sigma^2_a,\sigma^3_a) \neq (0,0)$, any cycle has contributions from three different fixed point orbits, two of them corresponding to class I and the third one to class II.
For example, for $(\sigma^2_a,\sigma^3_a) =(1,0)$, $\alpha=1$ has two contributions leading to \mbox{$\hat z^{(1)}_{1, a} =(-1)^{\tau^{\Z_2^{(1)}}_a} \hspace{-3mm}\bigl[1 + (-1)^{\tau_a^2} \bigr]$}.
The other two fixed points lead to contributions of class I with $z^{(1)}_{\alpha, a} \in \{ (-1)^{\tau^{\Z_2^{(1)}}_a \hspace{-2mm}+ \tau_a^3 }, (-1)^{\tau^{\Z_2^{(1)}}_a  \hspace{-2mm}+ \tau_a^2 + \tau_a^3}\}$ for two values of $\alpha \in \{3,4,5\}$,
which depend on the even-/oddness of $(n^2_a, m^2_a)$ for fixed choice of orbit representant $(n^3_a, m^3_a)=(\text{odd},\text{odd})$.
\\
For $(\sigma^2_a,\sigma^3_a)=(0,1)$, the same reasoning applies for $\alpha=2$ contributing with class II and $\tau_a^2 \leftrightarrow \tau_a^3$, while for $(\sigma^2_a,\sigma^3_a)=(1,1)$ all three orbit labels are in the range 
$\alpha \in \{3,4,5\}$.
\end{enumerate}

In the $\Z_2^{(2)}$ and $\Z_2^{(3)}$ twisted sector, the structure of the exceptional wrapping numbers is identical and can be explained simultaneously by exchanging two-torus labels $2\leftrightarrow 3$. For concreteness, we will discuss the exceptional wrapping numbers in the $\Z_2^{(2)}$ twisted sector. 
A generic three-cycle $\Pi_a^{\Z_2^{(2)}}$ receives contributions from two separate  $\Z_2^{(2)}$ twisted orbits. The value of the displacement parameter $\sigma_a^3$ determines whether these orbits contribute once or twice:
\begin{enumerate}
\item
When the displacement parameter is $\sigma^3_a = 0$,  the three-cycle $\Pi_a^{\Z_2^{(2)}}$ is generated by two $\Z_2^{(2)}$ twisted orbits from table~\ref{tab:Z2Z6ExceptionalCycles} that each contribute only once. For the choice of orbifold representant $(n^3_a, m^3_a)=(\text{odd},\text{odd})$, these correspond to the fixed point orbits of $\kappa 6$ for two values of $\kappa \in \{1,2,3,4\}$ on $T^2_{(1)}$, 
\begin{equation}
\Pi_a^{\Z_2^{(2)}} =(-1)^{\tau^{\Z_2^{(2)}}_a \hspace{-1mm} + \tau^3_a} \Bigl(\hat{\cal E}^{(n^2_a,m^2_a)}_{\kappa_1} + (-1)^{\tau^1_a} \,  \hat{\cal E}^{(n^2_a,m^2_a)}_{\kappa_2} \Bigr)
\quad
\text{with}
\quad
\hat{\cal E}^{(n^2_a,m^2_a)}_{\kappa} \equiv - (n^2_a+ m^2_a) \, \varepsilon^{(2)}_{\kappa} + n^2_a \, \tilde{\varepsilon}^{(2)}_{\kappa}
,
\end{equation}
i.e. twice class I with $\zeta_{\kappa,a}^{(2)} \in \{ (-)^{\tau^{\Z_2^{(2)}}_a \hspace{-2mm} + \tau_a^3},  (-1)^{\tau^{\Z_2^{(2)}}_a \hspace{-2mm} + \tau_a^1+ \tau_a^3}\}$ in the notation of table~\ref{tab:Z2Z6ExceptionalWrappingNumbers}.
\item
In case $\sigma^3_a = 1$, the two different $\Z_2^{(2)}$ twisted orbits will each contribute twice to the three-cycle $\Pi_a^{\Z_2^{(2)}}$, with exceptional wrapping numbers given by one of the three combinations in class II. For $(n^3_a, m^3_a)=(\text{odd},\text{odd})$, the exceptional wrapping numbers take the form $ \left( \hat\zeta^{(2)}_{\kappa,a} \; n^2_a+\zeta^{(2)}_{\kappa,a} \; m^2_a \; , \; -\zeta^{(2)}_{\kappa,a}\; n^2_a +(\hat \zeta^{(2)}_{\kappa,a} -\zeta^{(2)}_{\kappa,a} ) \; m^2_a \right)$ with $\zeta^{(2)}_{\kappa,a}$ as defined above due to the orbits of $\kappa 5$ and $\hat{\zeta}^{(2)}_{\kappa,a} \in \{(-1)^{\tau^{\Z_2^{(2)}}_a} \hspace{-3mm}, (-1)^{\tau^{\Z_2^{(2)}}_a \hspace{-2mm} + \tau_a^1} \}$  for contributions from orbits of $\kappa4$.
\end{enumerate}
Table \ref{tab:exceptional-wrappings-Z2Z6} in appendix~\ref{A:ClassBulkThreeCycles} provides a more detailed overview of the exceptional wrapping numbers per $\Z_2^{(i)}$ twisted sector, depending on the even/oddness of the torus wrapping numbers, the $\Z_2^{(i)}$ eigenvalue $(-)^{\tau_a^{\Z_2^{(i)}}}$, the displacement paremeters $\vec{\sigma}_a$ and the discrete Wilson lines $\vec{\tau}_a$.   

Using the definition of a fractional three-cyle in equation~(\ref{Eq:Z2Z6FractCycles}), the twisted RR tadpole cancellation conditions on the orientifold $T^6/(\Z_2\times\Z_6\times\OR)$ with discrete torsion were first written down in~\cite{Forste:2010gw} in terms of the exceptional wrapping numbers. The twisted RR tadpole cancellation conditions are summarized in table~\ref{tab:Z2Z6TwistedRRTadpoles} per lattice and per twisted sector, with an earlier typo in the $\Z_2^{(l=2,3)}$ twisted sector for $b=0$ and $\alpha=2,3$ corrected here.
\mathtabfix{
\begin{array}{|c||c||c|}
\hline
\multicolumn{3}{|c|}{\text{\bf Twisted RR tadpole cancellation conditions on } T^6/(\Z_2 \times \Z_6 \times \OR)\; \text{\bf with discrete torsion } (\eta = -1)}\\
\hline
\hline
\text{\bf lattice} & \text{\bf $\Z_2^{(1)}$ twisted sector} &  \text{\bf $\Z_2^{(l)}$ twisted sector with } l = 2, 3 \\
\hline
\hline
{\bf a/bAA} & 
\begin{array}{l}
\sum_a N_a (1-\eta_{(1)}) x^{(1)}_{\alpha,a} = 0, \hspace{31mm} \alpha = 0,1,2,3 \\
\sum_a N_a [(1+ \eta_{(1)}) y^{(1)}_{\alpha,a} +
  \eta_{(1)} 2b \, x^{(1)}_{\alpha,a} = 0, \qquad \alpha = 0,1,2,3 \\
  \sum_a N_a (x^{(1)}_{4,a} - \eta_{(1)} x^{(1)}_{5,a}) = 0,\\
  \sum_a N_a [y^{(1)}_{4,a} + \eta_{(1)}y^{(1)}_{5,a} + b \, (x^{(1)}_{4,a} + \eta_{(1)}x^{(1)}_{5,a})] = 0,
\end{array}
& 
\begin{array}{l}
\sum_a N_a  
[(1-\eta_{(l)})x^{(l)}_{\alpha,a} -\eta_{(l)} y^{(l)}_{\alpha,a} ] =0, \qquad \alpha = 1,4 \\
\sum_a N_a  (1+\eta_{(l)})y^{(l)}_{\alpha,a} = 0, \hspace{28mm} \alpha = 1,4\\
 \sum_a N_a [x^{(l)}_{2,a} - \eta_{(l)} x^{(l)}_{2+2b,a}  - \eta_{(l)} y^{(l)}_{2+2b,a})]  = 0,\\
 \sum_a N_a [x^{(l)}_{3,a} - \eta_{(l)} x^{(l)}_{3-2b,a}  - \eta_{(l)} y^{(l)}_{3-2b,a})]  = 0,\\
 \sum_a N_a (y^{(l)}_{3,a} + \eta_{(l)} y^{(l)}_{3-2b,a}) = 0,\\
 \sum_a N_a (y^{(l)}_{2,a} + \eta_{(l)} y^{(l)}_{2+2b,a}) = 0,
\end{array}
\\
\hline
{\bf a/bAB} &
\begin{array}{l}
\sum_a N_a (1-\eta_{(1)}) x^{(1)}_{\alpha,a} = 0,  \hspace{32mm}  \alpha = 0,1,2,5\\
\sum_a N_a [(1+ \eta_{(1)}) y^{(1)}_{\alpha,a} +  \eta_{(1)} 2b \, x^{(1)}_{\alpha,a} ] = 0,  \qquad \alpha = 0,1,2,5 \\
\sum_a N_a(x^{(1)}_{3,a} - \eta_{(1)} x^{(1)}_{4,a}) = 0, \\
\sum_a N_a[y^{(1)}_{3,a} + \eta_{(1)}y^{(1)}_{4,a} + b \, (x^{(1)}_{3,a} + \eta_{(1)}x^{(1)}_{4,a})]  = 0,
\end{array}
 & 
 \begin{array}{l}
 \sum_a N_a  (x^{(l)}_{\alpha,a} +(-1)^{l} \, \eta_{(l)} \, y^{(l)}_{\alpha,a} ) = 0,  \qquad \alpha = 1,4\\
 \sum_a N_a (x^{(l)}_{2,a} + (-1)^{l} \,\eta_{(l)}y^{(l)}_{2+2b,a} ) = 0, \\
 \sum_a N_a (x^{(l)}_{3,a} + (-1)^{l} \,\eta_{(l)}y^{(l)}_{3-2b,a} ) = 0, 
 \end{array}
 \\
\hline
{\bf a/bBB} &
\begin{array}{l}
\sum_a N_a (1-\eta_{(1)}) x^{(1)}_{\alpha,a}  = 0, \hspace{32mm}  \alpha = 0,1,2,4  \\
\sum_a N_a [(1+ \eta_{(1)}) y^{(1)}_{\alpha,a} +  \eta_{(1)} 2b \, x^{(1)}_{\alpha,a} ] = 0, \qquad \alpha = 0,1,2,4  \\
 \sum_a N_a(x^{(1)}_{3,a} - \eta_{(1)} x^{(1)}_{5,a}) = 0 , \\
 \sum_a N_a[y^{(1)}_{3,a} + \eta_{(1)}y^{(1)}_{5,a} + b \,  (x^{(1)}_{3,a} + \eta_{(1)}x^{(1)}_{5,a})] = 0,
\end{array}
 & 
 \begin{array}{l}
 \sum_a N_a (1-\eta_{(l)})x^{(l)}_{\alpha,a} = 0, \hspace{28mm}  \alpha = 1,4 \\
 \sum_a N_a [(1+ \eta_{(l)})y^{(l)}_{\alpha,a}  +  \eta_{(l)} x^{(l)}_{\alpha,a}  ] = 0, \qquad \alpha = 1,4 \\ 
 \sum_a N_a (y^{(l)}_{2,a} - \eta_{(l)} y^{(l)}_{2+2b,a}) = 0 ,\\ 
  \sum_a N_a (y^{(l)}_{3b,a} - \eta_{(l)} y^{(l)}_{3-2b,a}) = 0 ,\\ 
  \sum_a N_a [x^{(l)}_{2,a} + \eta_{(l)} x^{(l)}_{2+2b,a} + \eta_{(l)} y^{(l)}_{2+2b,a})] = 0, \\
  \sum_a N_a [x^{(l)}_{3,a} + \eta_{(l)} x^{(l)}_{3-2b,a} + \eta_{(l)} y^{(l)}_{3-2b,a})] = 0, 
 \end{array}
 \\
\hline
\end{array}
}{Z2Z6TwistedRRTadpoles}{Twisted RR tadpole cancellation conditions for the $\Z_2^{(i)}$ twisted sectors on all lattice backgrounds for $T^6/(\Z_2 \times \Z_6 \times \OR)$ with discrete torsion.}

The identification of physically equivalent background lattices can now be extended from the bulk contributions in equation~\eqref{Eq:bulk-identifications} to the exceptional contributions and one-cycle wrapping numbers.
We will do so in two steps: at first, we consider only those fixed point orbits which are not permuted under the orientifold projection, i.e. $\alpha=0,1,2$ for $\Z_2^{(1)}$ and $\alpha=1,4$ for $\Z_2^{(l=2,3)}$, and determine 
the transformation of the one-cycle wrapping numbers. In the second step, we verify that this mapping permutes the remaining fixed point orbits, $\alpha=3,4,5$ for $\Z_2^{(1)}$ and $\alpha=2,3$ for $\Z_2^{(l=2,3)}$ correctly.

We can start by an educated guess, which relies on the fact that the bulk supersymmetry conditions in table~\ref{tab:Bulk-RR+SUSY-Z2Z6} can be expressed as~\cite{Forste:2010gw,Honecker:2011sm,Blaszczyk:2014xla} (${\cal Z}_a^{\bf AA}  \propto [n^1_a + i \frac{R_2}{R_1} \tilde{m}^1_a] \prod_{k=2}^3 [n^k_a + e^{\frac{\pi i}{3}} m^k_a]$)
\begin{equation}
\Re ({\cal Z}_a) >0,
\qquad
\Im ({\cal Z}_a)=0,
\qquad
\text{with}
\quad
 {\cal Z}_a^{\bf BB} = e^{-\pi i /6}  {\cal Z}_a^{\bf AB} = e^{-\pi i /3}  {\cal Z}_a^{\bf AA} 
 ,
\end{equation}
and thus demanding that the orbit representant stays $(n^3_a,m^3_a)=(\text{odd},\text{odd})$ leads to the overall non-supersymmetric rotation $\pi (-\frac{1}{2},\frac{1}{3},0)$ for ${\bf a/bAA} \leftrightarrow {\bf a/bAB}$
and $\pi(0,-\frac{1}{3},0)$ for  ${\bf a/bAA} \leftrightarrow {\bf a/bBB}$. This first guess is in agreement with the permutation of the sets $(P_a,Q_a)$ and $(\frac{\tilde{U}_a}{1-b},\frac{\tilde{V}_a}{1-b})$
in equation~\eqref{Eq:bulk-identifications}, in particular  $(n^1_a,\frac{\tilde{m}^1_a}{1-b})_{\bf a/bAA} \leftrightarrow  (\frac{\tilde{m}^1_a}{1-b},-n^1_a)_{\bf a/bAB} \leftrightarrow (n^1_a,\frac{\tilde{m}^1_a}{1-b})_{\bf a/bBB}$.
Notice that in this paragraph, we use the range $-\frac{\pi}{2} \leqslant \phi^{(1)}_a \leqslant 0$ to single out one orientifold representant on {\bf a/bAB}, whereas on {\bf a/bAA} and {\bf a/bBB} we stick to the range 
$ 0 \leqslant \phi^{(1)}_a \leqslant \frac{\pi}{2}$ used in section~\ref{Sss:BulkCycles}.

Let us now focus on the map between the {\bf a/bAA} and {\bf a/bAB} lattices. The rotation by $\pi (-\frac{1}{2},\frac{1}{3},0)$ acts on the one-cycle wrapping numbers as
\begin{equation}\label{Eq:Trafo_nm_AA-AB}
\left(\begin{array}{cc} n^1_a & m^1_a \\ n^2_a & m^2_a \\ n^3_a & m^3_a \end{array}\right)_{\bf a/bAA}
\to
\left(\begin{array}{cc} \frac{\tilde{m}^1_a}{1-b} & -(1-b) \, n^1_a - \frac{b}{1-b} \tilde{m}^1_a \\ - m^2_a & n^2_a + m^2_a
\\ n^3_a & m^3_a \end{array}\right)_{\bf a/bAB}
,
\end{equation}
which is in agreement with the proposed transformation of the bulk wrapping numbers $(P_a,Q_a,U_a,V_a)$ in equation~\eqref{Eq:bulk-identifications}.
Using the expressions in table~\ref{tab:Z2Z6ExceptionalWrappingNumbers}, one can immediately read off that the exceptional wrapping numbers in the $\Z_2^{(1)}$ twisted sector transform as follows
($\tilde{y}^{(1)}_{\beta, a} \equiv y^{(1)}_{\beta, a} + b \,x^{(1)}_{\beta, a}$) 
\begin{equation}\label{Eq:relations_Z21_AA-AB}
\bigl(x^{(1)}_{\alpha, a} \; , \; y^{(1)}_{\alpha, a} \bigr)_{\bf a/bAA} \to 
\Bigl(\frac{\tilde{y}^{(1)}_{\beta, a} }{1-b} \; , \; -(1-b) \, x^{(1)}_{\beta, a} - \frac{b}{1-b} \tilde{y}^{(1)}_{\beta, a} \Bigr)_{\bf a/bAB}
\quad 
\text{for}
\quad
\left\{\begin{array}{c}
\alpha=\beta=0,1,2 \\ (\alpha,\beta) \in \{(3,5), (4,3), (5,4) \}
\end{array}\right.
,
\end{equation}
where for $\alpha \notin \{0,1,2\}$ we applied an {\it inverse} rotation by $-\frac{\pi}{3}$ on the fixed point labels in order to compensate the rotation of the basic one-cycles.
Using $\eta_{(1)} \to - \eta_{(1)}$, it is now obvious that all RR tadpole conditions in the $\Z_2^{(1)}$ twisted sector of table~\ref{tab:Z2Z6TwistedRRTadpoles} are mapped correctly from  {\bf a/bAA}  to {\bf a/bAB}.

In the transformation of the $\Z_2^{(2)}$ twisted sector, the fixed points along $T^2_{(3)}$ stay inert, whereas the fixed points $(2,3,4)$ along $T^2_{(1)}$ are permuted non-trivially by the $-\frac{\pi}{2}$ rotation. 
The exceptional wrapping numbers transform as
\begin{equation}\label{Eq:relations_Z22_AA-AB}
\Bigl( x^{(2)}_{\alpha, a} \; , \; y^{(2)}_{\alpha, a} \Bigr)_{\bf a/bAA} \to 
\Bigl(-y^{(2)}_{\beta, a} \; , \; x^{(2)}_{\beta, a} + y^{(2)}_{\beta, a} \Bigr)_{\bf a/bAB}
\quad
\text{for}
\quad
\left\{\begin{array}{cc}
\begin{array}{c} \alpha=\beta=1,3  \\ (\alpha,\beta) \in \{(2,4),(4,2)\} \end{array} \Biggr\} & {\bf a}\\
\begin{array}{c}\alpha=\beta=1,4 \\  (\alpha,\beta) \in \{(2,3), (3,2) \} \end{array} \Biggr\}  & {\bf b}
\end{array}\right.
,
\end{equation}
as can be checked by inserting the transformation of $(n^2_a,m^2_a)$ in the expressions of table~\ref{tab:Z2Z6ExceptionalWrappingNumbers}.
Using further $\eta_{(2)} \to \eta_{(2)}$ establishes the map of $\Z_2^{(2)}$ twisted RR tadpole cancellation conditions from the {\bf a/bAA} to {\bf a/bAB} lattice in table~\ref{tab:Z2Z6TwistedRRTadpoles}.

The $\Z_2^{(3)}$ twisted sector can be treated similarly by setting $\eta_{(3)} \to - \eta_{(3)}$, in other words
\begin{equation}\label{Eq:map-Oplanes}
\bigl( \OR, \OR\Z_2^{(1)},\OR\Z_2^{(2)},\OR\Z_2^{(3)} \bigr)_{\bf a/bAA}
\to
\bigl(\OR\Z_2^{(2)},\OR\Z_2^{(3)},\OR, \OR\Z_2^{(1)} \bigr)_{\bf a/bAB}
,
\end{equation}
 and by making use of the option of choosing a different representant of the $\Z_6$ orbit, cf. equation~\eqref{Eq:1-cycle-orbits}.
This completes the proof that at the level of RR tadpole cancellation and supersymmetry conditions, D6-brane models on {\bf a/bAA} and {\bf a/bAB} are equivalent.

The analogous discussion holds for relating {\bf a/bAB} to {\bf a/bBB}. It thus only remains to show that all other explicitly computable quantities like the matter spectrum and one-loop corrections to the gauge kinetic function are correctly identified under the above map.

\subsubsection{Intersection numbers between fractional three-cycles}\label{Sss:Intersections}

At the intersections of two fractional three-cycles $\Pi_a^{\text{frac}}$ and $\Pi_b^{\text{frac}}$, the net-chirality of matter in the bifundamental representation is counted by the intersection number,
\begin{equation}
\begin{aligned}
\chi^{(\N_a, \ov \N_b)} \equiv\Pi_a^{\text{frac}} \circ \Pi_b^{\text{frac}} &= \frac{1}{4} \Big( 2 \left( P_a U_b - P_b U_a + Q_a V_b - Q_b V_a \right) +  \left(  P_a V_b - P_b V_a +  Q_a U_b -  Q_b U_a \right)  \Big) \\
&- \frac{1}{4} \left( 3 \left[ x_{0,a}^{(1)} y_{0,b}^{(1)} -x_{0,b}^{(1)} y_{0,a}^{(1)} \right] + \sum_{\alpha=1}^5 \left[ x_{\alpha, a}^{(1)} y_{\alpha, b}^{(1)} - x_{\alpha, b}^{(1)} y_{\alpha, a}^{(1)} \right] \right)  \\
& - \frac{1}{4} \sum_{i=2}^3 \sum_{\alpha=1}^4 \left[ x_{\alpha,a}^{(i)}  y_{\alpha,b}^{(i)} - x_{\alpha,b}^{(i)}  y_{\alpha,a}^{(i)} \right], 
\end{aligned}
\end{equation}
while the net-chirality of symmetric and antisymmetric representations follows from the intersection numbers with orientifold images and O6-planes,
\begin{equation}\label{Eq:SymmAntiChiral}
\chi^{\Anti_a/\Sym_a} \equiv \frac{\Pi_a^{\text{frac}} \circ \Pi_{a'}^{\text{frac}} \pm \Pi_a^{\text{frac}} \circ \Pi_{O6}}{2},
\end{equation}
where the fractional three-cycle associated to the O6-planes only contains contributions from bulk three-cycles as the O6-planes do not carry twisted RR-charges, cf. table~\ref{tab:Z2Z6TwistedRRTadpoles}.

It can be explicitly checked that the transformations of bulk and exceptional wrapping number in equations~\eqref{Eq:bulk-identifications},~\eqref{Eq:relations_Z21_AA-AB} and~\eqref{Eq:relations_Z22_AA-AB} preserve the intersection numbers and thereby the chiral spectrum.
Following the elaborate discussion of symmetries on the other factorisable orientifold $T^6/(\Z_2 \times \Z_6' \times \OR)$ in~\cite{Honecker:2012qr}, the argument can be adjusted to the vector-like matter spectrum and one-loop correction to the gauge kinetic function as follows. The non-supersymmetric rotation by $\pi(-\frac{1}{2},\frac{1}{3},0)$ acts (up to rescaling of $\varrho$ on $T^2_{(1)}$) crystographically on the lattices, and therefore relative angles $(\vec{\phi}_{a(\omega^k b)})$ 
and toroidal intersection numbers $I_{a(\omega^k b)}$ between D6-branes $a$ and all orbifold images of $b$ are preserved. Due to the prescription of how to permute $\Z_2^{(l)}$ fixed points per two-torus, the same holds true for the $\Z_2^{(l)}$ invariant intersection numbers $I^{\Z_2^{(l)}}_{a(\omega^k b)}$ since both discrete Wilson lines $(\vec{\tau}_a)$ and displacements $(\vec{\sigma}_a)$ are preserved under the mapping between lattices discussed above.

In the $a(\omega^k b')$ sectors, the map between lattices permutes the power $k$, in analogy to the details given for $T^6/(\Z_2 \times \Z_6' \times \OR)$ in~\cite{Honecker:2012qr}. Since the total number of matter states is obtained
by summing over $k=0 ,1,2$, the permutation neither affects the low-energy spectrum nor the one-loop gauge threshold corrections, which depend - besides of the toroidal and $\Z_2$ invariant intersection numbers - also on the relative angles. 

In summary, upon the mapping of bulk~\eqref{Eq:bulk-identifications} and toroidal~\eqref{Eq:Trafo_nm_AA-AB} wrapping numbers and permutation of O6-planes~\eqref{Eq:map-Oplanes},  we have just shown the equivalences of 
${\bf aAA} \leftrightarrow {\bf aAB} \leftrightarrow {\bf aBB}$ and ${\bf bAA} \leftrightarrow {\bf bAB} \leftrightarrow {\bf bBB}$ at the level of all explicitly computable quantities, which include
\begin{itemize}
\item
the Hodge numbers $(h_{11}^+,h_{11}^-;h_{21})$ in table~\ref{tab:ClosedSpectrum_Z2Z6} counting closed string moduli,
\item
the RR tadpole cancellation and supersymmetry conditions in table~\ref{tab:Bulk-RR+SUSY-Z2Z6} implying the value of the tree-level gauge coupling in equation~\eqref{Eq:g_at_tree},
\item
the massless matter spectrum,
\item
the worldsheet areas among $n$ D-branes encoding the size of $n$-point couplings,
\item
the one-loop gauge thresholds, i.e. the full tower of massive matter states as well as the perturbatively exact holomorphic gauge kinetic function and leading order of the closed string K\"ahler potential and open string K\"ahler metrics.
\end{itemize}
The discussion can be transferred to the factorisable $T^6/(\Z_6' \times \OR)$ orientifold as briefly commented below.

\subsection{Reduction to symmetries on $T^6/\Z_6'$}\label{Ss:Reduction_Z6p}

The proof of equivalent lattices can be adjusted from $T^6/(\Z_2 \times \Z_6 \times \OR)$ with discrete torsion
to the $T^6/(\Z_6' \times \OR)$ case by setting $(\eta_{\OR},\eta_{\OR\Z_2^{(1)}}, \eta_{\OR\Z_2^{(2)}}, \eta_{\OR\Z_2^{(3)}}) \in \{(1,0,1,0),(1,0,0,1)\}$
in the bulk RR tadpole cancellation conditions in table~\ref{tab:Bulk-RR+SUSY-Z2Z6} and by truncating the two corresponding sets of $\Z_2^{(1)}$ and $\Z_2^{(2 \text{ or } 3)}$ twisted RR tadpole conditions
in table~\ref{tab:Z2Z6TwistedRRTadpoles}. 

For the {\bf a/bAA} and {\bf a/bBB} lattices, both choices of $k \in \{2,3\}$ of truncating the $\OR\Z_2^{(1)}$ plus one $\OR\Z_2^{(k)}$-invariant plane 
are equivalent. For the {\bf a/bAB} lattice, however, the RR tadpole cancellation conditions in tables~\ref{tab:Bulk-RR+SUSY-Z2Z6} and~\ref{tab:Z2Z6TwistedRRTadpoles} 
are not symmetric under the exchange $(\OR)\Z_2^{(2)} \leftrightarrow (\OR)\Z_2^{(3)}$. Upon the truncation  $(\eta_{\OR},\eta_{\OR\Z_2^{(1)}}, \eta_{\OR\Z_2^{(2)}}, \eta_{\OR\Z_2^{(3)}})=(1,0,1,0)$,
the equivalence ${\bf a/bAB} \leftrightarrow {\bf a/bAA}$ is preserved, whereas for $(\eta_{\OR},\eta_{\OR\Z_2^{(1)}}, \eta_{\OR\Z_2^{(2)}}, \eta_{\OR\Z_2^{(3)}})=(1,0,0,1)$
we find ${\bf a/bAB} \leftrightarrow {\bf a/bBB}$ as can be verified also using the original expressions for $T^6/(\Z_6' \times \OR)$  in~\cite{Gmeiner:2007zz}.
This completes and corrects the proof of the proposed pairwise symmetry between background lattices of $T^6/(\Z_6' \times \OR)$ in appendix~D of~\cite{Gmeiner:2008xq}.

\section{First Steps in D6-Brane Model Building}\label{S:StepsInD6BraneMB}

We have shown in the previous section that there exist only two physically inequivalent background lattices {\bf aAA} and {\bf bAA}, on which we will focus from now on. 
We will briefly discuss the conditions for an enhancement of the gauge group $U(N) \hookrightarrow USp(2N)$ or $SO(2N)$, classify D6-branes without matter in the adjoint representation
and provide constraints from the non-existence of matter in the symmetric representation on the {\it QCD} stack. Finally, we will search for intersection numbers $(\Pi_a \circ \Pi_b, \Pi_a \circ \Pi_{b'})$
leading to three particle generations.

In section~\ref{S:GlobalModels}, the above considerations will be combined to obtain {\it globally} defined D6-brane models with supersymmetric Standard Model or GUT spectrum.

\subsection{Gauge Group Enhancement}\label{Ss:Enhance}

Aiming towards model building, a full classification of fractional three-cycles supporting enhanced gauge groups of the type $USp(2N)$ or $SO(2N)$ is useful for various reasons. First of all, $USp(2)$ gauge factors can be used to account for the $SU(2)_L$ left stack in the MSSM gauge factor and/or for the $SU(2)_R$ stack in left-right symmetric models. Secondly, $USp(2)$ branes play the r\^ole of the probe branes when deriving the K-theory constraints~\cite{Uranga:2002vk,MarchesanoBuznego:2003hp}, which have to be satisfied to ensure the global consistency of the full Type II string theory compatification with D-branes. Furthermore, classifying rigid three-cycles supporting D-branes with enhanced $USp(2N)$ or $SO(2N)$ gauge factors forms the first step in studying Euclidean D-brane instanton corrections to the superpotential, see e.g.~\cite{Blumenhagen:2009qh} for a review.  Another application was advocated in~\cite{Honecker:2013hda,Honecker:2013kda}, where the shortest rigid three-cycles with an enhanced $USp(2N)$ or $SO(2N)$ gauge group were used to derive the sufficient conditions for the existence of discrete $\Z_n$ symmetries. 

The necessary and sufficient conditions for a gauge group $U(N)$ to enhance to an $USp(2N)$ or $SO(2N)$ gauge factor require the corresponding stack of D6-branes to wrap an orientifold-invariant three-cycle. This boils down to the geometric requirement that the bulk part of the D6-brane stack is parallel to one of the four O6-plane orbits, supplemented with a topological condition involving the discrete Wilson lines $(\vec{\tau})$, displacement parameters $(\vec{\sigma})$ and the individual tiltedness of the two-tori~\cite{Forste:2010gw}.
This topological condition is written out explicitly for the $T^6/(\Z_2 \times \Z_6 \times \OR)$ orientifold in the second column of table~\ref{Tab:Conditions-on_b+t+s-SOSp}. For the orientifold at hand, the first two-torus can be either untilted ($b=0$) or tilted ($b= \frac{1}{2}$), but the second and third two-torus are always tilted due to the $\Z_6$ orbifold action, as discussed in section~\ref{Ss:IIAonZ2Z6}. The remaining columns of table~\ref{Tab:Conditions-on_b+t+s-SOSp} provide an overview of the combinations $(\sigma^i; \tau^i)$ for which gauge group enhancement occurs, depending on the choice of the exotic O6-plane.
The full classification for the untilted $T_{(1)}^2$ is presented for the first time, whereas the classification for the tilted $T_{(1)}^2$ is identical to the one given in~\cite{Honecker:2012qr,Honecker:2013hda} for the $T^6/(\Z_2 \times \Z_6' \times \OR)$ orientifold. 
The gauge group enhancement and related matter content in the symmetric or antisymmetric representation - listed here for the first time for $T^6/(\Z_2 \times \Z_6 \times \OR)$- can be fully exposed for all cases by computing the beta-function coefficient of the respective gauge group as discussed e.g. in appendix~B of~\cite{Honecker:2012qr}.

An explicit number count of fractional three-cycles allowing for gauge group enhancement requires to combine the various choices of $(\sigma^i ; \tau^i)$ with the $2^2$ independent $\Z_2$ eigenvalues. For an untilted $T_{(1)}^2$, there are $144 + 2 \times 48=240$ combinations of $\Z_2$ eigenvalues, displacements and discrete Wilson lines leading to $USp(2N)$ gauge group enhancement and $16$ combinations leading to $SO(2N)$ gauge groups. For $b= \frac{1}{2}$, there exist $108 + 3 \times 36 = 216$ combinations yielding $USp(2N)$ gauge groups and $4 + 3 \times 12 = 40$ combinations giving rise to $SO(2N)$ gauge group enhancement. Observe that the overall count is not influenced by the choice of the exotic O6-plane, but the separate contributions do depend on the choice. For instance, the fractional three-cycles supporting $SO(2N)$ gauge groups on an untilted $T_{(1)}^2$ have a bulk part parallel to the $\OR\Z_2^{(1)}$-plane for $\eta_{\OR} = -1$ and parallel to the $\OR\Z_2^{(2)}$-plane for $\eta_{\OR\Z^{(3)}} = -1$.    

A closer look at table~\ref{Tab:Conditions-on_b+t+s-SOSp} reveals that the fractional three-cycles with enhanced gauge groups are in most cases accompanied by additional matter in the (anti-)symmetric representation. With regard to model building, we will allow at most matter in the antisymmetric representation, when fractional three-cycles with enhanced $USp(2)$ gauge groups are used to account for $SU(2)$ stacks. A thorough discussion about the presence of matter in the symmetric and/or antisymmetric representation is given in section~\ref{Ss:NoSyms}.

\begin{table}[h!]\hspace{-10mm}
\renewcommand{\arraystretch}{1.2}
\begin{minipage}[c]{0.79\textwidth}
  \begin{center}\hspace{-10mm}
\begin{equation*}\hspace{-20mm}
\begin{array}{|c|c||c|c||c|c|}\hline
\multicolumn{6}{|c|}{\text{\bf Existence of $\OR$ invariant three-cycles on } T^6/(\Z_2 \times \Z_6 \times \OR)}
\\\hline\hline
\pp & (\eta_{(1)},\eta_{(2)},\eta_{(3)}) \stackrel{!}{=}
& \muc{2}{|c||}{(1,1,-1)} & \muc{2}{|c|}{(-1,-1,-1)}
\\
\text{O6} &  & b=0 & b=\frac{1}{2} & b=0 & b=\frac{1}{2} 
\\\hline\hline
\OR  & \hspace{-3mm}\left(\hspace{-3mm}\begin{array}{c} -(-1)^{\sigma^2\tau^2 + \sigma^3 \tau^3} \\ - (-1)^{2b\sigma^1\tau^1 + \sigma^3 \tau^3} \\  -(-1)^{2b\sigma^1\tau^1 +\sigma^2 \tau^2} \end{array}\hspace{-3mm}\right)\hspace{-3mm}
& \left(\!\!\!\!\begin{array}{c} \sigma^1;\tau^1\\ \underline{0;1} \\ 1;1\end{array}\!\!\!\! \right)
&   \left(\!\!\!\!\begin{array}{c} 1;1 \\ 1;1 \\ \underline{0;1} \end{array} \!\!\!\!\right)
& \left(\!\!\!\!\begin{array}{c} \sigma^1;\tau^1\\ \underline{0;1} \\ \underline{0;1} \end{array}\!\!\!\! \right)
&   \left(\!\!\!\!\begin{array}{c} 1;1 \\ 1;1 \\ 1;1 \end{array} \!\!\!\!\right)
\\\cline{3-6}
&  & USp(2N)& SO(2N) & USp(2N) & SO(2N)
\\
& & +1 \, \Anti & + 1 \, \Sym & + \emptyset & + \emptyset
\\\hline\hline
\!\!\OR\Z_2^{(1)}\!\!\!    &\hspace{-3mm} \left(\hspace{-3mm}\begin{array}{c} -(-1)^{\sigma^2\tau^2 + \sigma^3 \tau^3} \\ (-1)^{2b\sigma^1\tau^1 + \sigma^3 \tau^3} \\  (-1)^{2b\sigma^1\tau^1 + \sigma^2 \tau^2}\end{array} \hspace{-3mm}\right)\hspace{-3mm}
& \left(\!\!\!\!\begin{array}{c} \sigma^1;\tau^1\\ 1;1\\ \underline{0;1}\end{array}\!\!\!\! \right)
&   \left(\!\!\!\!\begin{array}{c} 1;1 \\  \underline{0;1} \\ 1;1 \end{array} \!\!\!\!\right)
& \left(\!\!\!\!\begin{array}{c} \sigma^1;\tau^1\\1;1 \\ 1;1 \end{array}\!\!\!\! \right)
&   \left(\!\!\!\!\begin{array}{c} 1;1 \\  \underline{0;1} \\  \underline{0;1} \end{array} \!\!\!\!\right)
\\\cline{3-6}
&  & USp(2N) & SO(2N) & SO(2N) & USp(2N) \hspace{-2mm}
\\
& & + 5 \, \Anti & + 5 \, \Sym & + 4 \, \Anti  & + 4 \, \Sym 
\\\hline\hline
\!\!\OR\Z_2^{(2)}\!\!\!    & \hspace{-3mm} \left(\hspace{-3mm}\begin{array}{c} (-1)^{\sigma^2\tau^2 + \sigma^3 \tau^3} \\ - (-1)^{2b\sigma^1\tau^1 + \sigma^3 \tau^3} \\  (-1)^{2b\sigma^1\tau^1 + \sigma^2 \tau^2}\end{array}\hspace{-3mm} \right)\hspace{-3mm}
& \left(\!\!\!\!\begin{array}{c} \sigma^1;\tau^1\\ 1;1 \\ 1;1\end{array}\!\!\!\! \right)
&   \left(\!\!\!\!\begin{array}{c} 1;1 \\  \underline{0;1} \\  \underline{0;1} \end{array} \!\!\!\!\right)
& \left(\!\!\!\!\begin{array}{c} \sigma^1;\tau^1\\ 1;1\\ \underline{0;1}\end{array}\!\!\!\! \right)
&   \left(\!\!\!\!\begin{array}{c} 1;1 \\   \underline{0;1} \\ 1;1 \end{array} \!\!\!\!\right)
\\\cline{3-6}
&  & SO(2N)& USp(2N) & USp(2N) & SO(2N)
\\
& & + 1\, \Anti & + 1 \, \Sym & + 2 \, \Anti & + 2 \, \Sym
\\\hline\hline
\!\!\OR\Z_2^{(3)}\!\!\!  & \hspace{-3mm} \left(\hspace{-3mm}\begin{array}{c} (-1)^{\sigma^2\tau^2 + \sigma^3 \tau^3} \\  (-1)^{2b\sigma^1\tau^1 + \sigma^3 \tau^3} \\  -(-1)^{2b\sigma^1\tau^1 + \sigma^2 \tau^2} \end{array}\hspace{-3mm}\right)\hspace{-3mm}
& \left(\!\!\!\!\begin{array}{c} \sigma^1;\tau^1\\  \underline{0;1} \\ \underline{0;1}\end{array}\!\!\!\! \right)
&   \left(\!\!\!\!\begin{array}{c} 1;1 \\ 1;1 \\ 1;1 \end{array} \!\!\!\!\right)
& \left(\!\!\!\!\begin{array}{c} \sigma^1;\tau^1\\ \underline{0;1} \\ 1;1 \end{array}\!\!\!\! \right)
&   \left(\!\!\!\!\begin{array}{c} 1;1 \\ 1;1 \\  \underline{0;1} \end{array} \!\!\!\!\right)
\\\cline{3-6}
&  & USp(2N) & SO(2N) & USp(2N) & SO(2N)
\\
& & +1 \, \Sym & + 1 \, \Anti & + 2\, \Anti & + 2 \, \Sym
\\\hline
\end{array}
\end{equation*}
\end{center}
\end{minipage}\hfill
  \begin{minipage}[c]{0.25\textwidth}
\caption{Classification of $USp(2N)$ and $SO(2N)$ gauge groups and matter in the (anti)symmetric representation
 on $\OR$-invariant D6-branes. $\eta_{\OR\Z_2^{(1)}}=-1$ does not allow for supersymmetric solutions, while $\eta_{\OR\Z_2^{(2)}}=-1$ can be obtained from the listed case $\eta_{\OR\Z_2^{(3)}}=-1$ by exchanging two-torus labels $2 \leftrightarrow 3$.
Underlining denotes three choices, e.g. $(\sigma^2;\tau^2) \in \{(0;0);(1;0), (0;1)\}$ since only $\sigma^2\tau^2=0$ is required. In case of underlining of both $(\sigma^2;\tau^2)$ and $(\sigma^3;\tau^3)$, the choices are independent - in other words, there are $3^2=9$ options. For $b=\frac{1}{2}$,  the cases with $\sigma^1\tau^1=0$ coincide with those listed for $b=0$, but those with $\sigma^1\tau^1=1$ differ and are listed explicitly here.\label{Tab:Conditions-on_b+t+s-SOSp}
}
\end{minipage}
\end{table}

\subsection{Rigid D6-branes without Adjoint Matter}\label{Ss:NoAdjoints}

The bulk three-cycles from section~\ref{Sss:BulkCycles} are inherited from the underlying six-torus and therefore come with the usual three multiplets in the adjoint representation under the associated gauge groups, containing the position moduli of the D6-branes. As these massless multiplets correspond to flat directions in the superpotential, a spontaneous breaking of the gauge group by a brane displacement in the D6-brane stack cannot be prevented. The fractional three-cycles from section~\ref{Sss:FractionalCycles}, however, are located at $\Z_2 \times \Z_2$ fixed points along all three two-tori, such that these multiplets are projected out from the start. 

Nevertheless, the $\Z_2 \times \Z_2$ subgroup does not guarantee the total absence of matter in the adjoint representation for a D6-brane stack on $T^6/(\Z_2 \times \Z_6)$, as the intersections between a cycle $a$ and its orbifold images $(\omega^k a)_{k=1,2}$ might also give rise to matter in the adjoint representation. These multiplets in the adjoint representation contain the deformation moduli at the intersection points, by which a D-brane recombination of the orbifold images can be triggered in case these moduli acquire a non-vanishing {\it vev}. Hence, in order to ensure that the fractional three-cycles are completely rigid, one has to verify the absence of matter in the $a (\omega^k a)_{k=1,2}$ sectors as well.

Due to the invariance of $T_{(1)}^2$ under the $\Z_6$ orbifold action, the angles $\vec{\phi}_{a (\omega^k a)}  = \pm \pi (0, -\frac{1}{3}, \frac{1}{3})$ between cycle $a$ and its $\Z_6$ orbifold images $(\omega^k a)_{k=1,2}$ are always vanishing along the first two-torus. Hence, according to table 6 in~\cite{Honecker:2011sm} the absence of matter in the adjoint representation can be recast into the following condition on the number of multiplets  $\varphi^{a (\omega^k a)}$ per sector $(\omega^k a)_{k \in \{1,2\}}$,
\begin{equation}\label{Eq:ConditionAdjoint}
 \varphi^{a (\omega^k a)} \equiv | \chi^{a (\omega^k a)}| =  \frac{1}{2} \left(\left| I_{a(\omega^k a)}^{(2\cdot 3)} \right| -  I_{a(\omega^k a)}^{\Z_2^{(1)}, (2\cdot 3)} \right)  \stackrel{!}{=} 0 \qquad \text{for }\, k=1,2 ,
\end{equation}
where the toroidal intersection numbers $I_{a(\omega^k a)}^{(2\cdot 3)}  =  I_{a (\omega^k a)}^{(2)}  I_{a (\omega^k a)}^{(3)}$ along $T_{(2)}^2 \times T_{(3)}^2$ are given by:
\begin{equation}
I_{a (\omega^k a)}^{(l)} = (-)^{k+l} [ (n_a^l)^2 + n_a^l m_a^l + (m_a^l)^2] \in \Z_{\text{odd}},
\end{equation}
and the $\Z_2^{(1)}$ invariant intersection numbers $I_{a(\omega^k a)}^{\Z_2^{(1)}, (2\cdot 3)} = (-)^{\sigma_a^2 \tau_a^2 + \sigma_a^3 \tau_a^3} $ can be computed following appendix B in~\cite{Honecker:2012qr}. The sectors $a (\omega^k a)_{k=1,2}$ each provide half of the degrees of freedom to fill a full chiral multiplet in the adjoint representation, implying $\varphi^{a (\omega a)} = \varphi^{a (\omega^2 a)} $, and the total amount of matter in the adjoint representation at the intersections of the orbifold images is therefore given by:
\begin{equation}   
 \varphi^{\Adj_a} = \frac{1}{2} \sum_{k=1}^2  \varphi^{a (\omega^k a)}= \varphi^{a (\omega a)}.
\end{equation}
From condition~(\ref{Eq:ConditionAdjoint}), it follows directly that fractional three-cycles can only be truly rigid when they are composed of bulk two-cycles on $T_{(2)}^2 \times T_{(3)}^2$ that have minimal length, i.e.~with toroidal wrapping numbers $(n_a^2, m_a^2), (n_a^3, m_a^3) \in  \{ (\pm1,0), (0,\pm1), (\pm1, \mp1) \}$. Whether or not such a fractional three-cycle is free from matter in the adjoint representation still depends on the values of the displacement parameters and discrete Wilson lines along $T_{(2)}^2 \times T_{(3)}^2$ due to the form of $I_{a(\omega^k a)}^{\Z_2^{(1)}, (2\cdot 3)}$:  
\begin{equation} \label{Eq:NoAdjointClassII}
\varphi^{\Adj_a} =  \left\{ \begin{array}{cl} 0 & \sigma^2_a \tau^2_a = \sigma_a^3 \tau_a^3 \in \{ 0, 1 \} \\
1 & \sigma^2_a \tau^2_a \neq \sigma_a^3 \tau_a^3 \in \{ 0, 1 \} 
 \end{array} \right. .
\end{equation}
Yet, the amount of matter in the adjoint representation at the intersections of orbifold images is independent of several model building ingredients:
the $\Z_2$ eigenvalues, the discrete displacement and Wilson line along $T^2_{(1)}$, the choice of the exotic O6-plane, the lattice choice ({\bf aAA} or {\bf bAA}) and the complex structure modulus $\varrho$. 

In summary, fractional three-cycles without matter in the adjoint representation have on the {\bf a/bAA} lattice a bulk part that is either parallel to the $\OR$-plane or parallel to an orbit of the form $(n_a^1, m_a^1;1,0;1,-1)$, where the one-cycle torus wrapping numbers $(n_a^1, m_a^1)$ are set by the modulus $\varrho$ through the necessary supersymmetry condition, as further detailed in appendix~\ref{A:ClassBulkThreeCycles}. If a fractional three-cycle has a bulk orbit parallel to the $\OR$-plane or of the form $(n_a^1, m_a^1;1,0;1,-1)$, there exist $2^2\cdot4\cdot(3\cdot3+1\cdot1)=160$ combinations of $\Z_2$ eigenvalues, displacement parameters and discrete Wilson lines for which the fractional three-cycle is completely rigid, and $2^2 \cdot 4 \cdot (3 \cdot 1+1 \cdot 3 ) = 96 $ combinations with one chiral multiplet in the adjoint representation. Rigid three-cycles parallel to the $\OR$-plane exist for each lattice configuration and for each value of the modulus $\varrho$. Supersymmetric rigid three-cycles of the second type not overshooting the bulk RR tadpoles in table~\ref{tab:Bulk-RR+SUSY-Z2Z6}, however,  can only be found for 159 rational values of $\varrho$ on the {\bf aAA} lattice and for 79 rational values of $\varrho$ on the {\bf bAA} lattice.

Completely rigid fractional three-cycles form the most suitable candidates to accommodate the $SU(3)$ {\it QCD} stack or the $SU(4)$ Pati-Salam gauge group. For an $SU(5)$ GUT, however, a multiplet in the adjoint representation is conventionally introduced to spontaneously break the $SU(5)$ gauge group down to the Standard Model $SU(3)\times SU(2) \times U(1)_Y$ gauge factors. Moreover, fractional three-cycles with matter in the adjoint representation can also be used to include additional Abelian gauge factors completing the visible gauge sector, given that these states in the adjoint representation are mere singlets under any $U(1)$ gauge group. In this respect it makes sense to classify fractional three-cycles with at least one chiral multiplet in the adjoint representation as well.

A meticulous search of fractional three-cycles with one multiplet in the adjoint representation consists of the same steps as above, but now with the condition  $\varphi^{a (\omega^k a)} \stackrel{!}{=} 1$ for $k=1,2$. With and below equation (\ref{Eq:NoAdjointClassII}) we have already identified one category satisfying this constraint, namely those fractional three-cycles with bulk orbit parallel to the $\OR$-plane or to an orbit of the form $(n_a^1, m_a^1;1,0;1,-1)$ and with discrete parameters $\sigma_a^2 \tau_a^2 \neq \sigma_a^3 \tau_a^3 $.
A second category comprises three-cycles that are shortest on one two-torus and next-to-shortest on the other two torus, such as toroidal wrapping numbers $(n_a^2, m_a^2;n_a^3,m_a^3)$ 
in the orbits $(1,-1;1,1)$ or $(1,1;1,-1)$, in other words three-cycles for which the toroidal intersection numbers on $T_{(2)}^2 \times T_{(3)}^2$ read $|I_{a (\omega a)}^{(2.3)}|=|I_{a (\omega^2 a)}^{(2.3)}|=3$.
Fractional three-cycles with a bulk orbit parallel to the O6-plane orbits $\OR\Z_2^{(2)}$ or $\OR\Z_2^{(3)}$ (with representant $(0,1;2,-1;1,-1)$ and $(0,1;0,-1;1,1)$, respectively) 
fall within this second category for each possible value of the modulus $\varrho$. Also three-cycles with bulk orbits of the form $(n_a^1,  m_a^1;1,-1;1,1)$ or $(n_a^1,  m_a^1;1,1;1,-1)$ fall within this category, but  their supersymmetric existence relies on the modulus $\varrho$ whose value constrains the one-cycle torus wrapping numbers $(n_a^1,  m_a^1)$. 
Through a scan over the respective parameter space, one can show that only a restricted number of rational $\varrho$-values allows for the existence of the second class of three-cycles which do not overshoot the bulk RR tadpoles: 56 different $\varrho$-values for the {\bf aAA} lattice and 27 $\varrho$-values for the {\bf bAA} lattice. For all fractional three-cycles with bulk orbit parallel to the $\OR\Z_2^{(2,3)}$-plane, or of the form  $(n_a^1,  m_a^1;1,-1;1,1)$ or $(n_a^1,  m_a^1;1,1;1,-1)$, the total number of multiplets in the adjoint representation is again solely determined by the displacement parameters and discrete Wilson lines along $T_{(2)}^2 \times T_{(3)}^2$:
 \begin{equation}
\varphi^{\Adj_a} =  \left\{ \begin{array}{cl} 1 & \sigma^2_a \tau^2_a = \sigma_a^3 \tau_a^3 \in \{ 0, 1 \} \\
2 & \sigma^2_a \tau^2_a \neq \sigma_a^3 \tau_a^3 \in \{ 0, 1 \} 
 \end{array} \right. ,
\end{equation}
irrespective of the $\Z_2$ eigenvalues, the displacement $\sigma_a^1$, the Wilson line $\tau_a^1$, the lattice choice ({\bf aAA} or {\bf bAA}) or the value of the complex structure modulus $\varrho$.

Last but not least, we also recall that the fractional three-cycles parallel to an O6-plane and supporting an enhanced $USp$ gauge group correspond to rigid three-cycles, provided they are not accompanied by matter in the symmetric representation. Matter in the antisymmetric representation is allowed for a $USp(2)$ gauge group, given that these states transform as gauge singlets. Candidate fractional three-cycles satisfying these constraints can be read off from table~\ref{Tab:Conditions-on_b+t+s-SOSp}.

\subsection{The Presence of Symmetric and/or Antisymmetric Matter}\label{Ss:NoSyms}

One of the unavoidable consequences of the orientifold projection is the presence of an orientifold image three-cycle $\Pi_{a'}^{\text{frac}}$ for each fractional three-cycle $\Pi_a^{\text{frac}}$. At the intersections between a three-cycle and its orientifold image, massless matter can emerge in the symmetric and/or antisymmetric representation. 
Chiral matter in the antisymmetric representation happens to be particularly useful in model building to generate right-handed quarks in case of the $U(3)_a$ {\it QCD} gauge group or to account for the quarks and leptons in the {\bf 10} representation of some $U(5)_a$ GUT gauge group. Chiral matter in the symmetric representation on the other hand is phenomenologically acceptable only if it is charged under a mere $U(1)$ gauge factor, in which case such matter states can serve as candidates for the right-handed leptons. 

A first important observation in this respect concerns the topological intersection number between a three-cycle $\Pi_a^{\text{frac}}$ and the O6-planes, which can be written generically for the {\bf a/bAA} lattice as,
\begin{equation}
\Pi_a^{\text{frac}} \circ \Pi_{O6} = - \frac{\left(\eta_{\OR} + 3\eta_{\OR\Z_2^{(1)}}\right)}{2} \left[ 2 \tilde U_a + \tilde V_a \right] - \frac{3 (1-b) \left( \eta_{\OR\Z_2^{(2)}} + \eta_{\OR\Z_2^{(3)}}  \right) }{2}  Q_a.
\end{equation}
Focussing on the truly rigid fractional three-cycles from the previous section, this formula immediately indicates that three-cycles with a bulk part parallel to the $\OR$-plane (and thus $Q_a = \tilde U_a = \tilde V_a =0$) have vanishing intersection number with the O6-planes, $\Pi_a^{\text{frac}} \circ \Pi_{O6}  = 0$. For this class of cycles, the net-chirality of  matter in the antisymmetric and symmetric representation will always be the same, i.e. $\chi^{\Sym_a} = \chi^{\Anti_a} \stackrel{!}{=} 0$. Consequently, these special three-cycle cannot be used to accommodate a $SU(5)$-GUT D6-brane stack, and they are also not suited for the {\it QCD} D6-brane stack if some of the right-handed quarks ought to be realised as chiral states in the antisymmetric representation. 

The other class of rigid fractional three-cycles with bulk orbit $(n_a^1, m_a^1;1,0;1,-1)$ and bulk wrapping numbers $(P_a = n_a^1 = -Q_a, \tilde U_a = \tilde m^1_a = - \tilde V_a)$ can intersect  non-trivially with the O6-planes, depending on the one-cycle wrapping numbers $(n_a^1, m_a^1)$.\footnote{Although these bulk orbits with $\tilde{m}^1_a>0$ are perfectly legitimate for local model building purposes from the viewpoint of supersymmetry, they might obstruct the construction of global models in case the $\OR\Z_2^{(2)}$-plane or the $\OR\Z_2^{(3)}$-plane is chosen as the exotic O6-plane. This is most easily seen by checking the second bulk RR tadpole cancellation condition in table~\ref{tab:Bulk-RR+SUSY-Z2Z6} for the {\bf aAA} and {\bf bAA} lattice.} The amount of chiral matter in the symmetric and antisymmetric representation can be computed explicitly via equation~(\ref{Eq:SymmAntiChiral}), which yields for a rigid three-cycle with e.g.~bulk orbit $(1,1;1,0;1,-1)$ on the {\bf aAA} lattice:
\begin{equation}\label{Eq:ExampleNoAdjNoChiralSymm}
\begin{aligned}
\chi^{\Anti_a/\Sym_a} =& \frac{1}{4} \left[ 1 -  \eta_{(1)} \left( 2 (-)^{\sigma_a^2 \tau_a^2} +  2 (-)^{\sigma_a^3 \tau_a^3}  +1  \right) - \eta_{(2)} \left( 1 - 2 (-)^{\tau_a^3 \sigma_a^3}\right) - \eta_{(3)} \left( 1 - 2 (-)^{\tau_a^2 \sigma_a^2}\right) \right] \\
& \pm \frac{1}{4} \eta_{\OR} \left[ - 1 - 3 \eta_{(1)} + 3 \eta_{(2)} + 3 \eta_{(3)}  \right] \\
  =&\left\{ \begin{array}{lc} 
  \frac{1}{4} (1 \mp \eta_{\OR}) \left[ 1+ 3 \eta_{(1)} - 3 \eta_{(2)} - 3 \eta_{(3)}   \right] - 2 \eta_{(1)} + \eta_{(2)} + \eta_{(3)} & \sigma_a^i \tau^i_a = 0,\\
\frac{1}{4} (1 \mp \eta_{\OR}) \left[ 1+ 3 \eta_{(1)} - 3 \eta_{(2)} - 3 \eta_{(3)}   \right] & \sigma_a^i \tau^i_a = 1,
  \end{array}\right.
\end{aligned}
\end{equation}
with $i=2,3$. Recall that in the other cases where $\sigma^2_a \tau_a^2 \neq \sigma^3_a \tau_a^3$ the fractional three-cycles are accompanied by matter in the adjoint representation, which makes them unsuitable to support the {\em QCD}-stack or the {\it left} stack. For the example in equation (\ref{Eq:ExampleNoAdjNoChiralSymm}), the rigid fractional three-cycle is free of chiral matter in the symmetric representation when the $\OR$-plane fulfils the r\^ole of the exotic O6-plane, i.e.~$\eta_{\OR} = -1 = \eta_{(1)} = \eta_{(2)} = \eta_{(3)}$, irrespective of the values for the discrete parameters $\sigma_a^2 \tau_a^2 = \sigma^3_a\tau_a^3 \in\{0,1 \}$. In case the exotic O6-plane is taken to be a $\OR\Z_2^{(k)}$-plane with $k \in \{2,3\}$, chiral matter in the symmetric representation will only be absent for a rigid fractional three-cycle if the discrete parameters satisfy $\sigma_a^2 \tau_a^2=\sigma^3_a\tau_a^3=0$ .

As the example in (\ref{Eq:ExampleNoAdjNoChiralSymm}) clearly shows, the net-chirality of  matter in the symmetric and antisymmetric representation also depends on the choice of the exotic O6-plane, in addition to the displacement parameters and discrete Wilson lines along $T_{(2)}^2 \times T_{(3)}^2$. Furthermore, $\chi^{\Anti_a}$ and $\chi^{\Sym_a}$ also depend on the torus wrapping numbers $(n_a^1, m_a^1)$, and through an explicit scan over this parameter space it turns out that only a fraction of the bulk orbits $(n_a^1, m_a^1; 1,0;1,-1)$ allows for fractional three-cycles {\em without chiral matter in the symmetric representation}, i.e.~$\chi^{\Sym_a} = 0$. On the {\bf aAA} lattice, 31 combinations $(n_a^1, m_a^1)$ can be found allowing for rigid fractional three-cycles without chiral matter in the symmetric representation, whereas the {\bf bAA} lattice comes with 15 combinations $(n_a^1, m_a^1)$ as presented in table~\ref{tab:AdjointSymmetricFree}. Depending on the choice of the exotic O6-plane and the values of the displacement parameters $(\sigma_a^2, \sigma_a^3)$ and discrete Wilson lines $(\tau_a^2, \tau_a^3)$, one can distinguish three different configurations to realise rigid fractional three-cycles without chiral matter in the symmetric representation:
\begin{itemize}
\item[$(I)$] the $\OR$-plane plays the r\^ole of the exotic O6-plane ($\eta_{\OR} = -1$), and the discrete parameters satisfy $\sigma_a^2 \tau_a^2 = \sigma_a^3 \tau_a^3  =0$,
\item[$(II)$] the $\OR$-plane plays the r\^ole of the exotic O6-plane with discrete parameters $\sigma_a^2 \tau_a^2 = \sigma_a^3 \tau_a^3  =1$,
\item[$(III)$] the $\OR\Z_2^{(l)}$-plane with $l \in\{2,3 \}$ is the exotic O6-plane ($\eta_{\OR\Z_2^{(l)}} = -1$), with discrete parameters $\sigma_a^2 \tau_a^2 = \sigma_a^3 \tau_a^3  =0$.
\end{itemize}
For the configuration with one of the $\OR\Z_2^{(l=2,3)}$-planes as the exotic O6-plane and the discrete parameters satisfying $\sigma_a^2 \tau_a^2 = \sigma_a^3 \tau_a^3  =1$, all fractional three-cycles are accompanied by chiral matter in the symmetric representation. 

\mathtabfix{
\begin{array}{cc}
\hline 
\multicolumn{2}{|c|}{\text{\bf Classification of bulk orbits $(n_a^1, m_a^1; 1, 0; 1,-1)$ without chiral symmetrics on $T^6/(\Z_2\times\Z_6\times\OR)$}} \\
\hline \hline
\multicolumn{1}{|c||}{\text{\bf aAA lattice}} & \multicolumn{1}{|c|}{\text{\bf bAA lattice}} \\
\hline \hline
\begin{array}{|c|c|c||}
\hline
(n_a^1, m_a^1)& \varrho & \text{\bf Configuration}   \\
\hline \hline
(1,16)& 3/16&(II), (III)\\
(1,15)&1/5 & (II), (III)\\
(1,14)&3/14 & (II), (III)\\
(1,13)&3/13 & (II), (III)\\
(1,12)&1/4 &(II), (III) \\
(1,11)&3/11 &(II), (III)\\
(1,10)&3/10 &(II), (III)\\
(1,9)&1/3 &(II), (III)\\
(1,8)&3/8&(II), (III) \\
(1,7)&3/7&(II), (III)\\
(1,6)&1/2&(II), (III)\\
(1,5)&3/5&(II), (III)\\
(1,4)&3/4&(II), (III)\\
(1,3)&1&(II), (III)\\
(1,2)&3/2&(II), (III)\\
(1,1)&3&(I), (II), (III)\\
\hline
\end{array}
\begin{array}{|c|c|c|}
\hline
(n_a^1, m_a^1)& \varrho & \text{\bf Configuration} \\
\hline \hline
(2,1)&6& (I) \\
(3,1)&9&  (I) \\
(4,1)&12 &  (I) \\
(5,1)&15&  (I) \\
(6,1)&18&  (I) \\
(7,1)&21& (I)\\
(8,1)&24& (I) \\
(9,1)&27& (I)\\
(10,1)&30& (I) \\
(11,1)&33& (I) \\
(12,1)&36&  (I) \\
(13,1)&39& (I) \\
(14,1)&42& (I) \\
(15,1)&45& (I)\\
(16,1)&48& (I) \\
&&\\
\hline
\end{array}
&
\begin{array}{|c|c|c|}
\hline
(n_a^1, m_a^1)& \varrho & \text{\bf Configuration}   \\
\hline \hline
(1,7)&2/5&(II), (III)\\
(1,6)&6/13&(II), (III)\\
(1,5)&6/11&(II), (III)\\
(1,4)&2/3&(II), (III)\\
(1,3)&6/7&(II), (III)\\
(1,2)&6/5&(II), (III)\\
(1,1)&2&(II), (III)\\
(1,0)&6&(I), (II), (III) \\
(3,-1)&18&(I)\\
(5,-2)&30&(I)\\
(7,-3)&42&(I)\\
(9,-4)&54&(I)\\
(11,-5)&66&(I)\\
(13,-6)&78&(I)\\
(15,-7)&90&(I)\\
&&\\
\hline
\end{array}
\end{array}
}{AdjointSymmetricFree}{Full list of bulk orbits $(n_a^1,m_a^1;1,0;1,-1)$ allowing for completely rigid fractional three-cycles free of chiral matter in the symmetric representation for the {\bf aAA} lattice (left) and for the {\bf bAA} lattice (right) on the orientifold $T^6/(\Z_2\times\Z_6\times\OR)$ with discrete torsion $\eta = -1$. The complex structure modulus $\varrho$ is related to the one-cycle wrapping numbers $(n_a^1, m_a^1)$ through the equation $\varrho = 3 \frac{n_a^1}{\tilde m_a^1}$ resulting from the bulk supersymmetry conditions on three-cycles. The third, sixth and ninth columns indicate for which choice of the exotic O6-plane and discrete parameters the total amount of chiral matter in the symmetric representation vanishes per three-cycle, i.e.~according to the three distinct configurations $(I), (II)$ and $(III)$ defined in the main text.}

Furthermore, these rigid three-cycles are also not supposed to be accompanied by an abundant amount of chiral matter in the {\em antisymmetric} representation, when they are meant to be wrapped by the {\it QCD} stack. To this end, one has to impose the additional requirement $|\chi^{\Anti_a}|~\in~\{0,1,2,3\}$.
In practice, the total amount of chiral matter in the antisymmetric representation can be computed using formula~(\ref{Eq:SymmAntiChiral}), whereas the contributions per sector $(\omega^k a) (\omega^k a)' _{k=0,1,2}$ are computed separately using the toroidal intersection numbers, see e.g.~\cite{Honecker:2011sm,Honecker:2012qr}:
\begin{equation}\label{Eq:SymmAntiTorusExp}
\begin{aligned}
\chi^{\Anti_a/\Sym_a} 
&= - \sum_{k=0}^2 \frac{ \left(I_{(\omega^k a)(\omega^k a)'} + \sum_{i=1}^3 I_{(\omega^k a)(\omega^k a)'}^{\Z_2^{(i)}} \right)
\pm \left( \sum_{n=0}^3 \eta_{\OR\Z_2^{(n)}} \, \tilde{I}_{(\omega^k a)}^{\OR\Z_2^{(n)}} \right) }{8}\\
& \equiv \sum_{k=0}^2  \chi^{\Anti_a/\Sym_a}_{(\omega^k a)},
\end{aligned}
\end{equation}
where $\tilde{I}_{(\omega^k a)}^{\OR\Z_2^{(n)}} \equiv 2 (1-b) I_{(\omega^k a)}^{\OR\Z_2^{(n)}} $ represents the intersection numbers between a orbifold image three-cycle $(\omega^k a)$ and the O6-plane $\OR\Z_2^{(n)}$ on the underlying torus, with $2(1-b)$ the number of parallel O6-planes in dependence of the shape of $T^2_{(1)}$. On the {\bf aAA} lattice, only six bulk orbits give rise to fractional three-cycles fulfilling the criterium $|\chi^{\Anti_a}| \leqslant 3$, with one-cycle torus wrapping numbers $(n_a^1, m_a^1) = (1,1)$, $(1,2)$, $(1,3)$ $(1,4)$, $(1,5)$, or $(1,6)$. Considering a bulk orbit of the type $(1,m_a^1;1,0;1,-1)$, the net-chirality for matter in the (anti-)symmetric representation reads per sector:
\begin{equation}\label{Eq:RigidNoSymmAntiaAA}
\chi^{\Anti_a/\Sym_a}_{(\omega^k a)} = \left\{ \begin{array}{lc}  -\frac{1}{2}\left( (-)^{\sigma_a^2 \tau_a^2} \pm \eta_{\OR}   \right) \left(  m_a^1 \, \eta_{(1)} - \eta_{(3)} \right)  & k=0,  \\ 
 \frac{1}{4}(1 \mp \eta_{\OR}) \left( m_a^1 (1 - \eta_{(1)}) - \eta_{(2)} - \eta_{(3)} \right) & k=1, \\
-\frac{1}{2} \left( (-)^{\sigma_a^3 \tau_a^3} \pm \eta_{\OR} \right) \left(  m_a^1 \, \eta_{(1)} - \eta_{(2)} \right) & k=2.
   \end{array}  \right.
\end{equation}

On the {\bf bAA} lattice, only five bulk orbits $(n_a^1,m_a^1;1,0;1,-1)$ can be found for which the fractional three-cycles satisfy the constraint on the net-chirality $|\chi^{\Anti_a}| \leqslant 3$, namely $(n_a^1,m_a^1) = (1,0)$, $(1,1)$, $(1,2)$, $(1,3)$, or $(1,4)$. The amount of matter in the  (anti-) symmetric representation for a fractional three-cycle with bulk orbit $(1,m_a^1;1,0;1,-1)$ on the {\bf bAA} lattice  reads per sector:
\begin{equation}\label{Eq:RigidNoSymmAntibAA}
\chi^{\Anti_a/\Sym_a}_{(\omega^k a)} = \left\{ \begin{array}{lc}  -\frac{1}{4}\left( (-)^{\sigma_a^2 \tau_a^2} \pm \eta_{\OR}   \right) \left( 2 \tilde m_a^1 \, \eta_{(1)} - \eta_{(3)} \right)  & k=0,  \\ 
 \frac{1}{4}(1 \mp \eta_{\OR}) \left( \tilde m_a^1 (1 - \eta_{(1)}) - \frac{\eta_{(2)} + \eta_{(3)}}{2} \right) & k=1, \\
-\frac{1}{4} \left( (-)^{\sigma_a^3 \tau_a^3} \pm \eta_{\OR} \right) \left(  2 \tilde m_a^1 \, \eta_{(1)} - \eta_{(2)} \right) & k=2.
   \end{array}  \right.
\end{equation}
For each of these 11 bulk orbits, table~\ref{tab:RigidSymmFreeAntisymmetrics} summarises the number of chiral states in the antisymmetric representation per sector and per configuration allowing for rigid three-cycles without chiral matter in the symmetric representation. In some cases, the total net-chirality is $\chi^{\Anti_a} \in \{0,\pm 1, \pm 2 \}$, even though the total sum $\sum_{k=0}^2|\chi^{\Anti_a}_{(\omega^k a)}  |$ over all sectors counts more than three massless states in the antisymmetric representation. This means that the total net-chirality $\chi^{\Anti_a} = \sum_{k=0}^2 \chi^{\Anti_a}_{(\omega^k a)}$ gives the effective number of chiral multiplets in the antisymmetric representation, while the total sum $\sum_{k=0}^2|\chi^{\Anti_a}_{(\omega^k a)} |$ also takes into account pairs of multiplets $\Anti_a + \ov \Anti_a$ from the non-chiral sector.

\mathtabfix{
\begin{array}{|c|c||c|c|c|}
\hline \multicolumn{5}{|c|}{\text{\bf $\chi^{\Anti_a}$ for rigid fractional three-cycles with $\chi^{\Sym_a}=0$ on $T^6/(\Z_2\times\Z_6\times\OR)$ with $\eta = -1$}} \\
\hline
\hline 
\multicolumn{2}{|c||}{} & \multicolumn{3}{|c|}{(\chi^{\Anti_a}_{a}, \chi^{\Anti_a}_{(\omega a)}, \chi^{\Anti_a}_{(\omega^2 a)} )}\\\hline
(n_a^1, m_a^1)& \varrho &(I) & (II) & (III) \\
\hline \hline
\begin{picture}(0,0) \put(-20,-30){\begin{rotate}{90}\text{\bf aAA}\end{rotate} }\end{picture}
 (1,1) & 3 & (0,2,0) & (0,2,0) & (0,0,-2) \text{ or } (-2,0,0)   \\
(1,2) & 3/2 & - & (-1,3,-1) & (-1,0,-3) \text{ or } (-3,0,-1)  \\
(1,3) &1 & - & (-2,4,-2) & (-2,0,-4) \text{ or } (-4,0,-2)  \\
(1,4) &3/4  & - & (-3,5,-3) & (-3,0,-5) \text{ or } (-5,0,-3)  \\
(1,5) & 3/5 & - &(-4,6,-4) & (-4,0,-6) \text{ or } (-6,0,-4)  \\
(1,6) &1/2  & - &(-5,7,-5) & (-5,0,-7) \text{ or } (-7,0,-5)  \\
\hline \hline
\begin{picture}(0,0) \put(-20,-20){\begin{rotate}{90}\text{\bf bAA}\end{rotate} }\end{picture}
(1,0) &6 & (0,1,0) & (0,1,0) & (0,0,-1)  \text{ or } (-1,0,0)   \\
 (1,1) & 2& - & (-1,2,-1) & (-1,0,-2) \text{ or } (-2,0,-1)   \\
 (1,2) & 6/5 & - & (-2,3,-2)  & (-2,0,-3) \text{ or } (-3,0,-2) \\
 (1,3) & 6/7 & - & (-3,4,-3)  & (-3,0,-4) \text{ or } (-4,0,-3) \\ 
 (1,4) & 2/3 & - & (-4,5,-4)  & (-4,0,-5) \text{ or } (-5,0,-4) \\ 
\hline
\end{array}
}{RigidSymmFreeAntisymmetrics}{Overview of the net-chirality $(\chi^{\Anti_a}_{a}, \chi^{\Anti_a}_{(\omega a)}, \chi^{\Anti_a}_{(\omega^2 a)} )$ per sector $(\omega^k a)(\omega^k a)'$ for the rigid fractional three-cycles without chiral states in the symmetric representations on the {\bf aAA} and {\bf bAA} lattice, see eqs.~(\ref{Eq:RigidNoSymmAntiaAA}) and (\ref{Eq:RigidNoSymmAntibAA}). The first result in the column with configuration $(III)$ corresponds to the case where the $\OR\Z_2^{(2)}$-plane is the exotic O6-plane, while the second result gives the net-chirality with the $\OR\Z_2^{(3)}$-plane as the exotic O6-plane.}

Searching suitable fractional three-cycles to harbour an $SU(5)$ GUT gauge group requires three chiral multiplets in the antisymmetric representation and no chiral states in the symmetric representation, while the rigidity constraint can be loosened as discussed in the previous section. More concretely, chiral multiplets in the ${\bf 10}$ (antisymmetric) representation are required to embed the quarks and leptons into consistent representations under $SU(5)$, but massless chiral multiplets in the {\bf 15} symmetric representation are excluded based on phenomenological grounds. Allowing for a multiplet in the adjoint representation on the other hand could offer a canonical way to break the $SU(5)$ gauge group down to the Standard Model gauge group. Going through the list of fractional cycles with at least one chiral multiplet in the adjoint representation of the gauge group, one encounters the first category containing three-cycles with bulk orbits parallel to the $\OR$-plane or to an orbit of the form $(n_a^1, m_a^1;1,0;1,-1)$. However, none of these fractional three-cycles comes with three chiral multiplets in the antisymmetric representation, they are therefore a priori excluded as candidates for an $U(5)$ GUT stack. The second set of three-cycles one encounters are those cycles with a bulk orbit parallel to the $\OR\Z_2^{(2)}$- or to the $\OR\Z_2^{(3)}$-plane. In both cases, the corresponding bulk wrapping numbers ($Q_a = 0$, $2 \tilde U_a = - \tilde V_a = 2$) cause the intersection number with the O6-planes to vanish, such that the net-chiralities satisfy $\chi^{\Anti_a} = \chi^{\Sym_a} \stackrel{!}{=} 0$. These fractional three-cycles are therefore also excluded from accommodating an $U(5)$ GUT gauge group.

Turning to the last class of candidate fractional three-cycles, namely those with bulk orbits of the form $(n_a^1,  m_a^1;1,-1;1,1)$ or $(n_a^1,  m_a^1;1,1;1,-1)$, one infers from their bulk wrapping numbers ($P_a = - 2 Q_a = 2 n_a^1$, $ \tilde U_a = - 2 \tilde V_a = 2 \tilde m_a^1 $) that chiral matter in the antisymmetric representation can be realised depending on the choice of exotic O6-plane and the one-cycle wrapping numbers $(n_a^1,m_a^1)$. Requiring that the corresponding fractional three-cycles are free from chiral states in the symmetric representations and characterised by exactly three chiral multiplets in the antisymmetric representation, i.e.~$(\chi^{\Anti_a},\chi^{\Sym_a}) \stackrel{!}{=}( \pm 3,0)$, eliminates most of the candidate fractional three-cycles discussed in the previous section. 
On the  {\bf aAA} lattice, none of the 56 bulk orbits allow to construct fractional three-cycles satisfying the constraints with $\chi^{\Anti_a} = +3$, while only two candidate bulk orbits on the {\bf bAA} lattice match the requirements $(\chi^{\Anti_a},\chi^{\Sym_a})=( + 3,0)$: the orbits $(4,-1;1,1;1,-1)$ and $(4,-1;1,-1;1,1)$ provided that the complex structure modulus takes the value $\varrho=4$ and that the $\OR$-plane is the exotic O6-plane $(\eta_{\OR}=-1$) to avoid violations of the (bulk) RR tadpole cancellation conditions.
For a fractional three-cycle with bulk orbit $(4,-1;1,1;1,-1)$, the net-chirality of matter in the (anti-)symmetric representation per sector is given by:
\begin{equation}\label{Eq:NonRigidNoSymmAntibAA}
\chi^{\Anti_a/\Sym_a}_{(\omega^k a)} \stackrel{\eta_{\OR} = -1}{=} \left\{ \begin{array}{lc} \frac{1}{2} \left(    (-)^{\sigma_a^1 \tau_a^1} -  (4 \mp 1)  \right) & k=0,  \\ 
 \frac{1}{2} \left(  4 (-)^{\sigma_a^2 \tau_a^2} +  (-)^{\sigma_a^1 \tau_a^1 + \sigma_a^2 \tau_a^2}  \pm3  \right) & k=1, \\
\frac{1}{2} \left(  4 (-)^{\sigma_a^3 \tau_a^3} - 3 (-)^{\sigma_a^1 \tau_a^1 + \sigma_a^3 \tau_a^3}   \mp1   \right) & k=2,
   \end{array}  \right. 
\end{equation}
and for a fractional three-cycle with bulk orbit $(4,-1;1,-1;1,1)$ the net-chiralities read per sector:
\begin{equation}\label{Eq:NonRigidNoSymmAntibAA2}
\chi^{\Anti_a/\Sym_a}_{(\omega^k a)} \stackrel{\eta_{\OR} = -1}{=} \left\{ \begin{array}{lc}  \frac{1}{2} \left(    (-)^{\sigma_a^1 \tau_a^1} -  (4 \mp 1)  \right)  & k=0,  \\ 
\frac{1}{2} \left(  4 (-)^{\sigma_a^2 \tau_a^2} - 3 (-)^{\sigma_a^1 \tau_a^1 + \sigma_a^2 \tau_a^2}   \mp1   \right)& k=1, \\
 \frac{1}{2} \left(  4 (-)^{\sigma_a^3 \tau_a^3} +  (-)^{\sigma_a^1 \tau_a^1 + \sigma_a^3 \tau_a^3}  \pm3  \right) & k=2.
   \end{array}  \right. 
\end{equation}
Although the net-chirality for the matter in the symmetric representation is non-vanishing for separate $(\omega^k a)(\omega^k a)^{\prime}_{k=0,1,2}$ sectors, the sum over all sectors adds up to a total net-chirality $(\chi^{\Anti_a},\chi^{\Sym_a}) =(3,0)$ when the discrete displacements $\vec{\sigma}_a$ and Wilson lines $\vec{\tau}_a$ satisfy $\sigma_a^1 \tau_a^1 = \sigma_a^2 \tau_a^2 = \sigma_a^3 \tau_a^3 =0$.
Assuming one of these $2^2 \cdot 27$ configurations for the discrete parameters, a fractional three-cycle with e.g.~the second bulk orbit is characterised by the net-chiralities $(\chi^{\Anti_a}_{a}, \chi^{\Anti_a}_{(\omega a)}, \chi^{\Anti_a}_{(\omega^2 a)})=(-1,0,4)$ and $(\chi^{\Sym_a}_{a},\chi^{\Sym_a}_{(\omega a)}, \chi^{\Sym_a}_{(\omega^2 a)}) = (-2,1,1)$ as dictated by expression (\ref{Eq:NonRigidNoSymmAntibAA2}).  

Considering the opposite sign for the net-chirality $\chi^{\Anti_a}$, i.e.~$(\chi^{\Anti_a},\chi^{\Sym_a})=(-3,0)$, provides us with another pair of bulk orbit candidates of the form $(1,m_a^1;1,1;1,-1)$ and $(1,m_a^1;1,-1;1,1)$, where $m_a^1  = 2$ for the {\bf aAA} lattice with the value $\varrho = \frac{1}{2}$ of the complex structure modulus  and $m_a^1 = 1$ for the {\bf bAA} lattice with $\varrho = \frac{2}{3}$. Aiming for the construction of global GUT models, one has to assume that the $\OR$-plane is the exotic O6-plane, given that other choices for the exotic O6-plane imply a violation of the RR bulk tadpole conditions. Under these considerations the fractional three-cycles with bulk orbit   
$(1,m_a^1;1,1;1,-1)$ are characterised by the following net-chiralities of matter in (anti-)symmetric representations:
\begin{equation}\label{Eq:NonRigidNoSymmAntiNegChiral}
\chi^{\Anti_a/\Sym_a}_{(\omega^k a)} \stackrel{\eta_{\OR} = -1}{=} \left\{ \begin{array}{lc} -\frac{1}{2} \left(  1 \pm 1\right) \left( \tilde m_a^1 -  1 + b  \right)& k=0,  \\ 
\frac{1}{2} \left( \tilde m_a^1 + 1-b \right) \left( (-)^{\sigma_a^2 \tau_a^2} \pm 1\right) & k=1, \\
   \frac{1}{2} \left[  \tilde m_a^1 \left( (-)^{\sigma^3_a \tau_a^3} \mp 3 \right) - (1-b) \left( 3 (-)^{\sigma_a^3 \tau_a^3} \mp 1 \right)  \right] &  k=2,
   \end{array}  \right. 
\end{equation}
and for the fractional three-cycles with bulk orbit $(1,m_a^1;1,-1;1,1)$ the net-chirality per sector is given by:
\begin{equation}\label{Eq:NonRigidNoSymmAntiNegChiral2}
\chi^{\Anti_a/\Sym_a}_{(\omega^k a)} \stackrel{\eta_{\OR} = -1}{=} \left\{ \begin{array}{lc} -\frac{1}{2} \left(  1 \pm 1\right) \left( \tilde m_a^1 -  1 + b  \right)& k=0,  \\ 
 \frac{1}{2} \left[  \tilde m_a^1 \left( (-)^{\sigma^2_a \tau_a^2} \mp 3 \right) - (1-b) \left( 3 (-)^{\sigma_a^2 \tau_a^2} \mp 1 \right)  \right] & k=1, \\
\frac{1}{2} \left( \tilde m_a^1 + 1-b \right) \left( (-)^{\sigma_a^3 \tau_a^3} \pm 1\right) & k=2.
   \end{array}  \right. 
\end{equation}
These expressions for the net-chiralities indicate that the condition $(\chi^{\Anti_a},\chi^{\Sym_a})\stackrel{!}{=}(-3,0)$ is satisfied when the discrete Wilson lines and displacements are chosen such that $\sigma_a^2 \tau_a^2 = 1 = \sigma_a^3 \tau_a^3$. Subsequently, per bulk orbit there exist $2^2 \cdot 4$ discrete parameter configurations for which the corresponding fractional three-cycles are apt to support an $U(5)$ GUT stack.

Not only chiral matter but also non-chiral matter in the symmetric and antisymmetric representations should be taken into consideration when constructing phenomenologically appealing models. Determining the total amount of non-chiral matter in these sectors can be done by calculating the beta-function coefficients, see e.g.~table 7 in~\cite{Honecker:2011sm} or table 39 in~\cite{Honecker:2012qr}, depending on the angles $(\vec{\phi}_{(\omega^k a) (\omega^k a)'})$ between a cycle $(\omega^k a)$ and its orientifold image $(\omega^k a)'$. Three different configurations have to be distinguished:

{\bf (1) Three vanishing angles: $(\vec{\phi}_{(\omega^k a) (\omega^k a)'}) = (0,0,0)$}\\
This configuration is realised for three-cycles $a$ parallel to one of the O6-plane orbits, and we assume that the displacements $(\vec{\sigma}_a)$ and discrete Wilson lines $(\vec{\tau}_a)$ do not conspire to combinations matching the topological conditions for gauge group enhancement in table~\ref{Tab:Conditions-on_b+t+s-SOSp}. Taking for instance a fractional three-cycle parallel to the $\OR$-plane, the beta-function coefficient describing the contributions of matter in the symmetric and antisymmetric representation coming from the $aa'$ sector reads: 
\begin{equation}
\begin{aligned}
b_{aa'}^{\cal A} + b_{aa'}^{\cal M} =& N_a \left[ \eta_{(1)} (-)^{\sigma_a^2 \tau_a^2 + \sigma_a^3 \tau_a^3} + \eta_{(2)} (-)^{2b \sigma_a^1 \tau_a^1 + \sigma_a^3 \tau_a^3} + \eta_{(3)} (-)^{2b \sigma_a^1 \tau_a^1 + \sigma_a^2 \tau_a^2 } \right] \\
 &- 2 \eta_{\OR} \left[ \eta_{(1)} (-)^{2b \sigma_a^1 \tau_a^1} + \eta_{(2)} (-)^{ \sigma_a^2 \tau_a^2} +  \eta_{(3)} (-)^{ \sigma_a^3 \tau_a^3} \right],
\end{aligned}
\end{equation}
and it depends on the choice of the exotic O6-plane, the lattice configuration (through the discrete parameter $b$), and combinations of displacements $(\vec{\sigma}_a)$ and discrete Wilson lines $(\vec{\tau}_a)$. For the {\bf aAA} lattice, the contribution from the $aa'$ sector to the beta-function coefficient can be summarised as:
\begin{equation}
b_{aa'}^{\cal A} + b_{aa'}^{\cal M} = \left\{ \begin{array}{llll} N_a + 2 & (\eta_{\OR} = -1, \, \sigma_a^2 \tau_a^2 = \sigma_a^3 \tau_a^3=1), & \text{or} & (\eta_{\OR\Z_2^{(l)}} = -1, \, \sigma_a^l \tau_a^l = 0 \neq \sigma_a^k \tau_a^k),    \\ 
N_a - 2 & (\eta_{\OR} = -1,  \,\sigma_a^2 \tau_a^2 \neq  \sigma_a^3 \tau_a^3), & \text{or}  &  (\eta_{\OR\Z_2^{(l)}} = -1, \, \sigma_a^2 \tau_a^2 =  \sigma_a^3 \tau_a^3),     \end{array}\right.
\end{equation}
with $l, k \in \{ 2,3\}$ and the parameter combinations $(\sigma_a^k \tau_a^k)$ chosen such that gauge group enhancement is excluded. The results for the {\bf bAA} lattice configuration are completely analogous to the ones discussed in section 3.3 of~\cite{Honecker:2012qr}.  Hence, in case a fractional three-cycle with bulk orbit parallel to the $\OR$-plane does not support an enhanced gauge group, the $aa'$ sector yields a non-chiral pair $\Anti_a + \ov \Anti_a$ or $\Sym_a + \ov \Sym_a$ depending on the specific choice of the exotic O6-plane and the discrete parameters $(\vec{\sigma}_a)$ and $(\vec{\tau}_a)$. The same analysis can be done for the fractional three-cycles parallel to  one of the other three O6-plane orbits, leading to similar expressions.

{\bf (2) One vanishing angle: $(\vec{\phi}_{(\omega^k a) (\omega^k a)'}) = (0_i, \phi_j, -\phi_k)$}\\
This situation occurs if the bulk orbit of the three-cycle is parallel to two O6-planes along a single two-torus. As an example one can consider the $(\omega^k a)(\omega^k a)'_{k=1,2}$ sectors of the bulk orbit parallel to the $\OR$-plane, whose orbifold images $(\omega^k a)_{k=1,2}$ remain parallel to the $\OR$-plane and the $\OR\Z_2^{(1)}$-plane along $T_{(1)}^2$. Hence, the contribution to the beta-function coefficient from the $(\omega^k a)(\omega^k a)'_{k=1,2}$ sectors follows from the third case in table 7 of~\cite{Honecker:2011sm} and reads for $k=1,2$:  
\begin{equation}
b_{(\omega^ka) (\omega^ka)'}^{\cal A} + b_{(\omega^ka)(\omega^ka)'}^{\cal M} = \frac{1}{2} \left(  1 + \eta_{(1)}\right)  \left[\frac{N_a}{2} - (-)^{2b \sigma_a^1 \tau_a^1} \eta_{\OR} \right]. 
\end{equation}
For choices of exotic O6-planes where $\eta_{(1)} = + 1$, the sum of both sectors adds up to a non-chiral pair  $\Anti_a + \ov \Anti_a$ or a non-chiral pair $\Sym_a + \ov \Sym_a$, depending on the discrete parameter combination $ 2b \sigma_a^1 \tau_a^1$. 

Also the analysis of the $aa'$ and $(\omega^2 a) (\omega^2 a)'$ sectors for the eleven types of fractional three-cycles discussed in table~\ref{tab:RigidSymmFreeAntisymmetrics} falls into this category. Taking fractional three-cycles with bulk orbits $(1,m_a^1;1,0;1,-1)$ with $m_a^1 \in \{1,2,3, 4, 5, 6 \}$ for the {\bf aAA} lattice and $m_a^1 \in \{0,1, 2, 3, 4\}$ for the {\bf bAA} lattice, the contributions to the beta-function coefficient encoding the amount of  matter in the symmetric and antisymmetric representation read:
\begin{equation}\label{Eq:BetaFunctionSituation2RigidNoSymm}
b_{(\omega^ka) (\omega^ka)'}^{\cal A} + b_{(\omega^ka)(\omega^ka)'}^{\cal M} = \left\{ \begin{array}{ll} \left(  \tilde m_a^1 + (1-b) \eta_{(2)} \right)  \left[\frac{N_a}{2} - (-)^{ \sigma_a^2 \tau_a^2}  \eta_{\OR}  \right] & k=0, \\ \left(  \tilde m_a^1 + (1-b) \eta_{(3)}\right)  \left[\frac{N_a}{2} - (-)^{ \sigma_a^3 \tau_a^3}  \eta_{\OR}  \right] & k=2.  \end{array} \right.
\end{equation}
The amount of chiral matter in the (anti-)symmetric representation in these sectors is proportional to the torus wrapping number $m_a^1$ and only has a chance to vanish with the $\OR$-plane as the exotic O6-plane (i.e.~$\eta_{(2)} = \eta_{(3)} = -1$) and with $m_a^1 = 1$ ($m_a^1 = 0$) for the {\bf aAA} ({\bf bAA}) lattice. For the other values of $m_a^1$ and for other choices of the exotic O6-plane, the contributions to the beta-coefficient in these sectors do not vanish, and the amount of chiral matter in the antisymmetric representation depends on the lattice choice and the integer $m_a^1$. Note that the amount of matter in the antisymmetric representation per sector matches the net-chiralities listed in table~\ref{tab:RigidSymmFreeAntisymmetrics} under the considered configurations. Moreover, by combining the beta-function coefficients~(\ref{Eq:BetaFunctionSituation2RigidNoSymm}) per sector with the expressions for the net-chiralities per sector in equations~(\ref{Eq:RigidNoSymmAntiaAA}) and~(\ref{Eq:RigidNoSymmAntibAA}), one can deduce that no {\it non-chiral} states in the symmetric representation appear for configurations without {\it chiral} states in the symmetric representation.

{\bf (3) Three non-vanishing angles: $ {\phi}^{(i)}_{(\omega^k a) (\omega^k a)'} \neq 0, \, \forall\, i  $}\\
An example for the last configuration consists in the $(\omega a) (\omega a)'$ sector of the fractional three-cycles with bulk orbit listed in table~\ref{tab:RigidSymmFreeAntisymmetrics}. The beta-function coefficient is now computed following the last category in table 7 of~\cite{Honecker:2011sm} and reads (combining both lattice configurations): 
\begin{equation}
\begin{aligned}
b_{(\omega a) (\omega a)'}^{\cal A} + b_{(\omega a)(\omega a)'}^{\cal M} &=\left( \frac{N_a}{2} + \eta_{\OR} \right) \left[  \frac{ \tilde m_a^1 (1 - \eta_{(1)}) -  (1-b) (\eta_{(2)} + \eta_{(3)})}{2}  \right] \\
&= \left\{ \begin{array}{ll}  \left( \frac{N_a}{2} -1 \right) \left[  \tilde m_a^1 + 1-b\right]    & \eta_{\OR} = -1, \\ 0 & \text{else}. \end{array}\right.
\end{aligned}
\end{equation}
Chiral matter in the antisymmetric representation only appears in this sector if the $\OR$-plane is chosen to be the exotic O6-plane, in which case the number of  chiral multiplets transforming in the antisymmetric representation is determined by the torus wrapping number $m_a^1$ and the lattice configuration. Observe that also here the amount of matter in the antisymmetric representation matches the net-chirality in the $(\omega a) (\omega a)'$ sector given in table~\ref{tab:RigidSymmFreeAntisymmetrics}, and the possible appearance of non-chiral states in the symmetric representation is completely absent in this sector.

In conclusion, we have identified two classes of completely rigid fractional three-cycles free from chiral states in the symmetric representation: the first class consists of three-cycles with bulk orbit parallel to the $\OR$-plane and is also free from chiral multiplets in the antisymmetric representation, while the second class contains three-cycles with bulk orbit $(n_a^1, m_a^1; 1,0; 1,-1)$ as listed in table~\ref{tab:AdjointSymmetricFree}. Every fractional three-cycle in this list is accompanied by massless states in the antisymmetric representation, and their appropriateness to support the {\it QCD} D6-brane stack is constrained by the requirement that the total antisymmetric net-chirality $| \chi^{\Anti_a}|$ has to lie within $\{0,1,2,3\}$. For the second class of three-cycles, this consideration only leaves the eleven bulk orbits presented in  table~\ref{tab:RigidSymmFreeAntisymmetrics} as candidates to accommodate the {\it QCD} stack. Fractional three-cycles suitable for $SU(5)$ GUT model building have also been identified by imposing the conditions $| \chi^{\Anti_a}| = 3$ and $\chi^{\Sym_a} = 0$: the {\bf aAA} lattice harbours two bulk orbits satisfying these constraints, whereas the {\bf bAA} lattice allows for four candidate bulk orbits.

\subsection{Towards three generations}\label{Ss:3gens}
\begin{figure}
\centering
\begin{tikzpicture}[node distance=1cm, auto]  
\tikzset{
    mynode1/.style={rectangle,rounded corners,draw=black, top color=white, bottom color=yellow!60,very thick, inner sep=0.8em, minimum size=2.6em, text centered},
    mynode2/.style={rectangle,rounded corners,draw=black, top color=white, bottom color=orange!30,very thick, inner sep=0.8em, minimum size=2.6em, text centered},
    mynode3/.style={rectangle,rounded corners,draw=black, top color=white, bottom color=green!30,very thick, inner sep=0.8em, minimum size=2.6em, text centered},
    mynode4/.style={rectangle,rounded corners,draw=black, top color=white, bottom color=violet!30,very thick, inner sep=0.8em, minimum size=2.6em, text centered},
    myarrow1/.style={->, >=latex', shorten >=1pt, thick, color=black},
    myarrow2/.style={->, >=latex', shorten >=1pt, thick, color=black},
    myarrow3/.style={->, >=latex', shorten >=1pt, thick, color=black},
    mylabel/.style={text width=12em, text centered} 
}  
\node[mynode1] (rho-modulus) {Complex structure modulus $\varrho$};
\node[below = 3cm of rho-modulus] (dummy) {};
\node[mynode1,left= 1cm of dummy](rhoind){$\begin{array}{c} \varrho\text{-indep. configurations}\\ \text{(section~\ref{Sss:3GenRhoIndep})} \end{array}$ };
\node[mynode1,right= 2cm of dummy](rhodep){$\begin{array}{c}\varrho\text{-dep. configurations}\\ \text{(section~\ref{Sss:3GenRhoDep})} \end{array}$};
\node[mylabel, below= 1.2 cm of dummy] (label 1) {{\color{myorange}\bf $a$-stack: rigid, $\chi^{\Sym_a} =0$ and $|\chi^{\Anti_a}|\leq 3$}};
\node[mynode2,below= 2.5cm of rhoind](QCDrhoind){$a$-stack $\pp$ $\OR$};
\node[mynode3, below= 2.5cm of QCDrhoind] (QCDrhoindbstack) {$\begin{array}{ll}b\text{-stack} \pp \OR & {\color{red} \lightning}   \\ b\text{-stack} \pp \OR\Z_2^{(1)} & \text{\color{mygr} \checkmark} \\ b\text{-stack} \pp \OR\Z_2^{(2)} & {\color{red} \lightning}   \\ b\text{-stack} \pp \OR\Z_2^{(3)} & {\color{red} \lightning}    \end{array}$} ;
\node[below=3cm of rhodep] (dummy2) {};
\node[mynode2, left= 0.5cm of dummy2] (QCDrhodep1) {$a$-stack $\pp$ $\OR$};
\node[mynode3, below= 2.4cm of QCDrhodep1] (QCDrhodep1bstack) {$\begin{array}{c}b\text{-stack} \pp \\ (1,m_a^1;1,0;1,-1)  \\ \text{(table~\ref{tab:AdjointSymmetricFree})}\end{array}$};
\node[mynode4, below= 1cm of QCDrhodep1bstack] (QCDrhodep1bstack3Chiral) {$\begin{array}{c} \text{three chiral generations}\\ \text{(table~\ref{tab:NetChiralitiesRhoDependentAppOR})}  \end{array}$};
\node[mynode2, right= 1.5cm of dummy2] (QCDrhodep2) {$\begin{array}{c}a\text{-stack} \pp \\ (1,m_a^1;1,0;1,-1) \\ \text{(table~\ref{tab:RigidSymmFreeAntisymmetrics})}\end{array}$};
\node[mynode3, below= 1.5cm of QCDrhodep2] (QCDrhodep2bstack) {$\begin{array}{c}b\text{-stack} \pp \OR, \OR\Z_2^{(i)}, \text{or} \\ (1,m_a^1;1,0;1,-1)  \\ \text{(table~\ref{tab:AdjointSymmetricFree})}\end{array}$};
\node[mynode4, below= 1cm of QCDrhodep2bstack] (QCDrhodep2bstack3Chiral) {$\begin{array}{c} \text{three chiral generations}\\ \text{(table~\ref{tab:BulkOrbits3QuarkGenerations})}  \end{array}$};
\node[mylabel, below= 2.5 cm of label 1] (label 2) {{\color{mygr}\bf $b$-stack: rigid and $\chi^{\Sym_b} = 0$}};

NetChiralitiesRhoDependentAppOR

\draw[myarrow1] (rho-modulus.south) -- ++(0,0) -- ++(0,-1) -|  (rhoind.north);	
\draw[myarrow1] (rho-modulus.south) -- ++(0,0) -- ++(0,-1) -|  (rhodep.north);
\draw[myarrow2] (rhoind.south) -- ++(0,0) -- ++(0,0) -|  (QCDrhoind.north);
\draw[myarrow2] (rhodep.south) -- ++(-0,0) -- ++(0,-1) -|  (QCDrhodep1.north);	
\draw[myarrow2] (rhodep.south) -- ++(0,0) -- ++(0,-1) -|  (QCDrhodep2.north);	
\draw[myarrow3] (QCDrhoind.south) -- ++(0,0) -- ++(0,0) -|  (QCDrhoindbstack.north);
\draw[myarrow3] (QCDrhodep1.south) -- ++(0,0) -- ++(0,0) -|  (QCDrhodep1bstack.north);
\draw[myarrow1] (QCDrhodep1bstack.south) -- ++(0,0) -- ++(0,0) -|  (QCDrhodep1bstack3Chiral.north);
\draw[myarrow3] (QCDrhodep2.south) -- ++(0,0) -- ++(0,0) -|  (QCDrhodep2bstack.north);
\draw[myarrow1] (QCDrhodep2bstack.south) -- ++(0,0) -- ++(0,0) -|  (QCDrhodep2bstack3Chiral.north);

\end{tikzpicture} 
\medskip
\caption{Roadmap of the search for MSSM-like intersecting D6-brane models with three chiral generations on the orbifold $T^6/(\Z_2 \times \Z_6 \times \OR)$ with discrete torsion $\eta=-1$, both on the {\bf aAA} and {\bf bAA} lattice, as presented in sections~\ref{Sss:3GenRhoIndep} and \ref{Sss:3GenRhoDep}. The symbol {\color{mygr} \checkmark} (${\color{red} \lightning} $) indicates that three chiral generations can(not) be realised in the D6-brane configurations under consideration.\label{Fig:Roadmap3Generations}} 
\end{figure}
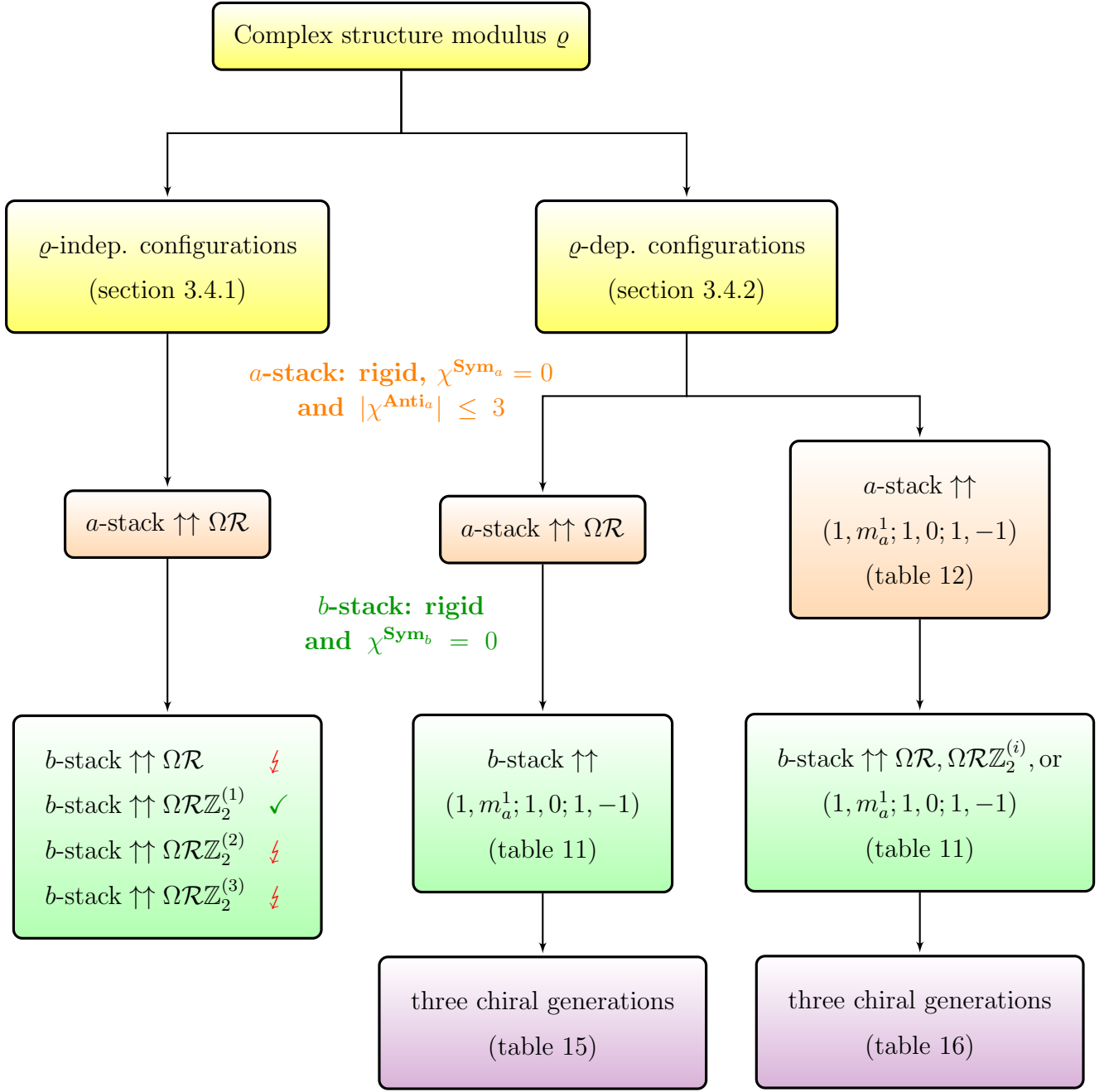

In the previous sections, we gave a rigorous classification of fractional three-cycles depending on the number of accompanying chiral matter in the adjoint, symmetric and/or antisymmetric representation. This classification allows us to identify suitable candidate three-cycles to accommodate the D6-brane stacks responsible for the gauge symmetries of the supersymmetric Standard Model or some GUT extension, as alluded to at the end of the previous section. In this section, we distinguish two ans\"{a}tze with $\varrho$-independent and $\varrho$-dependent configurations as depicted in figure~\ref{Fig:Roadmap3Generations}, and investigate under which conditions two fractional three-cycles can intersect to yield three left-handed quark generations.

\subsubsection{MSSM-like models with three generations: $\varrho$-independent configurations}\label{Sss:3GenRhoIndep}

Due to the absence of chiral matter in the adjoint representation, the completely rigid fractional three-cycles in section~\ref{Ss:NoAdjoints} represent the best candidates to embed the {\em QCD}-stack and the $SU(2)_L$-stack. In first instance, one might wonder whether a three-generation model can be found with two completely rigid fractional three-cycles {\it irrespective} of the value of the complex structure modulus $\varrho$. Under these assumptions, both fractional three-cycles are required to have a bulk orbit parallel to the $\OR$-plane, for which the gauge group can enhance to $USp(2N)$ or $SO(2N)$ for specific combinations of discrete displacements $(\vec{\sigma}_a)$ and discrete Wilson lines $(\vec{\tau}_a)$ as observed in section~\ref{Ss:Enhance}. An $USp(2)$ gauge group can be used to model the {\it left} stack, but gauge group enhancement should definitely be avoided for the {\em QCD} stack by excluding the displacements and Wilson lines along $T_{(2)}^2 \times T_{(3)}^2$ displayed in table~\ref{Tab:Conditions-on_b+t+s-SOSp}.
A schematic overview of the net-chiralities $(\chi^{ab}, \chi^{ab'})$ for matter in the bifundamental representation between the {\em QCD}-stack (D6$_a$-branes) and the {\it left} stack (D6$_b$-branes) is displayed in table~\ref{tab:NetChiral3GenRhoInd} taking into account the various background configurations and choices for the left stack. The candidate fractional three-cycles for the {\em QCD}-stack are selected from the set of completely rigid (with no matter in the adjoint representation) three-cycles with bulk orbit parallel to the $\OR$-plane and without gauge group enhancement. 

\mathtabfix{
\begin{array}{|c||c|c||c|c|}
\hline
\multicolumn{5}{|c|}{\text{\bf $(\chi^{ab}, \chi^{ab'})$ between D6-brane stacks $a$ and $b$ for $\varrho$-independent configurations on $T^6/(\Z_2\times\Z_6\times\OR)$}} \\
\hline 
\hline
&\multicolumn{2}{|c||}{\text{\bf $D6_a$ stack on aAA lattice}} &\multicolumn{2}{|c|}{\text{\bf $D6_a$ stack on bAA lattice}} \\
\text{\bf $D6_b$ stack} & \eta_{\OR} = -1 &  \eta_{\OR\Z_2^{(2,3)}} = -1  & \eta_{\OR} = -1 & \eta_{\OR\Z_2^{(2,3)}} = -1  \\
\hline \hline
\text{$U(2)_b$ without adjoints} & (0,0) & (0,0), (\pm 2, 0), (0, \pm 2) & (0,0) &(0,0), (\pm 2, 0), (0, \pm 2)\\
\text{$U(2)_b$ with adjoints} & (0,0), (\pm 1, \pm 1)  & (0,0), (\pm 1, \mp 1) & (0,0), (\pm 1, \pm 1)  &  (0,0), (\pm 1, \mp 1) \\
\text{enhanced $USp(2)_b$ } & (0,0), (\pm 2, \pm 2)  &  (0,0), (\pm 1, \pm 1) & (0,0), (\pm 2, \pm 2)  & (0,0), (\pm 1, \pm 1)  \\
\hline
\end{array}
}{NetChiral3GenRhoInd}{Overview of the net-chiralities $(\chi^{ab}, \chi^{ab'})$ between a completely rigid D6-brane stack $a$ (without matter in the adjoint representation) and a D6-brane stack $b$ for the two inequivalent lattice configurations {\bf aAA} and {\bf bAA} and the three choices for the exotic O6-plane ($\eta_{\OR} = -1$, $\eta_{\OR\Z_2^{(2)}} = -1$ or $\eta_{\OR\Z_2^{(3)}} = -1$ ) on the orientifold $T^6/(\Z_2 \times \Z_6 \times \OR)$ with discrete torsion $\eta = -1$. Both stacks have a bulk orbit parallel to the $\OR$-plane, implying that the results are valid for all values of the complex structure modulus $\varrho$. The stack $a$ also comes with two non-chiral matter pairs in the antisymmetric and/or symmetric representation, depending on the choice of the exotic O6-plane and the values of the displacements and Wilson lines. } 

Table~\ref{tab:NetChiral3GenRhoInd} shows that two fractional three-cycles with bulk orbits parallel to the $\OR$-plane do not give rise to models with three left-handed quark generations, i.e. $\left| \chi^{ab} + \chi^{ab'} \right| \neq 3$. However, this does not imply that all $\varrho$-independent configurations are thereby excluded. 
For a $USp(2)$ gauge group, massless states in the antisymmetric representation are gauge singlets, which do not constitute any obstruction to  model building, similar to the antisymmetric representations $(\1)_{\pm 2}$ of $U(2) \simeq SU(2) \times U(1)$.
This implies that the {\it left} stack can also be wrapped on a fractional three-cycle with bulk orbit parallel to one of the three $\OR\Z_2^{(i)}$-planes and supporting an enhanced $USp(2)_b$ gauge group, see table~\ref{Tab:Conditions-on_b+t+s-SOSp}. Under these considerations, the $b$-stack is identified with its orientifold image $b'$ and the three chiral quark generations have to be realised with net-chirality $\left| \chi^{ab} \right| = \left| \chi^{ab'} \right| = 3$. Table~\ref{tab:NetChiral3GenRhoIndAddO6} provides an overview of the realisable net-chiralities $\chi^{ab}$ for an $a$-stack parallel to the $\OR$-plane and a $b$-stack parallel to one of the $\OR\Z_2^{(i)}$-planes with enhanced $USp(2)$ gauge group. The immediate conclusion drawn from this table is that only a configuration with the $b$-stack parallel to the $\OR\Z_2^{(1)}$-plane allows for three generations of chiral quarks, both on the {\bf aAA}-lattice as well as on the {\bf bAA}-lattice. A closer investigation reveals that there are 192 combinations of $\Z_2^{(i)}$ eigenvalues, discrete Wilson lines and displacements for the $a$- and $b$-stack yielding three generations on the {\bf aAA}-lattice, and 144 combinations yielding three generations on the {\bf bAA}-lattice.

\mathtabfix{
\begin{array}{|c||c|c||c|c|}
\hline 
\multicolumn{5}{|c|}{\text{\bf $\chi^{ab} = \chi^{ab'}$ between D6-brane stacks $a$ and $b$ for $\varrho$-independent configurations on $T^6/(\Z_2\times\Z_6\times\OR)$}}\\
\hline \hline
& \multicolumn{2}{|c||}{\text{\bf $D6_a$ stack $\upuparrows \OR$ on aAA lattice}} & \multicolumn{2}{|c|}{\text{\bf $D6_a$ stack $\upuparrows \OR$ on bAA lattice}}\\
\text{\bf $D6_b$ stack} & \text{\bf exotic O6-plane} & \chi^{ab} & \text{\bf exotic O6-plane} & \chi^{ab} \\ 
\hline 
\hline
\upuparrows \OR\Z_2^{(1)} + 5\, \Anti&   \eta_{\OR\Z_2^{(2,3)}} = -1 & 0, \pm 1, \pm 3  &   \eta_{\OR\Z_2^{(2,3)}} = -1 &  0, \pm 1, \pm 3 \\
\upuparrows \OR\Z_2^{(2)} + 2\, \Anti&  \eta_{\OR} = -1  & \pm1, \pm2 &  \eta_{\OR} = -1 & \pm2, \pm4 \\
\upuparrows \OR\Z_2^{(3)} + 2\, \Anti&  \eta_{\OR} = -1 & \pm1, \pm2  &  \eta_{\OR} = -1 &\pm2, \pm4 \\
\hline
\end{array}
}{NetChiral3GenRhoIndAddO6}{Summary of the net-chiralitiy $\chi^{ab}= \chi^{ab'}$ between a completely rigid D6-brane stack $a$ (without matter in the adjoint rep.) and a D6-brane stack $b$ with enhanced $USp(2)$ gauge group on the orientifold $T^6/(\Z_2 \times \Z_6 \times \OR)$ with discrete torsion $\eta = -1$. The bulk orbit of the $a$-stack is parallel to the $\OR$-plane, while the $b$-stack has an orbit parallel to one of the $\OR\Z_2^{(i=1,2,3)}$-planes and is accompanied by massless states in the {\it antisymmetric} representation as indicated in the left column. The second and fourth column indicate the choice of the exotic O6-plane. The third and fifth column list the realisable net-chiralities $\chi^{ab}= \chi^{ab'}$ for the D6-brane configuration under consideration per lattice configuration. 
The configuration with $D6_b\upuparrows \OR$ has already been presented in table~\protect\ref{tab:NetChiral3GenRhoInd}.}

Note that in case the $\OR$-plane plays the r\^ole of the exotic O6-plane ($\eta_{\OR}=-1$), fractional three-cycles parallel to the $\OR\Z_2^{(1)}$-plane can a priori be excluded to accommodate the {\it left} stack, and the argument for this exclusion can be easily deduced from table~\ref{Tab:Conditions-on_b+t+s-SOSp}: on the {\bf aAA} lattice, there are no fractional three-cycles parallel to the $\OR\Z_2^{(1)}$-plane which support an enhanced $USp$ gauge group for $\eta_{\OR} = -1$, while on the {\bf bAA} lattice the discrete parameter configurations allowing for an enhanced $USp(2)$ gauge group also come with matter in the symmetric representation. Hence,
a crucial prerequisite to obtain three chiral generations with $b$-stack supporting a $USp(2)$ gauge group is tied to the choice of the exotic O6-plane, i.e.~$\eta_{\OR\Z_2^{(2)}}=-1$ or $\eta_{\OR\Z_2^{(3)}}=-1$. Yet, for both choices of the exotic O6-plane, the RR tadpole cancellation conditions in table~\ref{tab:Bulk-RR+SUSY-Z2Z6} show that global D6-brane model building can only be pursued for D6-branes with bulk wrapping number $\tilde V_a~\equiv~V_a~+~b\, Q_a = 0$. By consulting appendix~\ref{A:ClassBulkThreeCycles} it becomes clear that under these conditions only four bulk orbits can be taken into consideration to construct global D6-brane models: the bulk orbits parallel to the $\OR$-plane and the $\OR\Z_2^{(1)}$-plane and the bulk orbits $(\frac{1}{1-b},\frac{-b}{1-b};2,1;3,-1)$ and $(\frac{1}{1-b},\frac{-b}{1-b};4,-1;3,1)$  from table~\ref{tab:RhoIndependentaAA}. 

Considering such a configuration where the {\it QCD}-stack is wrapped on a fractional three-cycle $\Pi_a$ with bulk orbit parallel to the $\OR$-plane and the {\it left} stack on a fractional three-cycle $\Pi_b$ parallel to the $\OR\Z_2^{(1)}$-plane, one can deduce from the first RR tadpole cancellation condition in table~\ref{tab:Bulk-RR+SUSY-Z2Z6},
\begin{equation}\label{Eq:MSSM-like-bulk-constraint}
\sum_{x \in \{ a, b \}} N_x (2 P_x + Q_x) = N_a \frac{2}{1-b} + N_b  \frac{6}{1-b} \stackrel{!}{\leqslant} 32, \qquad \text{with } \eta_{\OR\Z_2^{(2)}} = -1 \text{ or } \eta_{\OR\Z_2^{(3)}} = -1 ,
\end{equation}
that on the {\bf bAA} lattice ($b=\frac{1}{2}$), models containing $U(3)_a \times U(2)_b$ are excluded. More explicitly, their contribution to the left hand side is 36, or even 40 for a Pati-Salam gauge structure containing $U(4)_a \times U(2)_b$. For $b=\frac{1}{2}$ and $U(3)_a \times USp(2)_b$, the contribution is 24 and does not overshoot the bulk RR tadpole. It remains to be investigated if global completions with MSSM-like spectrum can be found. 
However, left-right symmetric models with three generations and the minimal structure $U(3)_a \times USp(2)_b \times USp(2)_c$ (including extensions to Pati-Salam models)
can again be excluded from global supersymmetric model building since the contribution from stack $c$ is 12 and overshoots the bulk RR tadpole.

On the {\bf aAA} lattice on the other hand, the constraint~\eqref{Eq:MSSM-like-bulk-constraint} has only a contribution of 18 to the left hand side from $U(3)_a \times U(2)_b$, leaving a rich variety of options to search for MSSM-like or GUT spectra with  three left-handed quark generations in a $\varrho$-independent set-up.

\subsubsection{MSSM-like models with three generations: $\varrho$-dependent configurations}\label{Sss:3GenRhoDep}

To enhance the possibility of finding global D6-brane models on the orientifold $T^6/(\Z_2\times\Z_6\times\OR)$, our focus should turn to the other rigid fractional three-cycles identified in section~\ref{Ss:NoAdjoints}, in which case the complex structure modulus $\varrho$ is dynamically stabilised by the supersymmetry condition of the respective bulk orbit.

In first instance, the {\em QCD}-stack can still be wrapped along a rigid fractional three-cycle with bulk orbit parallel to the $\OR$-plane (without gauge enhancement), while the {\it left} stack is taken to lie along a rigid fractional three-cycle free of chiral states in the symmetric representation. Skimming through section~\ref{Ss:NoSyms} teaches us that fractional three-cycles with bulk orbits listed in table~\ref{tab:AdjointSymmetricFree} form the preferred candidate three-cycles to embed the $U(2)_L$ stack. The consideration that the {\it left} stack has to be wrapped on a fractional three-cycle with bulk orbit $(1,m_a^1;1,0;1,-1)$ forces the $\OR$-plane to be the exotic O6-plane. That is to say, the bulk orbits $(1,m_a^1;1,0;1,-1)$ in table~\ref{tab:AdjointSymmetricFree} are characterised by a non-vanishing bulk wrapping number $\tilde V_a$, from which we can immediately infer that the second bulk RR tadpole condition obstructs global model building with supersymmetric D6-branes for $\eta_{\OR \Z_2^{(2)}} = -1$ or $\eta_{\OR \Z_2^{(3)}} = -1$. Henceforth, the $\OR$-plane is assumed to be the exotic O6-plane, i.e. $\eta_{\OR} = -1$, for $\varrho$-dependent configurations.

However, only a subset of the bulk orbits in table~\ref{tab:AdjointSymmetricFree} do actually provide fractional three-cycles yielding three chiral generations at the intersections with the {\em QCD}-stack. A summary of their respective bulk orbits with the realisation of three generations is presented in table~\ref{tab:NetChiralitiesRhoDependentAppOR} for both lattice configurations. Most fractional three-cycles for the $b$-stack provide for a net-chirality $\chi^{ab} + \chi^{ab'} = \pm 3$, but give rise to additional non-chiral pairs in the bifundamental representation at the same time. The only bulk orbits in the list without additional non-chiral pairs in the bifundamental representation have torus wrapping numbers $(1, m_b^1) = (1,1)$ or $(1,3)$ along the first two-torus for the {\bf aAA} lattice and $(1, m_b^1) = (1,0)$ or $(1,1)$ for the {\bf bAA} lattice.\footnote{Strictly speaking one should also verify the absence of non-chiral matter in the bifundamental representation from the $ab$ and $ab'$ sector separately, from which additional vector-like massless states in the bifundamental representation might arise. In that respect, the statement has to be refined: only the bulk orbits with torus wrapping number $(1, m_b^1) = (1,1)$ along the first two-torus for the {\bf aAA} lattice and $(1, m_b^1) = (1,0)$ for the {\bf bAA} lattice are not accompanied by additional non-chiral pairs in the bifundamental representation.       
}

\mathtabfix{
\begin{array}{|c|c|c|c||c|c|c|c|}
\hline
\multicolumn{8}{|c|}{\text{\bf Bulk $D6_b$  orbits for three quark generations $\chi^{ab} + \chi^{ab'} = \pm 3$ with $a\pp \OR$ on $T^6/(\Z_2\times\Z_6\times\OR)$ with $\eta = -1$}}\\
\hline \hline
\multicolumn{4}{|c||}{\text{\bf aAA lattice}} &\multicolumn{4}{|c|}{\text{\bf bAA lattice}} \\
\hline \hline 
\text{\bf $D6_b$ orbit} & \varrho & (\chi^{ab}, \chi^{ab'}) & \text{\bf Occurrence frequency} & \text{\bf $D6_b$ orbit} & \varrho  & (\chi^{ab}, \chi^{ab'}) & \text{\bf Occurrence frequency} \\
\hline (1,1;1,0;1,-1) & 3&(1,2)  &  \text{64 out of $16 \times 160$ combinations}
& (1,0;1,0;1,-1) & 6 &(1,2)  & \text{48 out of $48 \times 160$ combinations}\\
& &(-2,-1)  &  \text{64 out of $16 \times 160$ combinations}
& & &(-2,-1)  & \text{48 out of $48 \times 160$ combinations} \\
 (1,3;1,0;1,-1) & 1 &(0,3)  & \text{64 out of $16 \times 16$ combinations} 
& (1,1;1,0;1,-1) & 2 &(0,3)  & \text{48 out of $48 \times 16$ combinations} \\
& &(-3,0)  &  \text{64 out of $16 \times 16$ combinations}
& & &(-3,0)  & \text{48 out of $48 \times 16$ combinations} \\
 (1,5;1,0;1,-1) & 3/5 &(-1,4)  &  \text{64 out of $16 \times 16$ combinations}
& (1,2;1,0;1,-1) & 6/5 &(-1,4)  & \text{48 out of $48 \times 16$ combinations} \\
& &(-4,1)  &  \text{64 out of $16 \times 16$ combinations}
& & &(-4,1)  & \text{48 out of $48 \times 16$ combinations} \\
 (1,7;1,0;1,-1) & 3/7 &(-2,5)  &  \text{64 out of $16 \times 16$ combinations}
& (1,3;1,0;1,-1) &6/7 &(-2,5)  &  \text{48 out of $48 \times 16$ combinations}\\
& &(-5,2)  &  \text{64 out of $16 \times 16$ combinations}
& & &(-5,2)  & \text{48 out of $48 \times 16$ combinations} \\
 (1,9;1,0;1,-1) &1/3 &(-3,6)  &  \text{64 out of $16 \times 16$ combinations}
& (1,4;1,0;1,-1) &2/3 &(-3,6)  & \text{48 out of $48 \times 16$ combinations} \\
& &(-6,3)  &  \text{64 out of $16 \times 16$ combinations}
& & &(-6,3)  & \text{48 out of $48 \times 16$ combinations} \\
 (1,11;1,0;1,-1) & 3/11 &(-4,7)  &  \text{64 out of $16 \times 16$ combinations}
& (1,5;1,0;1,-1) & 6/11 &(-4,7)  & \text{48 out of $48 \times 16$ combinations} \\
& &(-7,4)  &  \text{64 out of $16 \times 16$ combinations}
& & &(-7,4)  & \text{48 out of $48 \times 16$ combinations} \\
 (1,13;1,0;1,-1) & 3/13 &(-5,8)  &  \text{64 out of $16 \times 16$ combinations}
& (1,6;1,0;1,-1) & 6/13 &(-5,8)  & \text{48 out of $48 \times 16$ combinations} \\
& &(-8,5)  &  \text{64 out of $16 \times 16$ combinations}
& & &(-8,5)  & \text{48 out of $48 \times 16$ combinations} \\
 (1,15;1,0;1,-1) & 1/5 &(-6,9)  &  \text{64 out of $16 \times 16$ combinations}
& (1,7;1,0;1,-1) &2/5 &(-6,9)  & \text{48 out of $48 \times 16$ combinations} \\
& &(-9,6)  &  \text{64 out of $16 \times 16$ combinations}
& & &(-9,6)  &  \text{48 out of $48 \times 16$ combinations}\\
\hline
\end{array}
}{NetChiralitiesRhoDependentAppOR}{Bulk $D6_b$ orbits of fractional three-cycles yielding three left-handed quark generations on the orientifold $T^6/(\Z_2\times\Z_6\times\OR)$ with discrete torsion and exotic O6-plane $\eta_{\OR} = -1$. The $a$-brane is parallel to the $\OR$-plane, and the displacement parameters and discrete Wilson lines associated with the fractional three-cycles for both stacks are chosen such that chiral multiplets in the adjoint and in the symmetric representation are absent. The second and sixth column list the corresponding value of the complex structure modulus $\varrho$, for which stack $b$ is supersymmetric. The third and seventh column indicate how the three generations are realised by the net-chiralities $(\chi^{ab}, \chi^{ab'})$, and the fourth and eight column indicate how many combinations of $\Z_2^{(i)}$- eigenvalues, displacements and discrete Wilson lines give rise to the respective net-chiralities.} 

For phenomenological reasons the abundant appearance of vector-like quarks requires a mechanism by which their masses become much heavier than the three chiral generations, similarly to the set-up of a KSVZ-type invisible axion model~\cite{Kim:1979if,Shifman:1979if}. 
Instead of focusing on potential mechanisms by which redundant vector-like quarks acquire mass, we continue our search for three-generational intersecting D6-brane models and insist that the {\em QCD}-stack is characterised by one of bulk orbits $(1,m_a^1;1,0;1,-1)$ listed in table~\ref{tab:RigidSymmFreeAntisymmetrics}, with the $\OR$-plane playing the r\^ole of the exotic O6-plane.

Once the bulk orbit for the {\em QCD}-stack is fixed and thereby also the complex structure modulus $\rho$, we have to select a bulk orbit candidate for the {\it left} stack. Recalling that the {\it left} stack is supposed to be rigid and free of chiral matter in the symmetric representation, fractional three-cycles with bulk orbit parallel to the $\OR\Z_2^{(1)}$-plane can be excluded as candidates for the {\it left} stack, as discussed in the previous section. Fractional three-cycles parallel to the $\OR\Z_2^{(2,3)}$-plane and supporting a $USp(2)$ gauge group can also be excluded as candidates for the {\it left} stack, as they do not provide for three chiral quark generations at the intersections with an {\em QCD}-stack characterised by a bulk orbit of the form $(1,m_a^1;1,0;1,-1)$. Hence, the {\it left}-stack can only be wrapped on a fractional three-cycle with bulk orbit parallel to the $\OR$-plane or with the same bulk orbit as {\em QCD}-stack.

\mathtabfix{
\begin{array}{|c|c|c||c||c|}
\hline \multicolumn{5}{|c|}{\text{\bf Bulk orbits for three quark generations $\chi^{ab} + \chi^{ab'} = \pm 3$ on $T^6/(\Z_2\times\Z_6\times\OR)$ with $\eta = -1$}} \\
\hline
\hline
D6_a \text{\bf -orbit} & D6_b \text{\bf -orbit} & \varrho& (\chi^{ab}, \chi^{ab'}) & \text{\bf  Occurrence Frequency }\\
\hline \hline
\begin{picture}(0,0) \put(-20,-70){\begin{rotate}{90}\text{\bf aAA}\end{rotate} }\end{picture} 
(1,1;1,0;1,-1) & (1,1;1,0;1,-1) & 3 &(0,3) &  \text{864 out of $160 \times 160$ combinations}  \\
(1,3;1,0;1,-1) & \pp \OR: (1,0;1,0;1,0)  &1 &(3,0)&  \text{64 out of $16 \times 16$  combinations} \\
& &  & (0,3) &  \text{64 out of $16 \times 16$ combinations}  \\
(1,3;1,0;1,-1) & (1,3;1,0;1,-1) & 1 & (0,-3) &   \text{96 out of $16 \times 16$  combinations} \\
(1,4;1,0;1,-1)  & \pp \OR: (1,0;1,0;1,0) & 3/4  & (-3)&  \text{144 out of $16 \times 144$  combinations} \\
(1,5;1,0;1,-1)  & \pp \OR: (1,0;1,0;1,0) & 3/5  & (-3)&  \text{576 out of $16 \times 144$  combinations} \\
(1,6;1,0;1,-1)  & \pp \OR: (1,0;1,0;1,0) & 1/2  & (-3)&  \text{864 out of $16 \times 144$  combinations} \\
\hline \hline
\begin{picture}(0,0) \put(-15,-95){\begin{rotate}{90}\text{\bf bAA}\end{rotate} }\end{picture}
(1,0;1,0;1,-1) & (1,0;1,0;1,-1) & 6 &(2,1) &  \text{144 out of $160 \times 160$ combinations} \\
(1,1;1,0;1,-1) &\pp \OR: (2,-1;1,0;1,0)  & 2 &(3,0) &  \text{48 out of $16 \times 48$  combinations}  \\
& & & (0,3) &  \text{48 out of $16 \times 48$ combinations}  \\
&&& (-2,-1) & \text{144 out of $16 \times 48$ combinations} \\
&&& (-1,-2) & \text{144 out of $16 \times 48$ combinations} \\
(1,1;1,0;1,-1) & (1,1;1,0;1,-1) & 2 &(0,3) &  \text{128 out of $16 \times 16$ combinations} \\
 & & &(0,-3) &  \text{64 out of $16 \times 16$ combinations} \\
 (1,2;1,0;1,-1) &\pp \OR: (2,-1;1,0;1,0)  & 6/5 & (-3) &  \text{432 out of $16 \times 108$  combinations}  \\
  (1,3;1,0;1,-1) &\pp \OR: (2,-1;1,0;1,0)  & 6/7 & (-3) &  \text{432 out of $16 \times 108$  combinations}  \\
   (1,4;1,0;1,-1) &  (1,4;1,0;1,-1) & 2/3 & (0,-3)&  \text{48 out of $16 \times 48$  combinations} \\
\hline
\end{array}
}{BulkOrbits3QuarkGenerations}{Bulk orbits of fractional three-cycle pairs yielding three left-handed quark generations on the orientifold $T^6/(\Z_2\times\Z_6\times\OR)$ with discrete torsion and exotic O6-plane $\eta_{\OR} = -1$. The displacement parameters and discrete Wilson lines associated with the fractional three-cycles are chosen such that they are free of chiral multiplets in the adjoint and in the symmetric representation. The fourth column indicates how the three generations are realised by the net-chiralities $(\chi^{ab}, \chi^{ab'})$ and the fifth column how many combinations of $\Z_2^{(i)}$- eigenvalues, displacements and discrete Wilson lines give rise to the respective net-chiralities in the fourth column. For configurations where the $D6_b$-brane stack supports a $USp(2)$-type gauge group, the net-chirality reads $\chi^{ab} = \chi^{ab'} = - 3$.
}

An overview of all possible D6-brane combinations with three chiral generations can be found in table~\ref{tab:BulkOrbits3QuarkGenerations}. Most candidate three-cycles for the {\em QCD} stack are accompanied by chiral multiplets in the antisymmetric representation $(\left| \chi^{\Anti_a} \right| \leq 3)$, which account for right-handed quarks. Compatibility between the chirality of the right- and left-handed quarks translates into a relative sign between the net-chirality for the antisymmetrics $\chi^{\Anti_a}$ and for the bifundamentals $\chi^{ab}+\chi^{ab'}$: three-cycles $a$ with negative net-chirality $\chi^{\Anti_a} < 0$ have to realise the three generations of left-handed quarks by a positive net-chirality $\chi^{ab} +\chi^{ab'} > 0$, and vice versa. In case the net-chirality of antisymmetric representations vanishes, $\chi^{\Anti_a} = 0$, both signs for the net-chirality of the bifundamentals $\chi^{ab} +\chi^{ab'}$ are allowed, such as for the bulk orbit $(1,3;1,0;1,-1)$ on the {\bf aAA} lattice or the bulk orbit $(1,1;1,0;1,-1)$ on the {\bf bAA} lattice.  For various bulk orbit combinations three generations of left-handed quarks are realised by considering a fractional three-cycle parallel to the $\OR$-plane and supporting an enhanced $USp(2)$ gauge group for the {\it left} stack or $b$-stack. In these D-brane configurations, where the $b$-stack is orientifold-invariant, the sectors $ab$ and $ab'$ count the same number of degrees of freedom such that condition on the net-chirality reduces to the constraint $\chi^{ab} = \chi^{ab'} \stackrel{!}{=} \pm 3$.

\subsubsection{$SU(5)$ GUT models with three generations}\label{Sss:SU5GUT3Gen}

Our search for intersecting D6-brane models with three quark generations on the orientifold $T^6/(\Z_2\times\Z_6\times\OR)$ with discrete torsion has thus far focused on the realisation of three generations of chiral left-handed quarks charged under the {\it QCD} stack and the {\it left} stack in the same representation as they appear in the MSSM or Pati-Salam GUT models. In the case of a $U(5)$ GUT, the left handed quarks and leptons are embedded in the ${\bf 10}$ antisymmetric representation, which should come in three copies as discussed in section~\ref{Ss:NoSyms}. Furthermore, the massless chiral spectrum of an $SU(5)$ GUT should also contain three copies of the antifundamental state $ \ov{\bf 5}$, harbouring the three generations of down (or up) quarks in the bifundamental $(\ov{\3}, \1)$ representation and of leptons in the $(\1,\2)$ representation under the $SU(3)_{QCD}\times SU(2)_L$ gauge group. In this respect, the fractional three-cycles, identified in section~\ref{Ss:NoSyms} as candidates for $U(5)$ GUT D6-brane model building, have to be complemented with a second stack of $D6_b$-branes to generate three generations of antifundamental representations $ \ov{\bf 5}$. 

Phenomenologically, there are a priori no constraints to take into account for the $b$-stack with $U(1)_b$ gauge symmetry\footnote{Notice that the distinction of massless and massive Abelian gauge factors involving $U(1)_a \times U(1)_b$ can only be obtained for a global model, in which the non-Abelian representations satisfy all constraints discussed here.}, yet we need to ensure that the relative chirality between the antisymmetric ${\bf 10}$ and the antifundamental $\ov{ \bf 5}$ states is properly reflected in the relative signs of the net-chiralities, i.e. $\chi^{\Anti_a} = - (\chi^{ab}+\chi^{ab'})$.

More concretely, let us first focus on the fractional three-cycles from section~\ref{Ss:NoSyms} with $\chi^{\Anti_a} = + 3$, namely those with bulk orbit $(4,-1;1,-1;1,1)$ or  $(4,-1;1,1;1,-1)$ on the {\bf bAA} lattice.
Choosing one of these two bulk orbits for the $U(5)_a$-stack, one can a priori use all 17 supersymmetric bulk orbits on the {\bf bAA} lattice with value $\varrho = 4$ (cf. table~\ref{tab:ComplexStructureSUSYCycles2Lattices})
 of the complex structure modulus to construct an appropriate fractional three-cycle for the $b$-stack.
But a full parameter-scan using all 17 orbits shows that only three bulk orbits allow  for a (fractional) $b$-stack
whose intersections with the $U(5)_a$ stack add up to three generations of antifundamentals $\ov{\bf 5}$, i.e.~$\chi^{ab} + \chi^{ab'} = - 3$. A summary of the five D6-brane configurations is provided in table~\ref{tab:U53Generations}, where the third column contains the net-chiralities $(\chi^{ab}, \chi^{ab'})$ between D6-brane stack $a$ and $b$, and the fourth columns lists the amount of discrete parameter combinations, i.e.~$\Z_2^{(i)}$- eigenvalues, displacements and discrete Wilson lines, yielding the respective net-chirality. In the case where the $D6_b$-stack is parallel to the $\OR$-plane, one has to distinguish between discrete parameter configurations yielding enhancement of the gauge group and those that do not. In the latter case, the three generations of antifundamental states $ \ov{\bf 5}$ can be realised as $(\ov{\bf 5})_{1}$ and $(\ov{\bf 5})_{-1}$ of $U(5)_a \times U(1)_b$ with multiplicities given by
$(\chi^{ab}, \chi^{ab'}) = (-1,-2)$ or $(-2,-1)$, while for enhanced gauge groups the sectors $ab=ab'$ are identical, 
 and the net-chirality is constrained by $\chi^{ab} =  \chi^{ab'} = -3$, which is only found to be satisfied for configurations with $USp$ gauge group enhancement, i.e. the D-brane configuration contains the phenomenologically disfavoured non-Abelian factors $SU(5)_a \times USp(2)_b$. A similar consideration is valid for $D6_b$-stacks parallel to the $\OR\Z_2^{(1)}$-plane. Note, however, that when the $b$-stack is accommodated on a fractional three-cycle parallel to the $\OR\Z_2^{(1)}$-plane,
  the three generations are only obtained in case the $b$-stack supports an enhanced $USp$ gauge group, i.e.~for orientifold invariant $b$-stacks.
 Recall from section~\ref{Ss:NoSyms} that the {\bf aAA} lattice does not offer any $D6$-brane configuration for the $U(5)_a$-stack with a positive net-chirality $\chi^{\Anti} = +3$ for the states in the antisymmetric representation. As such, we have exhausted all plausible D6-brane configurations with positive net-chirality $\chi^{\Anti} = +3$ allowing for the construction of $U(5)$ GUT models.

\mathtabfix{
\begin{array}{|c|c||c||c|}
\hline
\multicolumn{4}{|c|}{\text{\bf Three generation $U(5)$ GUT models on $T^6/(\Z_2\times\Z_6\times\OR)$ with $\eta=-1$ (part I)}}\\
\hline \hline
 \text{\bf $U(5)_a$-orbit} & \text{\bf $D6_b$-orbit} &  (\chi^{ab}, \chi^{ab'})  & \text{\bf Occurrence Frequency} \\
\hline \hline
(4,-1;1,-1;1,1) &\pp \OR: (2,-1;1,0;1,0)& (-2,-1)  & \text{$3888$ out of $ 108 \times 144$ combinations} \\
& & (-1,-2)& \text{$3888$ out of $108 \times 144$ combinations}  \\
& \text{with $USp$ enhancement}& (-3) & \text{$1728$ out of $108 \times 108$ combinations}  \\
& \pp \OR\Z_2^{(1)}: (2,-1;-1,2;1,-2) & (-3) &\text{$864$ out of $108 \times 36$ combinations}\\
& \text{with $USp$ enhancement}&&\\
& (4,-1;1,-1;1,1) & (0,-3) & \text{$432$ out of $108 \times 256$  combinations} \\
& (4,-1;1,1;1,-1) & (0,-3) & \text{$432$ out of $108 \times 256$  combinations} \\
\hline \hline
(4,-1;1,1;1,-1) &\pp \OR:  (2,-1;1,0;1,0)& (-2,-1)  & \text{$3888$ out of $108 \times 144$ combinations} \\
& & (-1,-2)&\text{$3888$ out of $108 \times 144$ combinations}  \\
& \text{with $USp$ enhancement}& (-3) & \text{$1728$ out of $108 \times 108$ combinations}  \\
& \pp \OR\Z_2^{(1)}: (2,-1;-1,2;1,-2)  & (-3) &\text{$864$ out of $108 \times 36$ combinations}\\
& \text{with $USp$ enhancement}&&\\
& (4,-1;1,-1;1,1) & (0,-3) & \text{$432$ out of $108 \times 256$ combinations} \\
& (4,-1;1,1;1,-1) & (0,-3) & \text{$432$ out of $108 \times 256$ combinations}\\
\hline
\end{array}
}{U53Generations}{Bulk orbits of fractional three-cycle pairs allowing for three anti-fundamental representations ($\chi^{ab}+\chi^{ab'} = -\chi^{\Anti_a} = -3$) under $U(5)_a$ on the {\bf bAA} lattice of the orientifold $T^6/(\Z_2\times\Z_6\times\OR)$ with discrete torsion, exotic O6-plane $\eta_{\OR} = -1$ and complex structure $\varrho=4$. The net-chiralities $ (\chi^{ab}, \chi^{ab'})$ between the $U(5)_a$-stack and $D6_b$-stack are listed in the third column. The fourth column indicates how many combinations of $\Z_2^{(i)}$- eigenvalues, displacements and discrete Wilson lines conspire to the corresponding net-chiralities listed in the third column.}

Next, we turn to the fractional three-cycles from section~\ref{Ss:NoSyms} with $\chi^{\Anti_a} = - 3$, in which case the three antifundamentals $\ov{\bf 5}$ arise at the  intersection points of a $b$-stack satisfying the net-chirality condition $\chi^{ab} + \chi^{ab'} \stackrel{!}{=} 3$. On the {\bf aAA} lattice with the value $\varrho = \frac{1}{2}$ of the complex structure modulus,  there are 23 supersymmetric bulk orbits to our disposal  to construct an appropriate fractional three-cycle for the $b$-stack satisfying these net-chirality constraints. A summary of the various combinations is given in table~\ref{tab:U53GenerationspartII}, including the occurrence frequency based on the discrete parameter combinations in the last column. In case the $b$-stack is parallel to an O6-plane and supports an enhanced $USp$ gauge group, the net-chirality constraint again reduces to $\chi^{ab} = \chi^{ab'} = +3$. 

The {\bf bAA} lattice with complex structure modulus $\varrho = \frac{2}{3}$ offers 7+8=15 supersymmetric bulk three-cycles, but also here we find that only part of their corresponding fractional three-cycles satisfies the condition $\chi^{ab} + \chi^{ab'} \stackrel{!}{=} 3$ as summarised in table~\ref{tab:U53GenerationspartIII}. In this list. we encounter combinations ($\chi^{ab} = \chi^{ab'} = +3$) with three particle generation,
where the $D6_b$-stack is parallel to an O6-plane and supports an enhanced $SO$ gauge group, beyond the combinations with an enhanced $USp$ gauge group.

\mathtabfix{
\begin{array}{|c|c||c||c|}
\hline
\multicolumn{4}{|c|}{\text{\bf Three generation $U(5)$ GUT models on $T^6/(\Z_2\times\Z_6\times\OR)$ with $\eta=-1$ (part II)}}\\
\hline \hline
 \text{\bf $U(5)_a$-orbit} & \text{\bf $D6_b$-orbit} &  (\chi^{ab}, \chi^{ab'})  & \text{\bf Occurrence Frequency} \\
\hline \hline
(1,2;1,-1;1,1) &\pp \OR\Z_2^{(2)}: (0,1;1,0;1,-2)& (3,0)  & \text{$192$ out of $ 16 \times 208$ combinations} \\
& & (0,3)& \text{$192$ out of $16 \times 208$ combinations}  \\
& \pp \OR\Z_2^{(3)}: (0,1;1,-2;1,0) & (3,0) &  \text{$256$ out of $16 \times 208$ combinations}  \\
& & (2,1) &\text{$576$ out of $16 \times 208$ combinations}  \\
& & (1,2) &\text{$576$ out of $16 \times 208$ combinations}  \\
& & (0,3) &\text{$256$ out of $16 \times 208$ combinations}  \\
& \text{with $USp$ enhancement}& (3) & \text{$192$ out of $16 \times 48$ combinations}  \\
& (1,2;1,-1;1,1) & (3,0) & \text{$48$ out of $16 \times 256$  combinations} \\
&&(4,-1)&\text{$144$ out of $16 \times 256$  combinations} \\
& (1,2;1,1;1,-1) & (3,0) & \text{$96$ out of $16 \times 256$  combinations} \\
& (1,3;2,1;1,-1) & (3,0) & \text{$64$ out of $16 \times 256$  combinations} \\
& (1,6;1,0;1,-1) & (3,0) & \text{$48$ out of $16 \times 256$  combinations} \\
& (0,1;4,-5;3,-1) & (3,0) & \text{$64$ out of $16 \times 256$  combinations} \\
& (0,1;2,-3;5,-1) & (-3,6) & \text{$64$ out of $16 \times 256$  combinations} \\
\hline \hline
(1,2;1,1;1,-1) & \pp \OR\Z_2^{(2)}: (0,1;1,0;1,-2) & (3,0) &  \text{$256$ out of $16 \times 208$ combinations}  \\
& & (2,1) &\text{$576$ out of $16 \times 208$ combinations}  \\
& & (1,2) &\text{$576$ out of $16 \times 208$ combinations}  \\
& & (0,3) &\text{$256$ out of $16 \times 208$ combinations}  \\
& \text{with $USp$ enhancement}& (3) & \text{$192$ out of $16 \times 48$ combinations}  \\
&\pp \OR\Z_2^{(3)}: (0,1;1,-2;1,0)& (3,0)  & \text{$192$ out of $ 16 \times 208$ combinations} \\
& & (0,3)& \text{$192$ out of $16 \times 208$ combinations}  \\
& (1,2;1,1;1,-1) & (3,0) & \text{$48$ out of $16 \times 256$  combinations} \\
&&(4,-1)&\text{$144$ out of $16 \times 256$  combinations} \\
& (1,2;1,-1;1,1) & (3,0) & \text{$96$ out of $16 \times 256$  combinations} \\
& (1,3;0,1;1,-3) & (3,0) & \text{$64$ out of $16 \times 256$  combinations} \\
& (1,6;1,0;1,-1) & (3,0) & \text{$48$ out of $16 \times 256$  combinations} \\
& (0,1;4,-5;3,-1) & (-3,6) & \text{$64$ out of $16 \times 256$  combinations} \\
& (0,1;2,-3;5,-1) & (3,0) & \text{$64$ out of $16 \times 256$  combinations} \\
\hline
\end{array}
}{U53GenerationspartII}{Bulk orbits of fractional three-cycle pairs allowing for three anti-fundamental representations ($\chi^{ab}+\chi^{ab'} = -\chi^{\Anti_a} = +3$)  under $U(5)_a$ on the {\bf aAA} lattice of the orientifold $T^6/(\Z_2\times\Z_6\times\OR)$ with discrete torsion, exotic O6-plane $\eta_{\OR} = -1$ and complex structure $\varrho=\frac{1}{2}$. The net-chiralities $ (\chi^{ab}, \chi^{ab'})$ between the $U(5)_a$-stack and $D6_b$-stack are listed in the third column. The fourth column indicates how many combinations of $\Z_2^{(i)}$- eigenvalues, displacements and discrete Wilson lines conspire to the corresponding net-chiralities listed in the third column.}

\mathsidetabfix{
\begin{array}{|c|c||c||c|@{\hspace{0.2in}}|c|c||c||c|}
\hline
\multicolumn{8}{|c|}{\text{\bf Three generation $U(5)$ GUT models on $T^6/(\Z_2\times\Z_6\times\OR)$ with $\eta=-1$ (part III)}}\\
\hline \hline
 \text{\bf $U(5)_a$-orbit} & \text{\bf $D6_b$-orbit} &  (\chi^{ab}, \chi^{ab'})  & \text{\bf Occurrence Frequency} &  \text{\bf $U(5)_a$-orbit} & \text{\bf $D6_b$-orbit} &  (\chi^{ab}, \chi^{ab'})  & \text{\bf Occurrence Frequency} \\
\hline
(1,1;1,-1;1,1) &\pp \OR\Z_2^{(2)}: (0,1;1,0;1,-2)& (3,0)  & \text{$160$ out of $ 16 \times 208$ combinations} 
& (1,1;1,1;1,-1) & \pp \OR\Z_2^{(2)}: (0,1;1,0;1,-2) & (3,0) &  \text{$256$ out of $16 \times 208$ combinations}  \\
& & (2,1) &\text{$144$ out of $16 \times 208$ combinations} 
& & & (2,1) &\text{$576$ out of $16 \times 208$ combinations}  \\
& & (1,2) &\text{$144$ out of $16 \times 208$ combinations} 
& & & (1,2) &\text{$576$ out of $16 \times 208$ combinations}  \\
& & (0,3) &\text{$160$ out of $16 \times 208$ combinations}  
& & & (0,3) &\text{$256$ out of $16 \times 208$ combinations}  \\
& & (4,-1) &\text{$48$ out of $16 \times 208$ combinations} 
& & \text{with $USp$ enhancement}& (3) & \text{$144$ out of $16 \times 36$ combinations}  \\
& & (-1,4) &\text{$48$ out of $16 \times 208$ combinations} 
& & \text{with $SO$ enhancement}& (3) & \text{$48$ out of $16 \times 12$ combinations}  \\
 & \text{with $SO$ enhancement}& (3) & \text{$48$ out of $16 \times 12$ combinations}  
 & & & & \\
 & \pp \OR\Z_2^{(3)}: (0,1;1,-2;1,0) & (3,0) &  \text{$256$ out of $16 \times 208$ combinations} 
 & &\pp \OR\Z_2^{(3)}: (0,1;1,-2;1,0)& (3,0)  & \text{$160$ out of $ 16 \times 208$ combinations} \\
 & & (2,1) &\text{$576$ out of $16 \times 208$ combinations}  
 & & & (2,1) &\text{$144$ out of $16 \times 208$ combinations} \\
 & & (1,2) &\text{$576$ out of $16 \times 208$ combinations} 
 & & & (1,2) &\text{$144$ out of $16 \times 208$ combinations} \\
 & & (0,3) &\text{$256$ out of $16 \times 208$ combinations}  
 & & & (0,3) &\text{$160$ out of $16 \times 208$ combinations}  \\
 & \text{with $USp$ enhancement}& (3) & \text{$144$ out of $16 \times 36$ combinations}  
 &&  & (4,-1) &\text{$48$ out of $16 \times 208$ combinations}  \\
 &   \text{with $SO$ enhancement}& (3) & \text{$48$ out of $16 \times 12$ combinations} 
 & & & (-1,4) &\text{$48$ out of $16 \times 208$ combinations} \\
 & & & &
 &   \text{with $SO$ enhancement}& (3) & \text{$48$ out of $16 \times 12$ combinations}  \\
 & (1,1;1,1;1,-1) & (3,0) & \text{$96$ out of $16 \times 256$  combinations} 
 & & (1,1;1,-1;1,1) & (3,0) & \text{$96$ out of $16 \times 256$  combinations} \\
  & (1,1;1,-1;1,1) & (3,0) & \text{$48$ out of $16 \times 256$  combinations} 
 & & (1,1;1,1;1,-1) & (3,0) & \text{$48$ out of $16 \times 256$  combinations} \\
 &  & (1,2) & \text{$48$ out of $16 \times 256$  combinations} 
 & & & (1,2) & \text{$48$ out of $16 \times 256$  combinations} \\
  &  & (0,3) & \text{$64$ out of $16 \times 256$  combinations} 
 & & & (0,3) & \text{$64$ out of $16 \times 256$  combinations} \\
   &  & (4,-1) & \text{$144$ out of $16 \times 256$  combinations} 
 & & & (4,-1) & \text{$144$ out of $16 \times 256$  combinations} \\
   & (1,4;1,0;1,-1) & (1,2) & \text{$48$ out of $16 \times 256$  combinations} 
 & & (1,4;1,0;1,-1) & (1,2) & \text{$48$ out of $16 \times 256$  combinations} \\
 &  & (3,0) & \text{$192$ out of $16 \times 256$  combinations} 
 & & & (3,0) & \text{$192$ out of $16 \times 256$  combinations} \\
 &  (0,1;2,-3;5,-1)& (-3,6)&  \text{$64$ out of $16 \times 256$  combinations}
 &  & (0,1;4,-5;3,-1)& (-3,6)&  \text{$64$ out of $16 \times 256$  combinations} \\
  &(0,1;4,-5;3,-1)& (3,0)&  \text{$48$ out of $16 \times 256$  combinations}
 & &  (0,1;2,-3;5,-1) & (3,0)&  \text{$48$ out of $16 \times 256$  combinations} \\
  &  & (0,3) & \text{$16$ out of $16 \times 256$  combinations} 
 & & & (0,3) & \text{$16$ out of $16 \times 256$  combinations} \\
\hline
\end{array}
}{U53GenerationspartIII}{Bulk orbits of fractional three-cycle pairs allowing for three anti-fundamental representations ($\chi^{ab} + \chi^{ab'} =-\chi^{\Anti_a} = 3$)  under $U(5)_a$ on the {\bf bAA} lattice of the orientifold $T^6/(\Z_2\times\Z_6\times\OR)$ with discrete torsion, exotic O6-plane $\eta_{\OR} = -1$ and value of the complex structure parameter $\varrho=\frac{2}{3}$. The net-chiralities $ (\chi^{ab}, \chi^{ab'})$ between the $U(5)_a$-stack and $D6_b$-stack are listed in the third and seventh column. The fourth and eight column indicate how many combinations of $\Z_2^{(i)}$- eigenvalues, displacements and discrete Wilson lines conspire to the corresponding net-chiralities listed in the third and seventh column respectively.}

\clearpage
\section{Trailing down Global GUT Models}\label{S:GlobalModels}

The previous section contains a step-by-step analysis with the aim to identify phenomenologically appropriate fractional three-cycles for the {\em QCD} stack and $SU(2)_L$-stack on the one hand, or for the $SU(5)$-GUT-stack on the other hand. From that section, one can obviously conclude that the various phenomenological requirements, such as the absence of light exotic matter charged under the {\em QCD} gauge group or the presence of three generations of left-handed quarks, severely constrain the amount of appropriate D6-brane configurations. In a next phase, additional D6-brane stacks have to be added to complete the chiral spectrum, in order to accommodate all Standard Model particles and to render it free from gauge anomalies,
and to satisfy the bulk and exceptional RR tadpole cancellation conditions. In this section our search focuses on {\it global} intersecting D6-brane models, i.e. D-brane configurations that satisfy all RR tadpole conditions, realising the chiral spectra of $SU(5)$ GUT or Pati-Salam  models.

\subsection{Hunting for an $SU(5)$ GUT}\label{Ss:SU5-GUTS}

Section~\ref{Sss:SU5GUT3Gen} contains an exhaustive list of pairwise D6-brane configurations with $U(5)_a \times U(1)_b$ gauge group on the orientifold $T^6/(\Z_2\times\Z_6\times\OR)$  allowing for three-generations of ${\bf 10}$ and $\ov{\bf 5}$ of $SU(5) \subset U(5)_a$ as well as one matter multiplet in the adjoint representation.
A {\it global} model should also contain two Higgs-states  $H_u$ and $H_d$ in the fundamental ${\bf 5}$ and antifundamental $\ov{\bf 5}$ representation, respectively, at least three candidates for neutrinos and satisfy all bulk and exceptional RR tadpole cancellation conditions given in tables~\ref{tab:Bulk-RR+SUSY-Z2Z6} and~\ref{tab:Z2Z6TwistedRRTadpoles}. 

It is this last constraint in particular that presents the crucial obstruction in our search for global $SU(5)$ GUT models. Taking for instance the models from table~\ref{tab:U53Generations} where the $U(5)$-stack is characterised by a bulk orbit $(4,-1;1,-1;1,1)$ or $(4,-1;1,1;1,-1)$ on the {\bf bAA} lattice with value $\varrho = 4$ and exotic O6-plane $\eta_{\OR}=-1$, one notices immediately that the $U(5)_a$-stack contribution to the bulk RR tadpoles (see first relation in table~\ref{tab:Bulk-RR+SUSY-Z2Z6}) is larger than the RR-charges of the O6-planes:
\begin{equation}
N_a (2P_a + Q_a) = 5 \cdot (2 \cdot 8 -4) = 60 > 16.
\end{equation}
As all supersymmetric three-cycles on the {\bf bAA} lattice satisfy the condition $2P_a + Q_a \geq 0$ (see right column of table~\ref{tab:Bulk-RR+SUSY-Z2Z6}), it follows immediately that none of the models based on the bulk orbit $(4,-1;1,-1;1,1)$ or $(4,-1;1,1;1,-1)$ can be completed to a global model. 

Also the other set of bulk orbits $(1,1;1,-1;1,-1)$ and $(1,1;1,1;1,-1)$ on the {\bf bAA} lattice with value $\varrho= \frac{2}{3}$ of the complex structure modulus  and exotic O6-plane $\eta_{\OR}=-1$ allowing for three generations of ${\bf 10}$ and $\ov{\bf 5}$, as listed in table~\ref{tab:U53GenerationspartIII}, does not alleviate the obstruction posed by the bulk RR tadpole cancellation conditions. It is true that the RR-charges of the $U(5)$-stack, proportional to the bulk wrapping numbers for these orbits, do not overshoot the RR-charges of the O6-planes,
\begin{equation}
\begin{array}{ll}
\text{bulk RR tadpole 1:} &N_a  (2P_a + Q_a) = 5 \cdot ( 2 \cdot 2 - 1) = 15 < 16,\\
\text{bulk RR tadpole 2:} & -N_a  \frac{V_a + b Q_a}{1-b} = -5 \cdot 2 \cdot ( -1 + \frac{1}{2} ( - 1) ) = 15 < 16,\\
\end{array}
\end{equation}
but none of the 15 supersymmetric bulk orbits has small enough bulk wrapping numbers to ensure a cancellation of the remaining (bulk) RR-charges. Hence, also for the D6-brane configurations based on the bulk orbits $(1,1;1,-1;1,1)$ and $(1,1;1,1;1,-1)$ global supersymmetric $SU(5)$ GUT models cannot be realized on the {\bf bAA} lattice of the orientifold $T^6/(\Z_2 \times \Z_6 \times \OR)$ with discrete torsion. 

The only available playground to find global $SU(5)$ GUT models is thus located on the {\bf aAA} lattice with value $\varrho = \frac{1}{2}$ of the complex structure modulus  and exotic O6-plane $\eta_{\OR}=-1$, where the $U(5)$-stack is wrapped on a fractional three-cycle with bulk orbit $(1,2;1,-1;1,1)$ or $(1,2;1,1;1,-1)$, and a second fractional three-cycle is wrapped on one of the bulk orbits presented in table~\ref{tab:U53GenerationspartII}. The corresponding RR charges of the $U(5)_a$-stack for both bulk orbits $(1,2;1,-1;1,1)$ and $(1,2;1,1;1,-1)$ do not violate a priori the two (bulk) RR tadpole cancellation conditions,
 \begin{equation}
\begin{array}{ll}
\text{bulk RR tadpole 1:} &N_a  (2P_a + Q_a) = 5 \cdot ( 2 \cdot 2 - 1) = 15 < 16,\\
\text{bulk RR tadpole 2:} & - N_a  \, V_a  = -5 \cdot ( -2  ) =  10 < 16,\\
\end{array}
\end{equation}
and still leave room to be satisfied by adding a $U(1)_b$-stack.  Moreover, this $b$-stack should also allow for three generations of antifundamentals $\ov{\bf 5}$ at the intersections with the $U(5)_a$ stack, which implies we should look for a candidate bulk orbit in table~\ref{tab:U53GenerationspartII}. The only fractional three-cycles satisfying these two conditions are characterised by a bulk orbit $(1,6;1,0;1,-1)$. An explicit D6-brane configuration with cancelled bulk RR tadpoles is presented in table~\ref{tab:LocalSU5Example}:
  \begin{equation}\label{Eq:SU5BulkRRtadpoles}
\begin{array}{ll}
\text{bulk RR tadpole 1:} &\sum_{x\in \{a,b \}}N_x  (2P_x + Q_x) = N_a( 2 \cdot 2 - 1) + N_b ( 2 \cdot 1 - 1) \stackrel{!}{=}  16,\\
\text{bulk RR tadpole 2:} & - \sum_{x\in \{a,b \}} N_x  \, V_x  = - \left[ N_a ( -2  ) + N_b ( -6) \right] \stackrel{!}{=}   16.\\
\end{array}
\end{equation}  

\mathtabfix{
\begin{array}{|c||c|c||c|c|c||c|}\hline 
\muc{7}{|c|}{\text{\bf D6-brane configuration of a `local' $U(5)$ GUT model on the {aAA} lattice with $\varrho = \frac{1}{2}$}}
\\\hline \hline
&\text{\bf wrapping numbers} &\frac{\rm Angle}{\pi}&\text{\bf $\Z_2^{(i)}$ eigenvalues}  & (\vec \tau) & (\vec \sigma)&\text{\bf gauge group}\\
\hline \hline
 a&(1,2;1,-1;1,1)&(\frac{1}{6},-\frac{1}{3},\frac{1}{6})&(-+-)&(0,1,1) & (0,1,1)& U(5)\\
 b&(1,6;1,0;1,-1)&(\frac{1}{3},0,-\frac{1}{3})&(+++)&(0,0,1) & (0,0,1)&U(1)\\
 \hline
\end{array}
}{LocalSU5Example}{D6-brane configuration with two stacks of D6-branes yielding a local $SU(5)$ GUT model with gauge group 
$SU(5)_a\times U(1)_a\times U(1)_b$ on the {\bf aAA} lattice of the orientifold $T^6/(\Z_2 \times \Z_6 \times \OR)$ with discrete torsion and the $\OR$-plane the exotic O6-plane ($\eta_{\OR}=-1$).}
The chiral and non-chiral spectrum associated to the example in table~\ref{tab:LocalSU5Example} is presented in table~\ref{Tab:LocalSU5ExampleSpectrum}. The model contains three generations of chiral states in the antisymmetric ${\bf 10}$ representation, three generations of chiral states in the antifundamental $\ov{\bf 5}$ representation, and one multiplet in the adjoint representation of $SU(5)$, by construction. From the massless spectrum one immediately notes the absence of the Higgses $H_u$ and $H_d$ as bifundamental states in the (non-chiral) $ab$ or $ab'$ sector, which clearly disfavours this model on phenomenological grounds. Also the presence of vector-like pairs in the symmetric ${\bf 15}+ \ov{\bf 15}$ representation of $SU(5)$ imposes certain challenges in order to argue their phenomenologically required massiveness with respect to the chiral states in the ${\bf 10}$ representation.      

\begin{table}[h]
\begin{center}
\begin{tabular}{|c||c|c|c|}
\hline \multicolumn{3}{|c|}{\bf Massless spectrum of a `local' two-stack $U(5)$ GUT model}\\
\hline \hline
State & Sector & $(SU(5)_a)_{U(1)_a\times U(1)_b}$
 \\
\hline
${\bf 10}$&$aa'$ &  $3 \times ({\bf 10})_{(2,0)}$ 
\\
$\ov{\bf 5}$&$ab$ &  $3 \times (\ov{\bf 5})_{(-1,1)}$ 
\\
$\Phi_{\bf 24}, \Phi_{\1}$&$aa$&  $ ({\bf 24}_{\Adj})_{(0,0)} +  ({\bf 1})_{(0_{\Adj},0)}   $ 
\\
$B$& $bb$ & $(\1)_{(0,0_\Adj)} $ 
\\
$\Sigma_a^{i\in \{1,2,3\}}$, $\tilde \Sigma_a^{i\in \{1,2,3\}}$ & $aa'$ &$3 \times \left[ ({\bf 15})_{(2,0))}  + h.c.\right]$ 
\\
$\Sigma_b$&$bb'$&  $5 \times (\1)_{(0,2_\Sym)}$ 
\\
\hline
\end{tabular}
\caption{Chiral and non-chiral spectrum of the `local' $SU(5)$ GUT model with D6-brane configuration in table~\ref{tab:LocalSU5Example}. \label{Tab:LocalSU5ExampleSpectrum}}
\end{center}
\end{table}

Another concern deals with the spontaneous symmetry breaking of the GUT $SU(5)$ gauge group and the geometric characteristics of the $T^6/(\Z_2\times \Z_6)$ orbifold. The multiplet $\Phi_{\bf 24}$ in the adjoint ${\bf 24}$ representation is supposed to acquire a non-vanishing {\it vev} through gauge-invariant quadratic and cubic self-interactions in the superpotential without breaking supersymmetry: namely, by requiring that the F-terms for the chiral multiplet in the adjoint representation vanish for such a superpotential, a non-trivial vacuum configuration should exist allowing for the spontaneous breaking of the $SU(5)$ gauge group to $SU(3)\times~SU(2)$. In the language of intersecting D6-branes, cubic terms for the field $\Phi_{\bf 24}$ in the superpotential can be generated perturbatively through worldsheet-instantons with the topology of a disc and with an insertion on the boundary for each field $\Phi_{\bf 24}$ in the cubic coupling. Following \cite{Kachru:2000ih,Cremades:2003qj}, such worldsheet-instantons are holomorphically mapped to triangular shapes on each (ambient) two-torus $T_{(i)}^2$ in the factorisable orbifold $T^6/(\Z_2\times \Z_6)$. 
The toroidal cycles $a$, $(\omega a)$ and $(\omega^2 a)$  form the faces of the (closed) triangles, whose apexes correspond to the intersection points between the cycles. Observe, however, that the cycles $a$, $(\omega a)$ and $(\omega^2 a)$ are all parallel along $T_{(1)}^2$ due to the invariance of this two-torus under the $\Z_6$-action, such that the closed sequence $[a, (\omega a), (\omega^2 a)]$ does not correspond to a closed triangle on $T_{(1)}^2$.  
Or said otherwise, the closed sequence $[a, (\omega a), (\omega^2 a)]$ corresponds to a non-trivial closed one-cycle $\in H_1(T_{(1)}^2,\Z)$ on the ambient two-torus $T_{(1)}^2$, such that it cannot form the boundary of a triangle. Consequently, such a worldsheet-instanton - at least in models on a six-torus without orbifold, cf. e.g.~\cite{Cvetic:2003ch,Lust:2004cx} - 
does not contribute to the superpotential and cubic couplings of the form $\Tr (\Phi_{\bf 24}^3)$ are excluded. It remains an open question if new subtleties arise when the orbifold action is taken into account.
Hence, up to this caveat other mechanisms have to be invoked to argue for the spontaneous breaking of $SU(5)$ by a non-vanishing {\it vev} for $\Phi_{\bf 24}$ without breaking supersymmetry. 

A last concern about the model presented in tables~\ref{tab:LocalSU5Example} and \ref{Tab:LocalSU5ExampleSpectrum} brings us back to the global consistency conditions for the model. Even though the expressions in (\ref{Eq:SU5BulkRRtadpoles}) guarantee the cancellation of bulk RR-charges and the spectrum in table~\ref{Tab:LocalSU5ExampleSpectrum} is free of $SU(5)$ gauge anomalies, 
we still lack information about the twisted RR-charges per $\Z_2^{(i)}$-sector. More explicitly, calculating the twisted RR tadpoles using table~\ref{tab:Z2Z6TwistedRRTadpoles} for this model yields:
\begin{equation}\label{Eq:SU5GUTTwistedRRtadpoles}
\sum_{i=1}^3 \sum_{x \in \{a,b\}} N_x (\Pi_x^{\Z_2^{(i)}} + \Pi_{x'}^{\Z_2^{(i)}}) = \left\{ \begin{array}{l}
-8\, \varepsilon_3^{(1)} + 4\, \varepsilon_4^{(1)} +4 \, \varepsilon_5^{(1)} + 24\, \tilde \varepsilon_4^{(1)} - 24\, \tilde \varepsilon_5^{(1)}\\
+ 18\, \varepsilon_1^{(2)} + 18\, \varepsilon_2^{(2)}\\
 -14 \varepsilon_1^{(3)} -14  \varepsilon_2^{(3)} .
\end{array}  
\right.
\end{equation}
In order for the twisted RR tadpoles to vanish, one would have to add additional fractional three-cycles supporting D6-brane stacks, but in that case the bulk RR tadpole cancellation conditions would no longer be satisfied. Hence, the bulk RR tadpoles and twisted RR tadpoles mutually exclude each other from vanishing.

As a final remark, we note that the analysis and discussion given for the model in table~\ref{tab:LocalSU5Example} is also applicable to the 47 other models with bulk orbits $a=(1,2;1,-1;1,1)$ and $b=(1,6;1,0;1,-1)$ and the 48 models with bulk orbits $a=(1,2;1,1;1,-1)$ and $b=(1,6;1,0;1,-1)$ indicated in table~\ref{tab:U53GenerationspartII}. As a matter of fact 
one can easily check that all 96 models give rise to the same massless chiral spectrum presented in table~\ref{Tab:LocalSU5ExampleSpectrum}, and that all 96 models are also characterised by non-vanishing twisted RR tadpoles similar to equation (\ref{Eq:SU5GUTTwistedRRtadpoles}). As such, the existence of global $SU(5)$ GUT models is completely ruled out on the orientifold $T^6/(\Z_2\times\Z_6\times\OR)$ with discrete torsion.

\subsection{The Pursuit of Pati-Salam Models}\label{Ss:PS-Models}

Grand Unified Theories with an $SO(10)$ or an exceptional gauge structure are not attainable on the toroidal orientifold $T^6/(\Z_2\times\Z_6\times\OR)$ with discrete torsion, as expected for any intersecting D-brane construction in the perturbative regime of Type II string theory. The absence of global $SU(5)$ GUT models derived in section~\ref{Ss:SU5-GUTS}
leaves only the option of a Pati-Salam model to realise supersymmetric global GUT models on this orientifold. To construct the Pati-Salam gauge structure we assume the $SU(4)$ and $SU(2)_L$ stacks to be wrapped around completely rigid fractional three-cycles free from chiral matter in the symmetric representation. In addition, the $SU(4)$ stack is also taken to be free of chiral matter in the antisymmetric representation.\footnote{In principle this constraint can be relaxed by allowing one chiral state in the antisymmetric representation under $SU(4)$, in order to reproduce the supersymmetric spectrum of the models presented in~\cite{Antoniadis:1988cm,King:1997ia}. However, on the orientifold $T^6/(\Z_2\times \Z_6 \times \OR)$ the fractional three-cycles with bulk orbit $(1,2;1,0;1,-1)$ on the {\bf aAA} lattice and $(1,0;1,0;1,-1)$ on the {\bf bAA} lattice do not allow for three chiral generations of both left-handed and right-handed quarks and leptons, as can be deduced from tables~\ref{tab:RigidSymmFreeAntisymmetrics} and~\ref{tab:BulkOrbits3QuarkGenerations}.} Under these assumptions, the $SU(4)$ stack is restricted to have a bulk orbit parallel to the $\OR$-plane, or have the bulk orbit $(1,m_a^1;1,0;1,-1)$ with $m_a^1 = 3$ on the {\bf aAA} lattice and $m_a^1 = 1$ on the {\bf bAA} lattice. In the first D-brane configuration for the $a$-stack, the choice of the exotic O6-plane is still open, whereas the second, $\varrho$-dependent D6-brane configuration is only consistent if the $\OR$-plane fulfils the r\^ole of the exotic O6-plane ($\eta_{\OR}=-1$) for both lattices.

\subsubsection{Local Pati-Salam models on the bAA lattice}

Let us first consider the set-ups where the bulk orbit of the $U(4)_a$-stack is parallel to the $\OR$-plane, whose bulk wrapping numbers add up to $2 P_a + Q_a = 4$ on the {\bf bAA} lattice. Given that the rank of the gauge group is $N_a = 4$, the total contribution of the $a$-stack to the RR-charges already compensates a part of the negative RR-charges of the O6-planes, i.e. it saturates the first bulk RR tadpole condition $N_a (2 P_a + Q_a) = 4 \cdot 4 \stackrel{!}{\leqslant} 16$ in table~\ref{tab:Bulk-RR+SUSY-Z2Z6}, if the $\OR$-plane fulfils the r\^{o}le of the exotic O6-plane. Furthermore, in section~\ref{Sss:3GenRhoIndep} it was realised that the conditions imposed on the fractional three-cycle for the $SU(2)_L$ and $SU(2)_R$ stacks, namely rigidity and absence of chiral states in the symmetric representation, combined with the requirement of three chiral generations both left-handed and right-handed quarks, do not allow for suitable $\varrho$-independent D6-brane configurations with gauge group $U(4)_a \times USp(2)_L\times USp(2)_R$ to be completed into global models on the {\bf bAA} lattice. Therefore, $\varrho$-independent global Pati-Salam models (without the kinds of exotic matter discussed above) do not exist on the {\bf bAA} lattice.

Focusing instead on $\varrho$-dependent configurations, one can keep the bulk orbit of the $U(4)_a$-stack parallel to the $\OR$-plane and consider a bulk orbit from table~\ref{tab:NetChiralitiesRhoDependentAppOR} for the $SU(2)_L$ (or $b$-stack) to obtain three generations of left-handed quarks and leptons. However, for each bulk orbit from this table the bulk wrapping numbers of the $b$-stack add up to $2 P_b + Q_b =  n_b^1 \neq 0$:
\begin{equation}
\sum_{x \in \{a,b\}} N_{x} (2 P_x + Q_x) = 16 + N_b \,  n_b^1 > 16,
\end{equation} 
implying that three-generation models with an $a$-stack along the $\OR$-plane always violate the RR tadpole cancellation conditions on the {\bf bAA} lattice
with $\eta_{\OR}=-1$, both for $\varrho$-independent and for $\varrho$-dependent configurations. 

For lattice backgrounds with $\eta_{\OR} = 1$, a similar statement is valid: in section~\ref{Sss:3GenRhoIndep} we argued that D6-brane configurations with the $a$-stack parallel to the $\OR$-plane and the $b$- and $c$-stack along the $\OR\Z_2^{(1)}$-plane violate the bulk RR tadpole cancellation conditions. And given that these D6-brane configurations are the only $\varrho$-independent configurations with three chiral generations of left-and right-handed quarks, global $\varrho$-independent Pati-Salam models are excluded. Keeping the $a$-stack parallel to the $\OR$-plane, $\varrho$-dependent configurations are equally excluded based on the RR tadpole cancellation conditions. More explicitly, only those fractional three-cycles $\Pi_x^{\text{frac}}$ with bulk wrapping number $\tilde V_x = 0$ can be used for backgrounds with $\eta_{\OR} = 1$, as the second bulk RR tadpole cancellation condition in table~\ref{tab:Bulk-RR+SUSY-Z2Z6} would otherwise be violated. At the same time, there exist no bulk orbits in table~\ref{tab:NetChiralitiesRhoDependentAppOR} satisfying the constraint $\tilde V_x = 0$ while providing for three left-handed generations of quarks. Hence, also for $\varrho$-dependent configurations it is impossible to construct global Pati-Salam models when $\eta_{\OR} = 1$.

These considerations force the rigid $U(4)_a$-stack to lie along a bulk orbit $(1,m_a^1;1,0;1,-1)$, in which case $\eta_{\OR}=-1$ is automatically required, and imposing the absence of chiral matter in the symmetric and antisymmetric representation points straight to the bulk orbit with $m_a^1=1$, as can be verified from tables~\ref{tab:NetChiralitiesRhoDependentAppOR} and~\ref{tab:BulkOrbits3QuarkGenerations}. Choosing rigid fractional three-cycles free from chiral multiplets in the symmetric representation for the $SU(2)_L$ and $SU(2)_R$ stacks also constrains their respective bulk orbit to the following candidates on the {\bf bAA} lattice with value $\varrho = 2$ of the complex structure modulus: 
\begin{itemize}
\item the $SU(2)_L$ or $b$-stack can be parallel to the $\OR$-plane or have the same bulk orbit $(1,1;1,0;1,-1)$ as the $a$-stack, given that these configurations are the only ones providing three generations according to table~\ref{tab:BulkOrbits3QuarkGenerations};
\item the $SU(2)_R$ or $c$-stack can be parallel to the $\OR$-plane, or to the same orbit $(1,1;1,0;1,-1)$ as the $a$-stack, given that we also require three right-handed generations of quarks and leptons. 
\end{itemize}
One might wonder whether one of the $\OR\Z_2^{(i)}$-planes with $i\in \{1, 2, 3\}$ can be chosen to accommodate the $SU(2)_L$ or $SU(2)_R$ stack, provided that this $b$- or $c$-stack supports an enhanced $USp(2)$ gauge group through an appropriate choice of the discrete parameters as displayed in table~\ref{Tab:Conditions-on_b+t+s-SOSp}. However, this option is completely excluded by requiring the presence of three generations of chiral multiplets in the bifundamental representation at the intersections with the $a$-stack, as can be deduced from table~\ref{tab:BulkOrbits3QuarkGenerations}.  \footnote{Three right-handed generations can be obtained from a $c$-stack parallel to the $\OR\Z_2^{(1)}$-plane supporting an $SO(2)$ enhanced gauge group, but since our focus in the present article lies on genuinely left-right symmetric models, gauge configuration containing  $U(3)_a \times U(2)_L\times SO(2)_R$ will not be considered further at this point.
}

After having identified appropriate bulk orbits for the $a$-, $b$- and $c$-stacks, one ought to take the bulk RR tadpole cancellation conditions into consideration, in order to investigate whether or not a global completion of the models is attainable. Table~\ref{tab:PatiSalambAAOverviewRRtadpoles} provides a summary of the bulk orbits of the $a$-, $b$- and $c$-stack and their corresponding contributions to the bulk RR tadpoles for the local Pati-Salam model with three generations of left-handed and right-handed quarks and leptons on the {\bf bAA} lattice with complex structure modulus $\varrho = 2$. Recall from table~\ref{tab:Bulk-RR+SUSY-Z2Z6} there are two independent conditions ensuring vanishing bulk RR tadpoles, which taking into account $\eta_{\OR}=-1$ can be recast into two upper bounds on the RR tadpole contributions coming from the $a$-, $b$- and $c$-stack:
\begin{eqnarray}
 \sum_{x\in \{a, b, c \}} N_x (2 P_x + Q_x) \leq 16, \nonumber \\
  - 2 \sum _{x\in \{a, b, c \}} N_x \tilde V_x \leq 16.
\end{eqnarray} 
Comparing these upper bounds with the bulk RR tadpoles listed in table~\ref{tab:PatiSalambAAOverviewRRtadpoles} immediately reveals that for each local model one of the conditions is violated, implying that none of the local Pati-Salam models can be promoted to a global model with vanishing bulk and exceptional RR tadpoles.

\mathtabfix{
\begin{array}{|c|c|c||c|c|}
\hline
\multicolumn{5}{|c|}{\text{\bf Summary of local Pati-Salam models on the bAA lattice of $T^6/(\Z_2 \times \Z_6 \times \OR)$ with $\varrho = 2$ and $\eta = -1= \eta_{\OR}$}}\\
\hline \hline
\multicolumn{3}{|c||}{\text{\bf Bulk orbits for the Pati-Salam gauge groups}} & \multicolumn{2}{|c|}{\text{\bf bulk RR tadpoles}} \\
\hline \hline
\text{$a$-stack} & \text{$b$-stack} & \text{$c$-stack} & \sum_{x\in \{a, b, c \}} N_x (2 P_x + Q_x) & - 2 \sum _{x\in \{a, b, c \}} N_x \tilde V_x \\
\hline \hline
(1,1;1,0;1,-1)& \pp \OR: (2,-1;1,0;1,0) &(1,1;1,0;1,-1) &  N_a + 4 N_b +  N_c = 14  & 3 N_a + 3 N_c = 18 \\
(1,1;1,0;1,-1)& \pp \OR: (2,-1;1,0;1,0) &\pp \OR: (2,-1;1,0;1,0) &  N_a + 4 N_b + 4 N_c = 20   & 3 N_a = 12 \\
(1,1;1,0;1,-1)& (1,1;1,0;1,-1)& (1,1;1,0;1,-1)&  N_a +  N_b +  N_c = 8 & 3 N_a + 3 N_b + 3 N_c = 24\\
(1,1;1,0;1,-1)& (1,1;1,0;1,-1)&\pp \OR: (2,-1;1,0;1,0)&  N_a +  N_b + 4 N_c = 14  & 3 N_a + 3 N_b = 18 \\
\hline
\end{array}
}{PatiSalambAAOverviewRRtadpoles}{Overview of the local Pati-Salam models on the {\bf bAA} lattice of the toroidal orientifold $T^6/(\Z_2 \times \Z_6 \times \OR)$ with value $\varrho = 2$ of the complex structure modulus, discrete torsion $\eta = -1 $ and the $\OR$-plane as the exotic O6-plane. The bulk orbits for the $a$-, $b$- and $c$-stack are displayed in the first three columns, and their respective contributions to the bulk RR tadpoles are listed in the last two columns.}

Nevertheless, we can study two prototypes of local Pati-Salam models based on the bulk orbit configurations presented in table~\ref{tab:PatiSalambAAOverviewRRtadpoles}. Considering for instance the first D6-brane configuration from that table with the parameters for the corresponding fractional three-cycles spilled out explicitly in table~\ref{tab:PatiSalambAAPrototypeI}, one finds a first prototype of a Pati-Salam model with chiral and non-chiral spectrum written out in table~\ref{tab:PatiSalambAAPrototypeISpectrum}. A second prototype of a local Pati-Salam model is based on the third D6-brane configuration in table~\ref{tab:PatiSalambAAOverviewRRtadpoles}, and an explicit model is presented in tables~\ref{tab:PatiSalambAAPrototypeII} and \ref{tab:PatiSalambAAPrototypeIISpectrum}. The models resulting from the other two D6-brane configurations in table~\ref{tab:PatiSalambAAOverviewRRtadpoles} (row 2 and row 4) can be classified according to the chiral spectra of these two prototypes as well, up to an exchange of the net-chiralities in the $ac$ and $ac'$ sector, namely $\chi^{ac} \longleftrightarrow \chi^{ac'}$.

\mathtabfix{
\begin{array}{|c||c|c||c|c|c||c|}\hline 
\muc{7}{|c|}{\text{\bf D6-brane configuration of a local Pati-Salam model on the {bAA} lattice with $\varrho = 2$: prototype I}}
\\\hline \hline
&\text{\bf wrapping numbers} &\frac{\rm Angle}{\pi}&\text{\bf $\Z_2^{(i)}$ eigenvalues}  & (\vec \tau) & (\vec \sigma)& \text{\bf gauge group}\\
\hline \hline
 a&(1,1;1,0;1,-1)&(\frac{1}{3},0,-\frac{1}{3})&(--+)&(0,1,1) & (0,1,1)& U(4)\\
 b&(2,-1;1,0;1,0)&(0,0,0)&(+--)&(1,0,0) & (1,0,0)&U(2)\\
 c&(1,1;1,0;1,-1)&(\frac{1}{3},0,-\frac{1}{3})&(+++)&(0,1,1) & (0,1,1)&U(2)\\ 
 \hline
\end{array}
}{PatiSalambAAPrototypeI}{D6-brane configuration of a three-stack local Pati-Salam model with gauge group 
$SU(4)_a\times SU(2)_b \times SU(2)_c\times U(1)_a\times U(1)_b\times U(1)_c$ on the {\bf bAA} lattice of the orientifold $T^6/(\Z_2 \times \Z_6 \times \OR)$ with discrete torsion and the $\OR$-plane as the exotic O6-plane ($\eta_{\OR}=-1$).}
\mathtab{
\begin{array}{|c||c|c|c|}
\hline \multicolumn{3}{|c|}{\text{\bf Spectrum of the prototype I local Pati-Salam model on bAA}}\\
\hline
\hline
\text{State} & \text{Sector} & (SU(4)_a\times SU(2)_b\times SU(2)_c)_{U(1)_a\times U(1)_b \times U(1)_c} 
\\
\hline (Q_L, L) & ab & 2 \times (\4, \2, \1)_{(1,-1,0)} 
\\
 (Q_L, L) & ab' & (\4, \2, \1)_{(1,1,0)} 
 \\
 (Q_R, R) & ac' & 3 \times (\ov \4, \1, \2)_{(-1,0,-1)} 
  \\
A, \tilde A  &aa'& 2 \times [ ({\bf 6}_\Anti,\1,\1)_{(2,0,0)} + h.c.  ] 
\\
B, \tilde B    &bb'&  (\1,\1_\Anti, \1)_{(0,2,0)} + h.c.   
    \\
 C, \tilde C   &cc'& 2 \times [ (\1,\1, \1_\Anti)_{(0,0,2)} + h.c.  ] 
    \\
G_H, \tilde G_H  &ac & 2 \times \left[ ( \4, \1, \2)_{(1,0,-1)} + h.c. \right]  
\\
  \hline
\end{array}
}{PatiSalambAAPrototypeISpectrum}{Chiral and non-chiral massless matter spectrum of the prototype I local Pati-Salam model with D6-brane configuration in table~\ref{tab:PatiSalambAAPrototypeI}.}

A first difference between the two prototypes concerns the realisation of the three generations of left-handed quarks and leptons, which are scattered over the $ab$ and $ab'$ sector for the first prototype and arise purely from the $ab'$ sector in the second prototype. The absence of massless states in the $bc$ and $bc'$ sector indicates that the first prototype model is a Higgs-less model. In this respect, the second prototype offers a more promising scenario where Higgs-candidates arise both from the (chiral) $bc'$ sector and from the (non-chiral) $bc$ sector. In both models, the $ac$ sector contains two non-chiral pairs $\left[ ( \4, \1, \2)_{(1,0,-1)} + h.c.\right] $ suitable to act as the GUT-Higgses $G_H$ and $\tilde G_H$ responsible for the spontaneous breaking of the Pati-Salam gauge group $SU(4)\times SU(2)_R$ down to the Standard Model gauge group $SU(3)_{QCD}\times U(1)_Y$, following the scheme of~\cite{Antoniadis:1988cm,King:1997ia}. The antisymmetric states $\tilde A \equiv ({\bf 6}_\Anti,\1,\1)_{(-2,0,0)}$ arising from the (non-chiral) $aa'$ sector fit nicely within this scheme as well, as they allow for the existence of the cubic couplings $G_H  G_H \tilde A   + \tilde G_H  \tilde G_H \tilde A$ based on charge conservation. Upon GUT symmetry breaking, these cubic couplings reduce schematically to two supersymmetric mass terms with mass of the order ${\cal O}(M_{GUT})$ for the states in the antisymmetric representation and for those components of $G_H$ and $\tilde G_H$ that remain massless upon the spontaneous breaking of the GUT gauge group. :
\begin{equation} 
G_H  G_H \tilde A   + \tilde G_H  \tilde G_H  A \qquad \longrightarrow \qquad  \langle G_H\rangle  G_H \tilde A   + \langle \tilde G_H \rangle  \tilde G_H A.
\end{equation}
These considerations are purely field theoretic and in agreement with $SU(4)_a \times SU(2)_c$ gauge invariance. Observe, however, that the cubic couplings $G_H G_H \tilde A $ and $\tilde G_H  \tilde G_H  A$ violate $U(1)_c$ gauge invariance. There are two ways in which the couplings could emerge in intersecting D6-brane models. The first method consists of the nonrenormalisable quadratic couplings $C \,G_H  G_H \tilde A$ and $\tilde C\, \tilde G_H  \tilde G_H  A$ involving the states in the antisymmetric representation from the $cc'$ sector and suppressed by the string mass scale. Upon a supersymmetric stabilisation of the fields $C$ and $\tilde C$, the quadratic couplings reduce to the desired cubic couplings with coupling constant of the order ${\cal O}( \langle C \rangle / M_{\text{string}})$. A second method to compensate the $U(1)_c$ charge in the cubic couplings is provided by the potential presence of non-perturbative Euclidean D-brane instanton corrections to the superpotential, in which case the coefficient of the cubic coupling is exponentially suppressed by the volume of the three-cycle wrapped by the Euclidean instanton, see e.g.~\cite{Blumenhagen:2009qh}.

\mathtabfix{
\begin{array}{|c||c|c||c|c|c||c|}\hline 
\muc{7}{|c|}{\text{\bf D6-brane configuration of a local Pati-Salam model on the {bAA} lattice with $\varrho = 2$: prototype II}}
\\\hline \hline
&\text{\bf wrapping numbers} &\frac{\rm Angle}{\pi}&\text{\bf $\Z_2^{(i)}$ eigenvalues}  & (\vec \tau) & (\vec \sigma)& \text{\bf gauge group}\\
\hline \hline
 a&(1,1;1,0;1,-1)&(\frac{1}{3},0,-\frac{1}{3})&(+--)&(0,1,1) & (0,1,1)& U(4)\\
 b&(1,1;1,0;1,-1)&(\frac{1}{3},0,-\frac{1}{3})&(+++)&(0,1,1) & (0,1,1)&U(2)\\
 c&(1,1;1,0;1,-1)&(\frac{1}{3},0,-\frac{1}{3})&(-+-)&(0,1,1) & (0,1,1)&U(2)\\ 
 \hline
\end{array}
}{PatiSalambAAPrototypeII}{D6-brane configuration of a three-stack local Pati-Salam model with gauge group 
$SU(4)_a\times SU(2)_b \times SU(2)_c\times U(1)_a\times U(1)_b\times U(1)_c$ on the {\bf bAA} lattice of the orientifold $T^6/(\Z_2 \times \Z_6 \times \OR)$ with discrete torsion $\eta=-1$, complex structure modulus $\varrho=2$ and the $\OR$-plane as the exotic O6-plane ($\eta_{\OR}=-1$).}

\mathtab{
\begin{array}{|c||c|c|c|}
\hline \multicolumn{3}{|c|}{\text{\bf Spectrum of the prototype II local Pati-Salam model on bAA}}\\
\hline
\hline
\text{State} & \text{Sector} & (SU(4)_a\times SU(2)_b\times SU(2)_c)_{U(1)_a\times U(1)_b \times U(1)_c}  
\\
\hline 
 (Q_L, L) & ab' & 3 \times (\4, \2, \1)_{(1,1,0)} 
 \\
 (Q_R, R) & ac' & 3 \times (\ov \4, \1, \2)_{(-1,0,-1)} 
 \\
 (H_u, H_d) & bc' & 3 \times (\1,\2,\2)_{(0,-1,-1)} 
  \\
A, \tilde A &aa'& 2 \times [ ({\bf 6}_\Anti,\1,\1)_{(2,0,0)} + h.c.  ] 
\\
    &bb'& 2 \times [ (\1,\1_\Anti, \1)_{(0,2,0)} + h.c.  ]
    \\
    &cc'& 2 \times [ (\1,\1, \1_\Anti)_{(0,0,2)} + h.c.  ] 
    \\
 &ab&  (\4, \2, \1)_{(1,-1,0)} + h.c. 
  \\
  &ab'&  (\4, \2, \1)_{(1,1,0)} + h.c. 
  \\
G_H, \tilde G_H   &ac& 2 \times [ (\4, \1, \2)_{(1,0,-1)} + h.c.] 
\\
  (H_u, H_d) &bc& 2 \times [ (\1,\2,\2)_{(0,1,-1)} + h.c. ] 
  \\
    \hline
\end{array}
}{PatiSalambAAPrototypeIISpectrum}{Chiral and non-chiral massless matter spectrum of the prototype II local Pati-Salam model with D6-brane configuration in table~\ref{tab:PatiSalambAAPrototypeII}.}

Observe that the chiral spectrum in both local models is such that the non-Abelian gauge anomalies vanish, even though the RR tadpole cancellation conditions are violated. Mixed and purely Abelian anomalies on the other hand do not vanish, implying that the $U(1)$ symmetries are inherently anomalous. 
Given the presence of non-chiral pairs of antisymmetric representations in the $bb'$ sector for both prototypes, one might speculate about the Peccei-Quinn nature of the $U(1)_b$ symmetry and its r\^ole in a supersymmetric DFSZ axion model as proposed in~\cite{Honecker:2013mya}.

\subsubsection{Global Pati-Salam models on the aAA lattice}

Results concerning intersecting D6-branes model building on the {\bf aAA} lattice are visibly different from the ones on the {\bf bAA} lattice, as exemplified in the discussion on  D6-brane configurations with three generations for $\varrho$-independent set-ups in section~\ref{Sss:3GenRhoIndep}. On the {\bf aAA} lattice, it is possible to find pairs of fractional three-cycles yielding three generations at their mutual intersections for each value of the complex structure modulus $\varrho$ without overshooting the bulk RR tadpole cancellation conditions, see e.g.~table~\ref{tab:NetChiral3GenRhoIndAddO6}. A three-generational, supersymmetric Pati-Salam model with gauge group $U(4)\times SU(2)_L \times SU(2)_R$ can easily be constructed on the {\bf aAA} lattice by wrapping the $U(4)$ stack along a rigid fractional three-cycle parallel to the $\OR$-plane, while the $SU(2)_L$ and $SU(2)_R$ stacks are wrapped along rigid fractional three-cycles parallel to the $\OR\Z_2^{(1)}$-plane with enhanced $USp(2)$ gauge groups. In such a D6-brane configuration, either the $\OR\Z_2^{(2)}$-plane or the $\OR\Z_2^{(3)}$-plane is taken to be the exotic O6-plane, such that the bulk RR tadpole contributions from the D6-brane stacks do not overcompensate the O6-plane charge contributions:
\begin{equation}
\sum_{x\in \{ a, b, c \}} N_x (2 P_x + Q_x) =  2 N_a + 6 N_b + 6 N_c = 20 \stackrel{!}{\leq} 32.
\end{equation}     
The second bulk RR tadpole cancellation condition is trivially satisfied for three-cycles with bulk wrapping number $V_x = 0$, i.e.~with bulk orbit $(1,0;1,0;1,0)$, $(1,0;-1,2;1,-2)$, $(1,0;2,1;3,-1)$ or $(1,0;4,-1;3,1)$. The last two bulk orbits do not give rise to rigid fractional three-cycles and have bulk wrapping numbers for which $2 P_x + Q_x >12$. Hence, they are neither useful to account for a visible D6-brane stack nor for a hidden D6-brane stack. To complete the above D6-brane set-up into a global model, we must add (at least) a fourth D6-brane stack supporting a $U(6)$ gauge group or a $U(2)$ gauge group (or several stacks with smaller gauge groups), depending on whether the corresponding fractional three-cycle has a bulk orbit parallel to the $\OR$-plane or parallel to the $\OR\Z_2^{(1)}$-plane, respectively. 

\mathtabfix{
\begin{array}{|c||c|c||c|c|c||c|}\hline 
\muc{7}{|c|}{\text{\bf D6-brane configuration of a global Pati-Salam model on the {aAA} lattice: prototype I}}
\\\hline \hline
&\text{\bf wrapping numbers} &\frac{\rm Angle}{\pi}&\text{\bf $\Z_2^{(i)}$ eigenvalues}  & (\vec \tau) & (\vec \sigma)& \text{\bf gauge group}\\
\hline \hline
 a&(1,0;1,0;1,0)&(0,0,0)&(--+)&(0,1,1) & (0,1,1)& U(4)\\
 b&(1,0;-1,2;1,-2)&(0,\frac{1}{2},-\frac{1}{2})&(+++)&(0,0,1) & (0,0,1)&USp(2)\\
 c&(1,0;-1,2;1,-2)&(0, \frac{1}{2},-\frac{1}{2})&(--+)&(0,0,1) & (0,0,1)&USp(2)\\ 
  d&(1,0;1,0;1,0)&(0,0,0)&(+--)&(0,1,1) & (0,1,1)& U(6)\\
 \hline
\end{array}
}{PatiSalamaAAPrototypeI}{D6-brane configuration of a global four-stack Pati-Salam model with gauge group 
$SU(4)_a\times USp(2)_b \times USp(2)_c\times SU(6)_d\times U(1)_a\times U(1)_d $ on the {\bf aAA} lattice of the orientifold $T^6/(\Z_2 \times \Z_6 \times \OR)$ with discrete torsion $\eta=-1$ and the $\OR\Z_2^{(2)}$-plane as the exotic O6-plane ($\eta_{\OR\Z_2^{(2)}}=-1$).}

\mathtab{
\begin{array}{|c||c|c|c|}
\hline \multicolumn{3}{|c|}{\text{\bf Spectrum of the prototype I global Pati-Salam model on aAA}}\\
\hline
\hline
\text{State} & \text{Sector} & (SU(4)_a\times USp(2)_b\times USp(2)_c \times SU(6)_d)_{U(1)_a\times U(1)_d} 
\\
\hline 
 (Q_L, L) & ab=ab' & 3 \times (\4, \2, \1, \1)_{(1,0)} 
 \\
  (Q_R, R) & ac=ac' & 3 \times (\ov \4, \1, \2,\1)_{(-1,0)} 
  \\
      (H_u, H_d)  &bc = bc'&  10 \times (\1,\2,\2, \1)_{(0,0)} 
      \\
  & bd = bd'& 3 \times (\1, \2, \1,{\bf 6})_{(0,1)}   
  \\
    & cd = cd'& 3 \times (\1, \1, \2, \ov{\bf 6})_{(0,-1)} 
    \\
A, \tilde A  &aa'& 2 \times \left[  ({\bf 6}_\Anti,\1,\1,\1)_{(2,0)} + h.c. \right]  
\\
    &bb'& 5 \times  (\1,\1_\Anti, \1,\1)_{(0,0)}  
    \\
    &cc'& 5 \times  (\1,\1, \1_\Anti,\1)_{(0,0)}   
    \\
      &dd'& 2 \times [ (\1,\1,\1,{\bf15}_\Anti))_{(0,2} + h.c.  ] 
      \\
& ad & 2 \times [ (\4,\1,\1,\ov{\6})_{(1,-1)} + h.c. ]  
\\ 
& ad' &   (\4,\1,\1,\6)_{(1,1)} + h.c.   
\\ 
    \hline
\end{array}
}{PatiSalamaAAPrototypeISpectrum}{Chiral and non-chiral massless matter spectrum of the prototype I global Pati-Salam model with D6-brane configuration in table~\ref{tab:PatiSalamaAAPrototypeI}.}

The above ansatz with $U(4)_a \times USp(2)_b \times USp(2)_c \times U(N_d)$ gauge group amounts to two distinguishable prototypes of global four-stack Pati-Salam models. For each prototype, we can write down an explicit D6-brane configuration as given in tables~\ref{tab:PatiSalamaAAPrototypeI}  and \ref{tab:PatiSalamaAAPrototypeII}, with corresponding massless matter spectrum listed respectively in tables~\ref{tab:PatiSalamaAAPrototypeISpectrum} and \ref{tab:PatiSalamaAAPrototypeIISpectrum}. The major difference between the two prototype models is of course the gauge group in the hidden sector: $U(6)$ for the first prototype and $U(2)$ for the second prototype. Differences can also be found in the non-chiral part of the spectra for the two models, such as the presence of three non-chiral pairs in the $bd$ and $cd$ sector of the second prototype model. The chiral spectrum for both prototypes is realised in exactly the same way, including 10 Higgses, three `chiral exotics' in the $bd$ sector and three `chiral exotics' in the $cd$ sector,
where by `chiral exotics' we denote states arising from some non-vanishing intersection number with at least some charge under an (anomalous) $U(1)$ factor. 
Both the $b$-stack and the $c$-stack support states in the antisymmetric representation under $USp(2)\simeq SU(2)$, which appear as gauge singlets in the massless matter spectrum. 
The two $U(1)$ gauge factors in each prototype model are anomalous and acquire a St\"uckelberg mass of the order of the string mass scale when invoking the generalized Green-Schwarz mechanism, i.e. 
some closed string axion stemming from the RR 5-form in ten dimensions forms the longitudinal mode of the massive vector boson~\cite{Ibanez:2001nd}, see e.g. also section 5 of~\cite{Honecker:2012qr} 
for a similar computation for  one for models on the closely related $T^6/(\Z_2\times \Z_6' \times \OR)$ orbifold.
The remaining perturbative global $U(1)$ symmetry is generically  broken by non-perturbative effects, but some $\Z_n$ subgroup can survive in the full low-energy effective field theory~\cite{BerasaluceGonzalez:2011wy,Anastasopoulos:2012zu} for which the necessary and sufficient conditions have to be derived in analogy to the $T^6/\Z_{N=6,6'}$ and $T^6/(\Z_2\times \Z_6' \times \OR)$ orbifolds in~\cite{Honecker:2013hda,Honecker:2013kda}.

From the field theory side, the standard Yukawa couplings between the Higgses quarks and leptons are allowed by charge conservation:
\begin{equation}
 {\cal W} \supset f_{ijk}\,  (H_u, H_d)_i\, \cdot   (Q_L, L)_j\,  \cdot   (Q_R, R)_k, 
\end{equation}
where the indices $j$ and $k$ indicate the quark-lepton generations ($j,k \in\{1,2,3\}$) and the index $i$ refers to one of the 10 Higgs-doublets from the $bc$ sector. In order to see whether the Yukawa coefficients $f_{ijk}$ are non-vanishing, they have to be computed explicitly. Analogously to~\cite{Honecker:2012jd,Honecker:2012qr}, an estimate of the order of magnitude computation can be performed, following the methods  of~\cite{Cremades:2003qj} developed for the six-torus without orbifold, by calculating the area of the (triangular) worldsheet instanton spanned by three intersecting D-branes and with the resulting intersection points functioning as triangle apexes on the toroidal orbifold $T^6/(\Z_2 \times \Z_6)$.
 
The $ac=ac'$ sector does not come with additional non-chiral pairs, which would contain natural GUT-Higgses. Hence, the field theoretic scheme~\cite{Antoniadis:1988cm,King:1997ia} outlined for the local models in the previous section to spontaneously break the $SU(4)\times SU(2)_R$ gauge group down to the Standard Model gauge group $SU(3)_{QCD} \times U(1)_Y$ might not work for these global models, and an alternative method to break the Pati-Salam GUT gauge groups should be considered along with potential {\it vev}s of the scalar partners to the right-handed particles in the $(\ov \4, \1, \2,\1)_{(-1,0)}$ representation. We postpone a thorough field theory discussion of these global models, including the computation of the Yukawa couplings and minimum of the scalar potential, to future work~\cite{EckerHoneckerStaessens:2014}.   

\mathtabfix{
\begin{array}{|c||c|c||c|c|c||c|}\hline 
\muc{7}{|c|}{\text{\bf D6-brane configuration of a global Pati-Salam model on the {aAA} lattice: prototype II}}
\\\hline \hline
&\text{\bf wrapping numbers} &\frac{\rm Angle}{\pi}&\text{\bf $\Z_2^{(i)}$ eigenvalues}  & (\vec \tau) & (\vec \sigma)& \text{\bf gauge group}\\
\hline \hline
 a&(1,0;1,0;1,0)&(0,0,0)&(--+)&(0,1,1) & (0,1,1)& U(4)\\
 b&(1,0;-1,2;1,-2)&(0,\frac{1}{2},-\frac{1}{2})&(+++)&(0,0,1) & (0,0,1)&USp(2)\\
 c&(1,0;-1,2;1,-2)&(0, \frac{1}{2},-\frac{1}{2})&(--+)&(0,0,1) & (0,0,1)&USp(2)\\ 
  d&(1,0;-1,2;1,-2)&(0, \frac{1}{2},-\frac{1}{2})&(+++)&(0,0,0) & (0,0,0)& U(2)\\
 \hline
\end{array}
}{PatiSalamaAAPrototypeII}{D6-brane configuration of a four-stack global Pati-Salam model with gauge group 
$SU(4)_a\times USp(2)_b \times USp(2)_c\times SU(2)_d \times U(1)_a\times U(1)_d$ on the {\bf aAA} lattice of the orientifold $T^6/(\Z_2 \times \Z_6 \times \OR)$ with discrete torsion $\eta=-1$ and the $\OR\Z_2^{(2)}$-plane as the exotic O6-plane ($\eta_{\OR\Z_2^{(2)}}=-1$).}

\mathtab{
\begin{array}{|c||c|c|c|}
\hline \multicolumn{3}{|c|}{\text{\bf Spectrum of the prototype II global Pati-Salam model on aAA}}\\
\hline
\hline
\text{State} & \text{Sector} & (SU(4)_a\times USp(2)_b\times USp(2)_c \times SU(2)_d)_{U(1)_a\times U(1)_d}  
\\
\hline 
 (Q_L, L) & ab=ab' & 3 \times (\4, \2, \1, \1)_{(1,0)} 
 \\
  (Q_R, R) & ac=ac' & 3 \times (\ov \4, \1, \2,\1)_{(-1,0)}  
  \\
      (H_u, H_d)  &bc = bc'&  10 \times (\1,\2,\2, \1)_{(0,0)} 
      \\
  & bd = bd'& 3 \times (\1, \2, \1,{\2})_{(0,1)}  
  \\
    & cd = cd'& 3 \times (\1, \1, \2, \2)_{(0,-1)} 
    \\
A, \tilde A  &aa'& 2 \times \left[  ({\bf 6}_\Anti,\1,\1,\1)_{(2,0)} + h.c. \right]  
\\
    &bb'& 5 \times  (\1,\1_\Anti, \1,\1)_{(0,0)} 
    \\
    &cc'& 5 \times  (\1,\1, \1_\Anti,\1)_{(0,0)}   
    \\
      &dd'& 6 \times [ (\1,\1,\1,\1_\Anti)_{(0,2)} + h.c.  ] 
      \\
& ad & 2 \times [ (\4,\1,\1,\2)_{(1,-1)} + h.c. ]  
\\ 
& ad' &   (\4,\1,\1,\2)_{(1,1)} + h.c.    
\\ 
&bd = bd' & 3\times [(\1, \2, \1,{\2})_{(0,-1)} + h.c. ] 
\\
&cd = cd' & 3\times [(\1, \1, \2,{\2})_{(0,-1)} + h.c. ] 
\\
    \hline
\end{array}
}{PatiSalamaAAPrototypeIISpectrum}{Chiral and non-chiral massless matter spectrum of prototype II global Pati-Salam model with D6-brane configuration in table~\ref{tab:PatiSalamaAAPrototypeI}.}

After the successful identification of two prototypes of $\varrho$-independent global Pati-Salam models one could raise the question whether the {\bf aAA} lattice also allows for $\varrho$-dependent D6-brane configurations of global Pati-Salam models. For $\varrho$-dependent configurations, two scenarios for the D6-brane configurations can be considered:
\begin{itemize} 
\item the $U(4)_a$-stack can be taken to be parallel to the $\OR$-plane, while the $SU(2)_L$ and $SU(2)_R$ stacks are wrapped on the same bulk orbit taken from the list in table~\ref{tab:NetChiralitiesRhoDependentAppOR}. In this case, it is the supersymmetry requirement for the $b$- and $c$-stacks that dynamically stabilises the complex structure modulus $\varrho$;
\item the $U(4)_a$-stack is characterised by the bulk orbit $(1,3;1,0;1,-1)$, one of the $SU(2)$ gauge factors is also wrapped on the bulk orbit $(1,3;1,0;1,-1)$, while the other $SU(2)$ gauge factor is parallel to the $\OR$-plane. This is the only D6-brane configuration in table~\ref{tab:BulkOrbits3QuarkGenerations} which provides for three generations of left-handed and right-handed quarks, and for which the bulk orbit of the $U(4)_a$-stack is not parallel to the $\OR$-plane.  
\end{itemize}
Extended systematic searches for global Pati-Salam (as well as Standard Models) with one of these (analogous) D6-brane configurations on the {\bf aAA} lattice go well beyond the scope of the present article, but are currently under way~\cite{EckerHoneckerStaessens:2014}. The most difficult constraint to fulfil appears to be the (twisted) RR tadpole cancellation conditions.

\section{Discussion and Conclusions}\label{S:Conclusions}

In this article, we started the systematic exploration of phenomenologically interesting D-brane configurations on the $T^6/(\Z_2 \times \Z_6 \times \OR)$ orientifold with discrete torsion, which was expected to be the most profitable one, based on earlier work with intersecting D6-branes on orbifolds.  We gave a complete classification of rigid three-cycles, three-cycles on which D-branes exist without chiral matter in the symmetric representation and  which discrete parameter combinations lead to gauge group enhancements $U(N) \hookrightarrow USp(2N)$ or $SO(2N)$ with their respective amount of matter in the symmetric and antisymmetric representation.

Assuming the absence of exotic states in the adjoint or symmetric representation, we argued that  globally defined $SU(5)$ GUT models with three particle generations are completely excluded on $T^6/(\Z_2 \times \Z_6 \times \OR)$. This finding agrees with searches on all other orbifolds to date~\cite{Honecker:2004np,Gmeiner:2006vb,Cvetic:2006by,Gmeiner:2007zz,Honecker:2012qr}, where constraints on three generations without exotic matter also excluded $SU(5)$ GUTs.

We proceeded by presenting first some local, then also global Pati-Salam models with two different prototype massless spectra displayed in tables~\ref{tab:PatiSalamaAAPrototypeISpectrum} and~\ref{tab:PatiSalamaAAPrototypeIISpectrum}. Besides three particle generations and ten Higgs multiplets, both models display a mild amount of vector-like particles. 
These first limited searches support the claim that the $T^6/(\Z_2 \times \Z_6 \times \OR)$ orientifold with discrete torsion might be the most fertile orbifold for D6-brane model building.
A wider, or even complete, scan for Pati-Salam, left-right symmetric and MSSM vacua is beyond the scope of this paper and will be performed in future work~\cite{EckerHoneckerStaessens:2014}.

The scan for MSSM or GUT-like spectra needs to be supplemented by a thorough exploration of the corresponding low-energy effective field theory.
This includes discrete gauge symmetries extending the work presented in~\cite{BerasaluceGonzalez:2011wy,Ibanez:2012wg,Anastasopoulos:2012zu,Honecker:2013hda,Honecker:2013kda}\footnote{See also~\cite{Klevers:2014bqa,Garcia-Etxebarria:2014qua,Mayrhofer:2014haa} for recent investigations in the context of F-theory models.}, perturbative gauge~\cite{Blumenhagen:2007ip,Honecker:2011sm,Honecker:2011hm} and Yukawa couplings, non-perturbative effects from D-brane instantons~\cite{Blumenhagen:2009qh}, as well as the derivation of one-loop corrections to K\"ahler metrics for rigid D-branes in extension of toroidal models and special combinations of fractional D-branes with pairwise opposite $\Z_2$ eigenvalues~\cite{Berg:2011ij,Berg:2014ama}. 
These field theoretical considerations require new computational tools to be developed and refined in the future, and  they will further constrain phenomenologically viable models beyond the demands on the massless matter spectrum presented in this article.

Global string vacua with MSSM or GUT-like spectra constitute an important corner stone for ultra-violet complete models of Beyond the Standard model physics. It is thus of particular importance to study their 
low-energy phenomenology in future, and in particular to test the validity of low-string scale scenarios in our class of global D-brane models proposed originally in e.g.~\cite{Antoniadis:2002qm,Lust:2008qc,Anchordoqui:2008di,Feng:2010yx,Anchordoqui:2011eg,Berenstein:2014wva,Anchordoqui:2014wha}, and to investigate the interplay between phenomenological and cosmological implications \cite{Kiritsis:2003mc,Anchordoqui:2012wt}, which includes e.g. the further exploitation of axion models~\cite{Svrcek:2006yi,Berenstein:2012eg,Honecker:2013mya}.

\noindent
{\bf Acknowledgements:} 
This work is partially supported by the {\it Cluster of Excellence `Precision Physics, Fundamental Interactions and Structure of Matter' (PRISMA)} DGF no. EXC 1098,
the DFG research grant HO 4166/2-1 and the DFG Research Training Group {\it `Symmetry Breaking in Fundamental Interactions'} GRK 1581.

\appendix
\section{Classification of bulk three-cycles on a/bAA lattice}\label{A:ClassBulkThreeCycles}

As pointed out in section~\ref{Sss:BulkCycles}, a full classification of supersymmetric bulk three-cycles on $T^6/(\Z_2\times \Z_6 \times \OR)$ not overshooting the bulk RR tadpoles in table~\ref{tab:Bulk-RR+SUSY-Z2Z6} depends, besides on the lattice configuration, on the complex structure modulus $\varrho$ encoding the ratio between the two basic one-cycle lengths on $T_{(1)}^2$. The modulus $\varrho$ is dynamically stabilised by one D6-brane wrapping a special Lagrangian three-cycle at non-trivial angle along $T^2_{(1)}$, i.e. with 
$(n^1,\tilde{m}^1) \notin \{(\frac{1}{1-b},0),(0,1)\}$, as can be seen from the necessary bulk supersymmetry condition in table~\ref{tab:Bulk-RR+SUSY-Z2Z6} and the definition of bulk wrapping numbers in equation~(\ref{Eq:BulkWrappingNumbers}).
As a first step, one can classify all four-torus wrapping numbers $(X_a, Y_a)$, whose lower bounds are derived from the supersymmetric calibration conditions, while their upper bounds are deduced from the bulk RR tadpole cancellation conditions as detailed at the end of section~\ref{Sss:BulkCycles}. 

Due to the symmetry of rotating the full D6-brane set-up by the non-supersymmetric angle $\pm (\frac{\pi}{2},-\frac{\pi}{3},0)$ or $\pm (0,\frac{\pi}{3},0)$ from {\bf a/bAA} to {\bf a/bAB} or {\bf a/bBB} as demonstrated in section~\ref{Sss:FractionalCycles},  it suffices to focus on the {\bf a/bAA} lattice configurations.
Three cases can be distinguished here as pointed out at the end of section~\ref{Sss:BulkCycles}. The special cases with $Y_a = 0$ or $-Y_a = 2 X_a > 0$ represent cycles whose calibration conditions are independent of the $\varrho$-modulus. In both cases, the condition on $Y_a$ requires $Y_a \in 2 \Z$, which using the definition in equation~(\ref{Eq:BulkWrappingNumbers}) implies that $(n_a^2,m_a^2)=$ (even, odd) due to the previously imposed condition $(n_a^3,m_a^3)=$~(odd,~odd). The full list of two-cycles on $T_{(2)}^2\times T_{(3)}^2$ satisfying these two different special conditions on $(X_a, Y_a)$ is given in the first two blocks of the left column in table \ref{tab:XYlistAA1}. 

Adding the appropriate torus wrapping numbers $(n_a^1,m_a^1) \in \{(\frac{1}{1-b},\frac{-b}{1-b}),(0,1)\}$ to the above $(X_a,Y_a)$ then allows to identify eight bulk three-cycles which are supersymmetric for all values of $\varrho$, after having eliminated orbifold and orientifold images (e.g.~by imposing $|n_a^3| \geq |m_a^3|$).
As an example one can consider the bulk three-cycles $(\frac{1}{1-b},\frac{-b}{1-b};2,1;3,-1)$ and $(\frac{1}{1-b},\frac{-b}{1-b};-2,3;1,-3)$, which are characterised by the same bulk wrapping numbers $(X_a,Y_a)$ and $(P_a,Q_a,U_a,V_a)$, and are related to each other through a combination of orbifold and orientifold transformations. By imposing the condition $|n_a^3| \geq |m_a^3|$, the first orbit is selected, and double-counting of three-cycles belonging in the same orbifold/orientifold orbit is avoided. 
 Four of these eight bulk three-cycles are parallel to an orientifold $\OR(\Z_2^{(i)})$-plane (see table~\ref{tab:Z2Z6-Oplanes-torus}), while the four other independent bulk three-cycles are listed in table~\ref{tab:RhoIndependentaAA}.   
\mathtab{
\begin{array}{|c||c||c|c||c|c|c|c|}
\hline \multicolumn{8}{|c|}{\text{\bf SUSY bulk three-cycles on {\bf a/bAA} with $Y_a = 0$ or $-Y_a = 2 X_a > 0$}}\\
\hline
\hline (n^i,m^i)_{i \in\{1,2,3\}} & \varrho& X_a & Y_a & P_a & Q_a & U_a & V_a \\
\hline \hline
(\frac{1}{1-b},\frac{-b}{1-b};2,1;3,-1)&\forall \, \varrho&7&0&7&0&0&0\\
(\frac{1}{1-b},\frac{-b}{1-b};4,-1;3,1)&\forall \, \varrho&13&0&13&0&0&0\\
(0,1;4,-5;3,-1)&\forall \, \varrho&7&-14&0&0&7&-14\\
(0,1;2,-3;5,-1)&\forall \, \varrho&7&-14&0&0&7&-14\\
\hline
\end{array}
}{RhoIndependentaAA}{Supersymmetric bulk three-cycles on {\bf a/bAA} with either $X_a> Y_a=0$ or $2 X_a = -Y_a >0$, which do not overshoot the bulk RR tadpole cancellation conditions of table~\protect\ref{tab:Bulk-RR+SUSY-Z2Z6}.
In addition to the four cycles displayed here, the four bulk three-cycles parallel to some O6-plane orbit in table~\protect\ref{tab:Z2Z6-Oplanes-torus}  satisfy the same constraints.}

In the third and more generic case with $n_a^1 >0$ and $\tilde m_a^1>0$, one of the supersymmetry conditions explicitly relates the four-torus wrapping numbers to the modulus $\varrho$:
\begin{equation}
Y_a = -\frac{\varrho}{3} \frac{\tilde m_a^1}{n^1_a} [2 X_a + Y_a], \qquad 2X_a + Y_a >0,
\end{equation}
while the second one puts a lower bound on the bulk wrapping numbers. Including the upper bound deduced from the bulk RR tadpoles, one can first list all torus wrapping numbers $(n_a^2, m_a^2; n_a^3, m_a^3)$ based on the even- and/or oddness on $T_{(2)}^2$ for $(n^3_a,m^3_a)=(\text{odd},\text{odd})$ with $n^3_a>0$. An exhaustive list can be found in tables \ref{tab:XYlistAA1}, \ref{tab:XYlistAA2} and \ref{tab:XYlistAA3}.

Secondly, the torus wrapping numbers on $T_{(1)}^2$ have to be added such that the supersymmetry condition involving the modulus $\varrho$ is satisfied. For specific rational values of $\varrho$, one expects to encounter additional supersymmetric bulk three-cycles beyond those in tables~\ref{tab:Z2Z6-Oplanes-torus} and~\ref{tab:RhoIndependentaAA}. Furthermore, the presence of the discrete parameter $b$ in $\tilde m_a^1$ forces us to distinguish between an untilted ($b=0$) and tilted ($b=1/2$) two-torus $T_{(1)}^2$ when classifying the supersymmetric bulk three-cycles. More explicitly, for the {\bf aAA} lattice there exist 409 different values of $\varrho$ ranging between $ \frac{1}{80} \leq \varrho \leq 720$ for which at least one additional supersymmetric bulk three-cycle exists, while the {\bf bAA} lattice comes with additional supersymmetric bulk three-cycles for 181 different values of $\varrho$ lying within the interval $\frac{2}{75}\leq \varrho \leq 1350$. 

\begin{figure}[h]
\hspace*{-1in}\begin{tabular}{c}
\vspace*{-1.2in}\includegraphics[width=20cm]{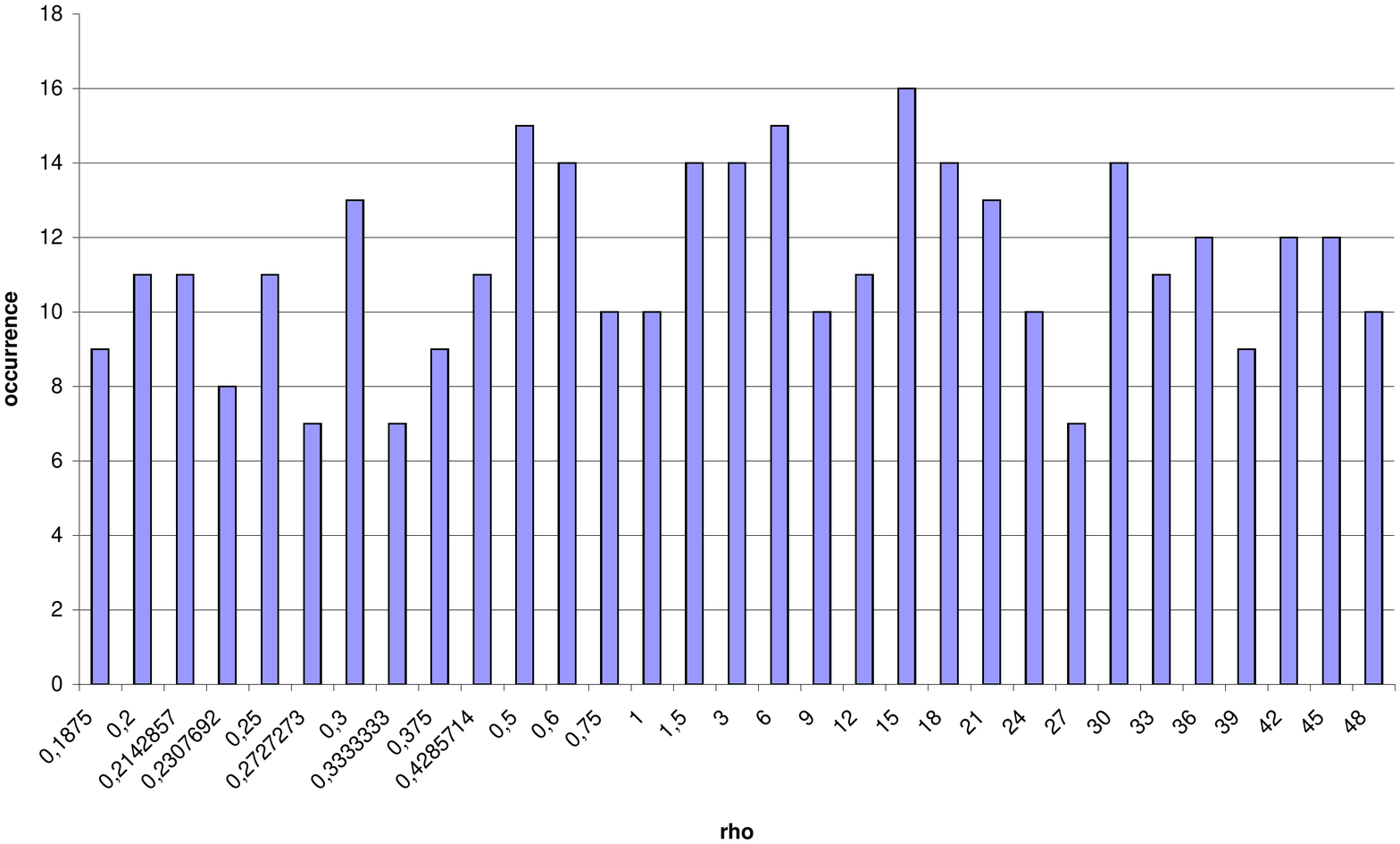}  \begin{picture}(0,0) \put(-280,340){\bf aAA}\end{picture} \\
\vspace*{-0.6in} \includegraphics[width=15cm]{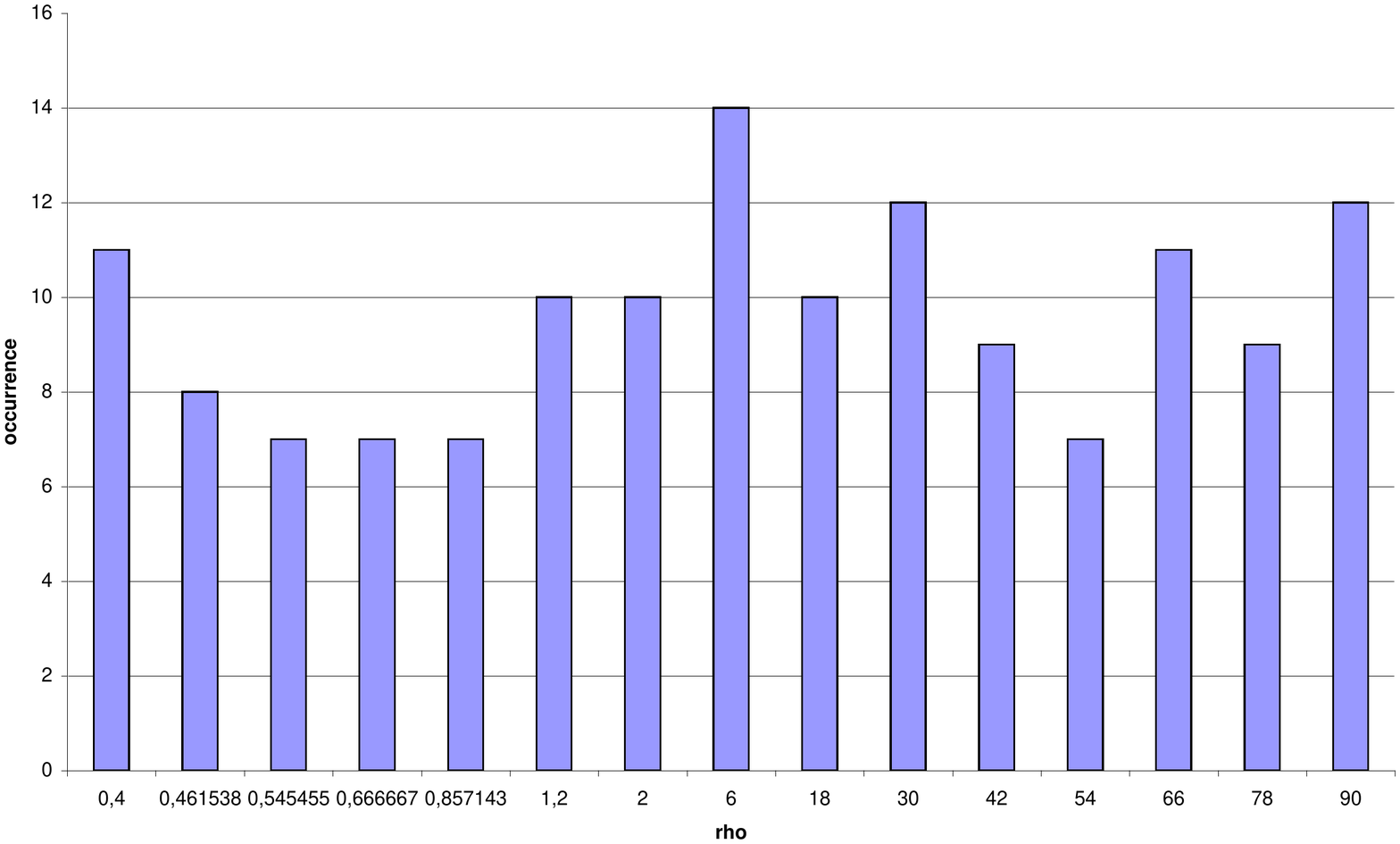}  \begin{picture}(0,0) \put(-200,250){\bf bAA}\end{picture} 
\end{tabular}
\begin{center}
\caption{Distribution of supersymmetric bulk three-cycles for values of the complex structure modulus $\varrho$ suitable for D6-brane model building with rigid three-cycles free of chiral matter in the symmetric
representation for the {\bf aAA} lattice (upper) and the {\bf bAA} lattice (lower).\label{Fig:DistributionRhoSUSYcycles}}
\end{center}
\end{figure}

We refrain from listing all 1760 (917) additional supersymmetric bulk three-cycles at non-trivial angle on $T^2_{(1)}$
 for the {\bf aAA} ({\bf bAA}) lattice and present instead some qualitative observations regarding the $\varrho$-dependent classification:
\begin{itemize} 
\item A majority of the values for the $\varrho$-modulus allows at most four additional supersymmetric bulk three-cycles. The values of $\varrho$ allowing for 9 or more supersymmetric bulk three-cycles
besides the eight existing for any value of $\varrho$ are listed in table~\ref{tab:ComplexStructureSUSYCycles2Lattices} for both lattices.  
\item The number of phenomenologically attractive $\varrho$-values can be reduced through the model building considerations discussed in section \ref{S:StepsInD6BraneMB}. It turns out that most of the fractional D6-branes on $T^6/(\Z_2 \times \Z_6 \times \OR)$ come with matter in the adjoint representation making them unsuitable to accommodate the {\em QCD}-stack or the $SU(2)_L$-stack. Requiring the presence of at least one completely rigid three-cycle besides the ones parallel to the $\OR$-plane lowers the number of appropriate $\varrho$-values to 159 for the {\bf aAA} lattice and to 79 for the {\bf bAA} lattice. As it turns out, for each value of $\varrho$ there exists only one supersymmetric bulk three-cycle (apart from the bulk cycles parallel to an O6-plane) suitable to construct rigid D6-branes. Its orbit takes the generic form $(n_a^1,\tilde m_a^1;1,0;1,-1)$ with $n_a^1$ and $\tilde m_a^1$ fixed by the value of $\varrho$, in other words, the orbit can be represented by the choice of angles $\pi(\frac{1}{3},0,-\frac{1}{3})$ w.r.t. the $\OR$-invariant plane.    
\item The absence of chiral multiplets in the symmetric representation for the {\em QCD}- and the $SU(2)_L$-stack forms an additional model building constraint on the rigid three-cycles. This consideration eliminates another chunk of $\varrho$'s, leaving only 31 appropriate values of the complex structure modulus for the {\bf aAA} lattice and 15 for the {\bf bAA} lattice, as discussed in section~\ref{Ss:NoSyms}. 
\end{itemize}
Table~\ref{tab:ComplexStructureSUSYCycles2Lattices} provides an overview of the complex structure moduli values $\varrho$ (for both lattices) allowing for 9 or more additional supersymmetric bulk three-cycles, apart from the 8 supersymmetric bulk three-cycles present for any value of the complex structure modulus as discussed above. For each value of the complex structure modulus, it is also indicated whether one of the $\varrho$-dependent supersymmetric bulk three-cycles allows for completely rigid fractional three-cycles ($\varrho$-values in between parentheses do not allow for $\varrho$-dependent completely rigid fractional three-cycles). Complex structure parameters $\varrho$ allowing also for $\varrho$-dependent rigid fractional three-cycles free of chiral matter in the symmetric representation are highlighted in bold font.
Figure~\ref{Fig:DistributionRhoSUSYcycles} provides a schematic distribution for all $\varrho$-values allowing for supersymmetric, rigid fractional three-cycles free of chiral matter in the symmetric representation for the {\bf aAA} lattice (upper) and for the {\bf bAA} lattice (lower). In this figure, all relevant $\varrho$-values are represented, and we do not impose a lower bound on the number of supersymmetric bulk three-cycles as in table~\ref{tab:ComplexStructureSUSYCycles2Lattices}. The $\varrho$-values printed in bold in table~\ref{tab:ComplexStructureSUSYCycles2Lattices} are self-evidently included in figure~\ref{Fig:DistributionRhoSUSYcycles}.

 \mathtab{
\begin{array}{|c||c|c|}
\hline \multicolumn{3}{|c|}{\text{\bf Complex structure moduli $\varrho_{\bf a/b AA}$ with $\geq$ 9 additional SUSY 3-cycles}}\\
\hline
\hline \#\, \text{\bf of 3-cycles} & \varrho_{\bf aAA} & \varrho_{\bf bAA}  \\
\hline \hline
16 & 2, \frac{15}{2}, {\bf 15} & - \\
15 &{\bf  \frac{1}{2}}, {\bf 6} & - \\
14 &  {\bf \frac{3}{5}}, {\bf \frac{3}{2}}, {\bf 3}, 4, {\bf 18}, {\bf  30} & {\bf 6} \\
13 & {\bf  \frac{3}{10}}, \frac{21}{2}, {\bf  21}& - \\
12 & \left(\frac{1}{15}\right), \left(\frac{1}{10}\right), \frac{6}{5},  \frac{9}{2}, {\bf  36}, {\bf  42}, {\bf 45} & \left(\frac{2}{15}\right), {\bf  30}, {\bf 90} \\
11 & {\bf  \frac{1}{5}}, {\bf \frac{3}{14}} , {\bf \frac{1}{4}}, {\bf \frac{3}{7}}, \frac{6}{7} ,\frac{5}{3}, \frac{5}{2}, 8, {\bf 12}, {\bf  33} & {\bf \frac{2}{5}}, \frac{10}{3}, {\bf 66} \\
10 & \left(\frac{1}{12}\right), \left(\frac{1}{8}\right), \left(\frac{1}{6}\right), {\bf  \frac{3}{4}}, {\bf  1}, \frac{9}{5}, \frac{9}{4}, 5, {\bf 9}, {\bf 24}, {\bf  48} & {\bf \frac{6}{5}}, {\bf 2}, \frac{18}{5}, 10, {\bf 18} \\
9 &{\bf \frac{3}{16}}, {\bf\frac{3}{8}}, \frac{9}{8}, \frac{7}{2}, \frac{18}{5}, 10, {\bf 39}, (72), (90), (135) & 1, 4, {\bf 42}, {\bf 78}, (270) \\
\hline
\end{array}
}{ComplexStructureSUSYCycles2Lattices}{Overview of the values of the complex structure modulus $\varrho_{\bf a/b AA}$ with 9 or more additional supersymmetric bulk three-cycles, besides the four cycles parallel to an O6-plane in table~\protect\ref{tab:Z2Z6-Oplanes-torus}
 and the four three-cycles listed in table~\ref{tab:RhoIndependentaAA}. The $\varrho$-values in brackets do not allow for fractional three-cycles without matter in the adjoint representation. The $\varrho$-values in bold allow for fractional three-cycles without chiral matter in the symmetric representation (in addition to the absence of matter in the adjoint representation).}

\mathtabfix{
\begin{array}{ccc}
\multicolumn{3}{c}{\text{\bf \large Systematic classification of bulk 2-cycles on $T_{(2)}^2 \times T_{(3)}^2$ with lattice AA (part I)}}\\
\hline \hline
\begin{array}[h]{|c|c|c|c|}
\hline
(n^2_a,m^2_a;n^3_a,m^3_a) & X_a & Y_a & 2X_a+Y_a\\
\hline
\hline
(0,1;1,-1)&1&0&2\\
(2,-1;1,1)&3&0&6\\
(2,1;3,-1)&7&0&14\\
(-2,3;1,-3)&7&0&14\\
(4,-1;3,1)&13&0&26\\
(4,-3;1,3)&13&0&26\\
\hline
\hline
(2,-1;1,-1)&1&-2&0\\
(0,-1;1,1)&1&-2&0\\
(4,-5;3,-1)&7&-14&0\\
(4,1;1,-3)&7&-14&0\\
(2,-3;5,-1)&7&-14&0\\
(2,1;1,-5)&7&-14&0\\
\hline
\hline
(0,1;1,-3)&3&-2&4\\
(2,1;1,-1)&3&-2&4\\
(0,1;3,-5)&5&-2&8\\
(2,3;1,-1)&5&-2&8\\
(2,-1;3,1)&7&-2&12\\
(0,1;5,-7)&7&-2&12\\
(4,-3;1,1)&7&-2&12\\
(2,5;1,-1)&7&-2&12\\
(0,1;7,-9)&9&-2&16\\
(2,7;1,-1)&9&-2&16\\
(0,-1;1,3)&3&-4&2\\
(4,-1;1,-1)&3&-4&2\\
(2,-1;3,-1)&5&-4&6\\
(0,1;1,-5)&5&-4&6\\
(2,-3;1,1)&5&-4&6\\
(4,1;1,-1)&5&-4&6\\
(0,1;3,-7)&7&-4&10\\
(4,3;1,-1)&7&-4&10\\
(4,5;1,-1)&9&-4&14\\
(0,1;5,-9)&9&-4&14\\
(6,-1;1,-1)&5&-6&4\\
\hline
\end{array}
&
\begin{array}[h]{|c|c|c|c|}
\hline
(n^2_a,m^2_a;n^3_a,m^3_a) & X_a & Y_a & 2X_a+Y_a\\
\hline
\hline
(0,-1;1,5)&5&-6&4\\
(0,1;1,-7)&7&-6&8\\
(6,1;1,-1)&7&-6&8\\
(4,-5;1,1)&9&-6&12\\
(2,-1;5,-1)&9&-6&12\\
(2,-3;1,3)&11&-6&16\\
(0,1;5,-11)&11&-6&16\\
(6,5;1,-1)&11&-6&16\\
(4,-1;3,-1)&11&-6&16\\
(0,-1;3,5)&5&-8&2\\
(8,-3;1,-1)&5&-8&2\\
(2,1;1,-3)&5&-8&2\\
(0,-1;1,7)&7&-8&6\\
(8,-1;1,-1)&7&-8&6\\
(2,-1;5,-3)&7&-8&6\\
(2,-5;1,1)&7&-8&6\\
(8,1;1,-1)&9&-8&10\\
(0,1;1,-9)&9&-8&10\\
(0,1;3,-11)&11&-8&14\\
(8,3;1,-1)&11&-8&14\\
(0,-1;3,7)&7&-10&4\\
(10,-3;1,-1)&7&-10&4\\
(2,-3;3,1)&9&-10&8\\
(10,-1;1,-1)&9&-10&8\\
(0,-1;1,9)&9&-10&8\\
(4,-3;3,-1)&9&-10&8\\
(4,-7;1,1)&11&-10&12\\
(2,-1;7,-3)&11&-10&12\\
(0,1;1,-11)&11&-10&12\\
(10,1;1,-1)&11&-10&12\\
(0,1;3,-13)&13&-10&16\\
(10,3;1,-1)&13&-10&16\\
&&&\\
\hline
\end{array}
&
\begin{array}[h]{|c|c|c|c|}
\hline
(n^2_a,m^2_a;n^3_a,m^3_a) & X_a & Y_a & 2X_a+Y_a\\
\hline
\hline
(0,-1;5,7)&7&-12&2\\
(12,-5;1,-1)&7&-12&2\\
(2,-7;1,1)&9&-12&6\\
(2,-1;7,-5)&9&-12&6\\
(2,3;1,-3)&11&-12&10\\
(12,-1;1,-1)&11&-12&10\\
(0,-1;1,11)&11&-12&10\\
(2,1;3,-5)&11&-12&10\\
(0,1;1,-13)&13&-12&14\\
(12,1;1,-1)&13&-12&14\\
(0,-1;5,9)&9&-14&4\\
(14,-5;1,-1)&9&-14&4\\
(0,-1;3,11)&11&-14&8\\
(14,-3;1,-1)&11&-14&8\\
(0,-1;1,13)&13&-14&12\\
(2,-1;9,-5)&13&-14&12\\
(4,-9;1,1)&13&-14&12\\
(14,-1;1,-1)&13&-14&12\\
(0,1;1,-15)&15&-14&16\\
(14,1;1,-1)&15&-14&16\\
(0,-1;7,9)&9&-16&2\\
(16,-7;1,-1)&9&-16&2\\
(0,-1;5,11)&11&-16&6\\
(2,-1;9,-7)&11&-16&6\\
(16,-5;1,-1)&11&-16&6\\
(2,-9;1,1)&11&-16&6\\
(16,-3;1,-1)&13&-16&10\\
(6,-5;3,-1)&13&-16&10\\
(2,-3;5,1)&13&-16&10\\
(0,-1;3,13)&13&-16&10\\
(0,-1;1,15)&15&-16&14\\
(16,-1;1,-1)&15&-16&14\\
&&&\\
\hline
\end{array}
\end{array}
}{XYlistAA1}{Full list of the torus wrapping numbers on $T^2_{(2)} \times T^2_{(3)}$, characterised by $(n^2_a,m^2_a)=${\bf (even,~odd)},  $(n^3_a,m^3_a)=$(odd,~odd) and  $n^3_a>0$, and their corresponding bulk wrapping numbers $(X_a,Y_a)$, which satisfy the constraints for supersymmetry \{$X_a>Y_a=0$ or $2X_a = -Y_a >0$ or $2X_a + Y_a >0>Y_a$\} on the {\bf AA} lattice at the end of section~\ref{Sss:BulkCycles} and do not overshoot the bulk RR tadpoles \{$ 0 < (2X_a +Y_a), -Y_a  \leqslant 16$ or $0=Y_a < X_a \leqslant 16$ \}. 
}

\mathtabfix{
\begin{array}{ccc}
\multicolumn{3}{c}{\text{\bf \large Systematic classification of bulk 2-cycles on $T_{(2)}^2 \times T_{(3)}^2$ with lattice AA (part II)}}\\
\hline \hline
\begin{array}[h]{|c|c|c|c|}
\hline
(n^2_a,m^2_a;n^3_a,m^3_a) & X_a & Y_a & 2X_a+Y_a\\
\hline
\hline
(1,-1;1,1)&2&-1&3\\
(1,1;1,-1)&2&-1&3\\
(1,-1;1,3)&4&-1&7\\
(1,3;1,-1)&4&-1&7\\
(1,-1;1,5)&6&-1&11\\
(1,5;1,-1)&6&-1&11\\
(1,1;5,-3)&8&-1&15\\
(1,-1;1,7)&8&-1&15\\
(5,-3;1,1)&8&-1&15\\
(1,7;1,-1)&8&-1&15\\
(3,-1;1,-1)&2&-3&1\\
(1,-1;3,-1)&2&-3&1\\
(3,1;1,-1)&4&-3&5\\
(1,-1;3,1)&4&-3&5\\
(-1,3;1,-3)&8&-3&13\\
(3,5;1,-1)&8&-3&13\\
(1,-1;3,5)&8&-3&13\\
(1,1;1,-3)&4&-5&3\\
(1,-1;5,-1)&4&-5&3\\
(5,-1;1,-1)&4&-5&3\\
(1,-3;1,1)&4&-5&3\\
(5,1;1,-1)&6&-5&7\\
(1,-1;5,1)&6&-5&7\\
(1,-1;5,3)&8&-5&11\\
(3,-1;3,-1)&8&-5&11\\
(5,3;1,-1)&8&-5&11\\
\hline
\end{array}
&
\begin{array}[h]{|c|c|c|c|}
\hline
(n^2_a,m^2_a;n^3_a,m^3_a) & X_a & Y_a & 2X_a+Y_a\\
\hline
\hline
(1,-1;7,-3)&4&-7&1\\
(7,-3;1,-1)&4&-7&1\\
(7,-1;1,-1)&6&-7&5\\
(1,-1;7,-1)&6&-7&5\\
(1,-1;7,1)&8&-7&9\\
(7,1;1,-1)&8&-7&9\\
(1,1;3,-5)&8&-7&9\\
(3,-5;1,1)&8&-7&9\\
(7,3;1,-1)&10&-7&13\\
(1,-1;7,3)&10&-7&13\\
(1,-5;1,1)&6&-9&3\\
(1,1;1,-5)&6&-9&3\\
(9,-1;1,-1)&8&-9&7\\
(1,-1;9,-1)&8&-9&7\\
(1,-1;9,1)&10&-9&11\\
(1,-3;1,3)&10&-9&11\\
(1,3;1,-3)&10&-9&11\\
(9,1;1,-1)&10&-9&11\\
(5,-7;1,1)&12&-9&15\\
(1,1;5,-7)&12&-9&15\\
(3,1;1,-3)&6&-11&1\\
(1,-3;3,1)&6&-11&1\\
(1,-1;11,-5)&6&-11&1\\
(11,-5;1,-1)&6&-11&1\\
(11,-3;1,-1)&8&-11&5\\
(1,-1;11,-3)&8&-11&5\\
\hline
\end{array}
&
\begin{array}[h]{|c|c|c|c|}
\hline
(n^2_a,m^2_a;n^3_a,m^3_a) & X_a & Y_a & 2X_a+Y_a\\
\hline
\hline
(3,-7;1,1)&10&-11&9\\
(1,1;3,-7)&10&-11&9\\
(11,-1;1,-1)&10&-11&9\\
(1,-1;11,-1)&10&-11&9\\
(1,-1;11,1)&12&-11&13\\
(3,-1;5,-3)&12&-11&13\\
(5,-3;3,-1)&12&-11&13\\
(11,1;1,-1)&12&-11&13\\
(1,-1;13,-5)&8&-13&3\\
(1,1;1,-7)&8&-13&3\\
(1,-7;1,1)&8&-13&3\\
(13,-5;1,-1)&8&-13&3\\
(1,-1;13,-3)&10&-13&7\\
(13,-3;1,-1)&10&-13&7\\
(1,-1;13,-1)&12&-13&11\\
(13,-1;1,-1)&12&-13&11\\
(5,-9;1,1)&14&-13&15\\
(1,-1;13,1)&14&-13&15\\
(1,1;5,-9)&14&-13&15\\
(13,1;1,-1)&14&-13&15\\
(-1,-3;1,3)&8&-15&1\\
(15,-7;1,-1)&8&-15&1\\
(1,-1;15,-7)&8&-15&1\\
(1,-1;15,-1)&14&-15&13\\
(15,-1;1,-1)&14&-15&13\\
&&&\\
\hline
\end{array}
\end{array}
}{XYlistAA2}{Full list of the torus wrapping numbers on $T^2_{(2)} \times T^2_{(3)}$, characterised by $(n^2_a,m^2_a)=${\bf (odd,~odd)},  $(n^3_a,m^3_a)=$(odd,~odd) and  $n^3_a>0$, and their corresponding bulk wrapping numbers $(X_a,Y_a)$, which satisfy the constraints for supersymmetry on the {\bf AA} lattice at the end of section~\ref{Sss:BulkCycles} and do not overshoot the bulk RR tadpole cancellation conditions.}

\mathtabfix{
\begin{array}{ccc}
\multicolumn{3}{c}{\text{\bf \large Systematic classification of bulk 2-cycles on $T_{(2)}^2 \times T_{(3)}^2$ with lattice AA (part III)}}\\
\hline \hline
\begin{array}[h]{|c|c|c|c|}
\hline
(n^2_a,m^2_a;n^3_a,m^3_a) & X_a & Y_a & 2X_a+Y_a\\
\hline
\hline
(1,0;1,-1)&1&-1&1\\
(1,0;3,-1)&3&-1&5\\
(1,2;1,-1)&3&-1&5\\
(1,0;5,-1)&5&-1&9\\
(-1,2;1,-3)&5&-1&9\\
(3,-2;1,1)&5&-1&9\\
(1,4;1,-1)&5&-1&9\\
(1,0;7,-1)&7&-1&13\\
(1,6;1,-1)&7&-1&13\\
(1,-2;1,1)&3&-3&3\\
(1,0;5,-3)&5&-3&7\\
(3,2;1,-1)&5&-3&7\\
(3,4;1,-1)&7&-3&11\\
(1,0;7,-3)&7&-3&11\\
(-1,2;1,-5)&9&-3&15\\
(5,-4;1,1)&9&-3&15\\
(5,-2;1,-1)&3&-5&1\\
(1,0;3,-5)&3&-5&1\\
(5,2;1,-1)&7&-5&9\\
(1,0;7,-5)&7&-5&9\\
(3,-4;1,1)&7&-5&9\\
(1,-2;1,3)&7&-5&9\\
(5,4;1,-1)&9&-5&13\\
(1,0;9,-5)&9&-5&13\\
(1,0;5,-7)&5&-7&3\\
\hline
\end{array}
&
\begin{array}[h]{|c|c|c|c|}
\hline
(n^2_a,m^2_a;n^3_a,m^3_a) & X_a & Y_a & 2X_a+Y_a\\
\hline
\hline
(1,-4;1,1)&5&-7&3\\
(7,-2;1,-1)&5&-7&3\\
(1,-2;3,1)&5&-7&3\\
(3,-2;3,-1)&7&-7&7\\
(1,2;1,-3)&7&-7&7\\
(1,0;9,-7)&9&-7&11\\
(7,2;1,-1)&9&-7&11\\
(1,0;11,-7)&11&-7&15\\
(1,-2;1,5)&11&-7&15\\
(7,4;1,-1)&11&-7&15\\
(5,-6;1,1)&11&-7&15\\
(1,0;5,-9)&5&-9&1\\
(9,-4;1,-1)&5&-9&1\\
(9,-2;1,-1)&7&-9&5\\
(1,0;7,-9)&7&-9&5\\
(9,2;1,-1)&11&-9&13\\
(1,0;11,-9)&11&-9&13\\
(1,0;7,-11)&7&-11&3\\
(1,-6;1,1)&7&-11&3\\
(1,-2;5,1)&7&-11&3\\
(11,-4;1,-1)&7&-11&3\\
(11,-2;1,-1)&9&-11&7\\
(1,0;9,-11)&9&-11&7\\
(5,-8;1,1)&13&-11&15\\
(1,4;1,-3)&13&-11&15\\
\hline
\end{array}
&
\begin{array}[h]{|c|c|c|c|}
\hline
(n^2_a,m^2_a;n^3_a,m^3_a) & X_a & Y_a & 2X_a+Y_a\\
\hline
\hline
(1,-2;3,5)&13&-11&15\\
(3,-2;5,-1)&13&-11&15\\
(1,0;13,-11)&13&-11&15\\
(11,2;1,-1)&13&-11&15\\
(13,-6;1,-1)&7&-13&1\\
(1,0;7,-13)&7&-13&1\\
(1,0;9,-13)&9&-13&5\\
(3,2;1,-3)&9&-13&5\\
(3,-2;5,-3)&9&-13&5\\
(13,-4;1,-1)&9&-13&5\\
(1,0;11,-13)&11&-13&9\\
(3,-8;1,1)&11&-13&9\\
(5,-4;3,-1)&11&-13&9\\
(1,-2;5,3)&11&-13&9\\
(13,-2;1,-1)&11&-13&9\\
(1,2;1,-5)&11&-13&9\\
(3,-4;3,1)&13&-13&13\\
(1,-4;1,3)&13&-13&13\\
(1,-2;7,1)&9&-15&3\\
(1,-8;1,1)&9&-15&3\\
(1,0;11,-15)&11&-15&7\\
(15,-4;1,-1)&11&-15&7\\
(1,0;13,-15)&13&-15&11\\
(15,-2;1,-1)&13&-15&11\\
&&&\\
\hline
\end{array}
\end{array}
}{XYlistAA3}{Full list of the torus wrapping numbers on $T^2_{(2)} \times T^2_{(3)}$, characterised by $(n^2_a,m^2_a)=$ {\bf (odd,~even)},  $(n^3_a,m^3_a)=$(odd,~odd) and  $n^3_a>0$, and their corresponding bulk wrapping numbers $(X_a,Y_a)$, which satisfy the constraints for supersymmetry on the {\bf AA} lattice at the end of section~\ref{Sss:BulkCycles} and do not overshoot the bulk RR tadpole cancellation conditions.}

\mathsidetabfix{
\begin{array}{|c||c||c||c||c|}\hline
& \muc{4}{|c|}{\text{\bf $\Z_2^{(1)}$ Exceptional wrapping numbers $(x^{(1)}_{\alpha}, y^{(1)}_{\alpha})$ in dependence of torus-wrappings $(n^1,m^1)$, 
$\Z_2$ eigenvalues $(-1)^{\tau^{\Z_2^{(1)}}}$, displacements  $(\vec{\sigma})$ and Wilson lines $(\vec{\tau})$}}
\\
(n^2,m^2;n^3,m^3) & (\sigma^2;\sigma^3)=(0;0) & (1;0) & (0;1) & (1;1)
\\\hline\hline
\text{(odd,odd;odd,odd)} & (z^{(1)}_{\alpha} n^1 \,, \, z^{(1)}_{\alpha} m^1 )_{\alpha=0,1,2,4}
&  \begin{array}{c}  (\hat z^{(1)}_{\alpha} n^1 \,, \, \hat z^{(1)}_{\alpha} m^1 )_{\alpha=1} \\
 (z^{(1)}_{\alpha} n^1 \,, \, z^{(1)}_{\alpha} m^1 )_{\alpha=5,3}
 \end{array}
&  \begin{array}{c} (\hat z^{(1)}_{\alpha} n^1 \,, \, \hat z^{(1)}_{\alpha} m^1 )_{\alpha=2} \\
 (z^{(1)}_{\alpha} n^1 \,, \, z^{(1)}_{\alpha} m^1 )_{\alpha=5,3}
\end{array}
&  \begin{array}{c}  (\hat z^{(1)}_{\alpha} n^1 \,, \, \hat z^{(1)}_{\alpha} m^1 )_{\alpha=4} \\
 (z^{(1)}_{\alpha} n^1 \,, \, z^{(1)}_{\alpha} m^1 )_{\alpha=3,5}   \end{array}
\\\hline
\text{(odd,even;odd,odd)} & (z^{(1)}_{\alpha} n^1 \,, \, z^{(1)}_{\alpha} m^1 )_{\alpha=0,1,2,5}
&   \begin{array}{c}  (\hat z^{(1)}_{\alpha} n^1 \,, \, \hat z^{(1)}_{\alpha} m^1 )_{\alpha=1} \\
 (z^{(1)}_{\alpha} n^1 \,, \, z^{(1)}_{\alpha} m^1 )_{\alpha=3,4}
 \end{array}
&  \begin{array}{c}  (\hat z^{(1)}_{\alpha} n^1 \,, \, \hat z^{(1)}_{\alpha} m^1 )_{\alpha=2} \\
 (z^{(1)}_{\alpha} n^1 \,, \, z^{(1)}_{\alpha} m^1 )_{\alpha=3,4}
\end{array}
&  \begin{array}{c} (\hat z^{(1)}_{\alpha} n^1 \,, \, \hat z^{(1)}_{\alpha} m^1 )_{\alpha=5} \\
 (z^{(1)}_{\alpha} n^1 \,, \, z^{(1)}_{\alpha} m^1 )_{\alpha=4,3}  \end{array}
\\\hline
\text{(even,odd;odd,odd)} &(z^{(1)}_{\alpha} n^1 \,, \, z^{(1)}_{\alpha} m^1 )_{\alpha=0,1,2,3}
&   \begin{array}{c} (\hat z^{(1)}_{\alpha} n^1 \,, \, \hat z^{(1)}_{\alpha} m^1 )_{\alpha=1} \\
 (z^{(1)}_{\alpha} n^1 \,, \, z^{(1)}_{\alpha} m^1 )_{\alpha=4,5} 
 \end{array}
&  \begin{array}{c}  (\hat z^{(1)}_{\alpha} n^1 \,, \, \hat z^{(1)}_{\alpha} m^1 )_{\alpha=2} \\
 (z^{(1)}_{\alpha} n^1 \,, \, z^{(1)}_{\alpha} m^1 )_{\alpha=4,5}  \end{array}
&  \begin{array}{c}  (\hat z^{(1)}_{\alpha} n^1 \,, \, \hat z^{(1)}_{\alpha} m^1 )_{\alpha=3} \\
 (z^{(1)}_{\alpha} n^1 \,, \, z^{(1)}_{\alpha} m^1 )_{\alpha=5,4} \end{array}\\
 \hline  & 
 z^{(1)}_\alpha = (-)^{\tau^{\Z_2^{(1)}}} \left\{ \begin{array}{lc}
 1 & \alpha = 0 \rightarrow 0 \rightarrow 0 \\
 (-)^{\tau^2} & \alpha = 1 \rightarrow 1 \rightarrow 1  \\
 (-)^{\tau^3} & \alpha = 2\rightarrow 2 \rightarrow 2   \\ 
  (-)^{\tau^2 + \tau^3} & \alpha = 4 \rightarrow 5 \rightarrow 3 \\
\end{array} \right. &
 \begin{array}{lc}
 \hat z^{(1)}_\alpha = (-)^{\tau^{\Z_2^{(1)}}} (1 + (-)^{\tau^2}) & \alpha =1 \rightarrow 1 \rightarrow 1 \\
 z^{(1)}_{\alpha} = (-)^{\tau^{\Z_2^{(1)}}+ \tau^3} & \alpha =5 \rightarrow 3 \rightarrow 4\\
 z^{(1)}_{\alpha} = (-)^{\tau^{\Z_2^{(1)}}+ \tau^2 + \tau^3} & \alpha =3 \rightarrow 4 \rightarrow 5
\end{array} &
 \begin{array}{lc}
 \hat z^{(1)}_\alpha = (-)^{\tau^{\Z_2^{(1)}}} (1 + (-)^{\tau^3}) & \alpha =2 \rightarrow 2 \rightarrow 2 \\
 z^{(1)}_{\alpha} = (-)^{\tau^{\Z_2^{(1)}}+ \tau^2} & \alpha =5 \rightarrow 3 \rightarrow 4\\
 z^{(1)}_{\alpha} = (-)^{\tau^{\Z_2^{(1)}}+ \tau^2 + \tau^3} & \alpha =3 \rightarrow 4 \rightarrow 5
\end{array} &
 \begin{array}{lc}
 \hat z^{(1)}_\alpha = (-)^{\tau^{\Z_2^{(1)}}} ((-)^{\tau^2} + (-)^{\tau^3}) & \alpha =4 \rightarrow 5 \rightarrow 3 \\
 z^{(1)}_{\alpha} = (-)^{\tau^{\Z_2^{(1)}}} & \alpha =3 \rightarrow 4 \rightarrow 5\\
 z^{(1)}_{\alpha} = (-)^{\tau^{\Z_2^{(1)}}+ \tau^2 + \tau^3} & \alpha =5 \rightarrow 3 \rightarrow 4
\end{array} \\
\hline
\hline
& \muc{4}{|c|}{\text{\bf $\Z_2^{(2)}$ Exceptional wrapping numbers $(x^{(2)}_{\alpha}, y^{(2)}_{\alpha})$ in dependence of torus-wrappings $(n^2,m^2)$, 
$\Z_2$ eigenvalues $(-1)^{\tau^{\Z_2^{(2)}}}$, displacements  $(\vec{\sigma})$ and Wilson lines $(\vec{\tau})$}}\\
(n^3,m^3) & (\sigma^1, \sigma^3) = (0,0) & (1,0)  & (0,1) & (1,1)
\\\hline \hline
\text{(odd,odd)} &\begin{array}{c} (- \zeta_\alpha^{(2)} (n^2 + m^2),  \zeta_\alpha^{(2)} n^2  )_{\alpha = 1} \\ (- \zeta_\alpha^{(2)} (n^2 + m^2),  \zeta_\alpha^{(2)} n^2  )_{\alpha\, \in\, \{3,2,4 \} } \end{array} & \begin{array}{c} (- \zeta_\alpha^{(2)} (n^2 + m^2),  \zeta_\alpha^{(2)} n^2  )_{\alpha\, \in\, \{2,3 \}}\\
(- \zeta_\alpha^{(2)} (n^2 + m^2),  \zeta_\alpha^{(2)} n^2  )_{\alpha\, \in\, \{3,4 \}}
\end{array}&
\begin{array}{c}
 \left( \hat\zeta^{(2)}_{\alpha} \; n^2+\zeta^{(2)}_{\alpha} \; m^2 \; , \; -\zeta^{(2)}_{\alpha}\; n^2 +(\hat \zeta^{(2)}_{\alpha} -\zeta^{(2)}_{\alpha} ) \; m^2_a \right)_{\alpha=1} \\
  \left( \hat\zeta^{(2)}_{\alpha} \; n^2+\zeta^{(2)}_{\alpha} \; m^2 \; , \; -\zeta^{(2)}_{\alpha}\; n^2 +(\hat \zeta^{(2)}_{\alpha} -\zeta^{(2)}_{\alpha} ) \; m^2 \right)_{\alpha\, \in\, \{ 3,2,4\}}
\end{array}
  & 
  \begin{array}{c}
 \left( \hat\zeta^{(2)}_{\alpha} \; n^2+\zeta^{(2)}_{\alpha} \; m^2 \; , \; -\zeta^{(2)}_{\alpha}\; n^2 +(\hat \zeta^{(2)}_{\alpha} -\zeta^{(2)}_{\alpha} ) \; m^l \right)_{\alpha\, \in\, \{ 2,3\}} \\
  \left( \hat\zeta^{(2)}_{\alpha} \; n^2+\zeta^{(2)}_{\alpha} \; m^2 \; , \; -\zeta^{(2)}_{\alpha}\; n^2 +(\hat \zeta^{(2)}_{\alpha} -\zeta^{(2)}_{\alpha} ) \; m^2 \right)_{\alpha\, \in\, \{ 3,4\}}
\end{array}
  \\
  \hline
& \zeta_\alpha^{(2)} = \left\{ \begin{array}{lc} (-)^{\tau^{\Z_2^{(2)}}+ \tau^3 } & \alpha=1 \rightarrow 1 \rightarrow 1 \\  (-)^{\tau^{\Z_2^{(2)}}+ \tau^1+ \tau^3 } & \alpha=3 \rightarrow 2 \rightarrow 4 \end{array} \right. & 
 \zeta_\alpha^{(2)} = \left\{ \begin{array}{lc} (-)^{\tau^{\Z_2^{(2)}}+ \tau^3 } & \alpha=2 \rightarrow 3 \rightarrow 2 \\  (-)^{\tau^{\Z_2^{(2)}}+ \tau^1 + \tau^3 } & \alpha=4 \rightarrow 4 \rightarrow 3 \end{array} \right.
& \zeta_\alpha^{(2)} = \hat \zeta_\alpha^{(2)} (-)^{\tau^3}, \quad  \hat \zeta_\alpha^{(2)} = \left\{ \begin{array}{lc} (-)^{\tau^{\Z_2^{(2)}} } & \alpha=1 \rightarrow 1 \rightarrow 1 \\  (-)^{\tau^{\Z_2^{(2)}}+ \tau^1} & \alpha=3 \rightarrow 2 \rightarrow 4 \end{array} \right.
 &  \zeta_\alpha^{(2)} = \hat \zeta_\alpha^{(2)} (-)^{\tau^3}, \quad  \hat \zeta_\alpha^{(2)} = \left\{ \begin{array}{lc} (-)^{\tau^{\Z_2^{(2)}} } & \alpha=2 \rightarrow 3 \rightarrow 2 \\  (-)^{\tau^{\Z_2^{(2)}}+ \tau^1} & \alpha=4 \rightarrow 4 \rightarrow 3 \end{array} \right. \\
\hline
\hline
& \muc{4}{|c|}{\text{\bf $\Z_2^{(3)}$ Exceptional wrapping numbers $(x^{(3)}_{\alpha}, y^{(3)}_{\alpha})$ in dependence of torus-wrappings $(n^3,m^3)$, 
$\Z_2$ eigenvalues $(-1)^{\tau^{\Z_2^{(3)}}}$, displacements  $(\vec{\sigma})$ and Wilson lines $(\vec{\tau})$}}\\
(n^2,m^2) & (\sigma^1, \sigma^2) = (0,0) & (1,0)  & (0,1) & (1,1)
\\\hline \hline
\text{(odd,odd)} &\begin{array}{c} (- \zeta_\alpha^{(3)} (n^3 + m^3),  \zeta_\alpha^{(3)} n^3  )_{\alpha = 1} \\ (- \zeta_\alpha^{(3)} (n^3 + m^3),  \zeta_\alpha^{(3)} n^3  )_{\alpha\, \in\, \{3,2,4 \} } \end{array} & \begin{array}{c} (- \zeta_\alpha^{(3)} (n^3 + m^3),  \zeta_\alpha^{(3)} n^3  )_{\alpha\, \in\, \{2,3 \}}\\
(- \zeta_\alpha^{(3)} (n^3 + m^3),  \zeta_\alpha^{(3)} n^3  )_{\alpha\, \in\, \{3,4 \}}
\end{array}&
\begin{array}{c}
 \left( \hat\zeta^{(3)}_{\alpha} \; n^3+\zeta^{(3)}_{\alpha} \; m^3 \; , \; -\zeta^{(3)}_{\alpha}\; n^3 +(\hat \zeta^{(3)}_{\alpha} -\zeta^{(3)}_{\alpha} ) \; m^3_a \right)_{\alpha=1} \\
  \left( \hat\zeta^{(3)}_{\alpha} \; n^3+\zeta^{(3)}_{\alpha} \; m^3 \; , \; -\zeta^{(3)}_{\alpha}\; n^3 +(\hat \zeta^{(3)}_{\alpha} -\zeta^{(3)}_{\alpha} ) \; m^3 \right)_{\alpha\, \in\, \{ 3,2,4\}}
\end{array}
  & 
  \begin{array}{c}
 \left( \hat\zeta^{(3)}_{\alpha} \; n^3+\zeta^{(3)}_{\alpha} \; m^3 \; , \; -\zeta^{(3)}_{\alpha}\; n^3 +(\hat \zeta^{(3)}_{\alpha} -\zeta^{(3)}_{\alpha} ) \; m^3 \right)_{\alpha\, \in\, \{ 2,3\}} \\
  \left( \hat\zeta^{(3)}_{\alpha} \; n^3+\zeta^{(3)}_{\alpha} \; m^3 \; , \; -\zeta^{(3)}_{\alpha}\; n^3 +(\hat \zeta^{(3)}_{\alpha} -\zeta^{(3)}_{\alpha} ) \; m^3 \right)_{\alpha\, \in\, \{ 3,4\}}
\end{array}
  \\
\hline
\text{(odd, even)} & 
\begin{array}{c}
(\zeta_\alpha^{(3)} m^3  ,- \zeta_\alpha^{(3)} (n^3 + m^3))_{\alpha = 1}\\
(\zeta_\alpha^{(3)} m^3  ,- \zeta_\alpha^{(3)} (n^3 + m^3))_{\alpha\, \in\, \{3,2,4\} }
\end{array}
&
\begin{array}{c}
(\zeta_\alpha^{(3)} m^3  ,- \zeta_\alpha^{(3)} (n^3 + m^3))_{\alpha\, \in\, \{2,3 \}}\\
(\zeta_\alpha^{(3)} m^3  ,- \zeta_\alpha^{(3)} (n^3 + m^3))_{\alpha\, \in\, \{3,4 \}}
\end{array}
& 
\begin{array}{c}
( - \zeta_\alpha^{(3)} n^3 + (\hat \zeta_\alpha^{(3)} - \zeta_\alpha^{(3)} ) m^3, (\zeta_\alpha^{(3)} - \hat \zeta_\alpha^{(3)} ) n^3 - \hat \zeta_\alpha^{(3)} m^3 )_{\alpha = 1}\\
( - \zeta_\alpha^{(3)} n^3 + (\hat \zeta_\alpha^{(3)} - \zeta_\alpha^{(3)} ) m^3, (\zeta_\alpha^{(3)} - \hat \zeta_\alpha^{(3)} ) n^3 - \hat \zeta_\alpha^{(3)} m^3 )_{\alpha\, \in\, \{2,3,4\} }
\end{array}
&
\begin{array}{c}
( - \zeta_\alpha^{(3)} n^3 + (\hat \zeta_\alpha^{(3)} - \zeta_\alpha^{(3)} ) m^3, (\zeta_\alpha^{(3)} - \hat \zeta_\alpha^{(3)} ) n^3 - \hat \zeta_\alpha^{(3)} m^3 )_{\alpha\, \in\, \{2,3 \}}\\
( - \zeta_\alpha^{(3)} n^3 + (\hat \zeta_\alpha^{(3)} - \zeta_\alpha^{(3)} ) m^3, (\zeta_\alpha^{(3)} - \hat \zeta_\alpha^{(3)} ) n^3 - \hat \zeta_\alpha^{(3)} m^3 )_{\alpha\, \in\, \{3,4 \}}
\end{array} \\
\hline
\text{(even,odd)} & 
\begin{array}{c}
(\zeta_\alpha^{(3)} n^3 , \zeta_\alpha^{(3)} m^3)_{\alpha = 1} \\
(\zeta_\alpha^{(3)} n^3 , \zeta_\alpha^{(3)} m^3)_{\alpha \, \in \, \{3,2,4 \}}
\end{array}
&
\begin{array}{c}
(\zeta_\alpha^{(3)} n^3 , \zeta_\alpha^{(3)} m^3)_{\alpha\, \in\, \{2,3 \}} \\
(\zeta_\alpha^{(3)} n^3 , \zeta_\alpha^{(3)} m^3)_{\alpha\, \in\, \{3,4 \}}
\end{array}
& 
\begin{array}{c}
(( \zeta_\alpha^{(3)} -\hat \zeta_\alpha^{(3)}) n^3 - \hat \zeta_\alpha^{(3)} m^3, \hat \zeta_\alpha^{(3)} n^3 +  \zeta_\alpha^{(3)} m^3 )_{\alpha = 1} \\
(( \zeta_\alpha^{(3)} -\hat \zeta_\alpha^{(3)}) n^3 - \hat \zeta_\alpha^{(3)} m^3, \hat \zeta_\alpha^{(3)} n^3 +  \zeta_\alpha^{(3)} m^3 )_{\alpha \, \in \, \{2,3,4 \}}
\end{array}
& 
\begin{array}{c}
(( \zeta_\alpha^{(3)} -\hat \zeta_\alpha^{(3)}) n^3 - \hat \zeta_\alpha^{(3)} m^3, \hat \zeta_\alpha^{(3)} n^3 +  \zeta_\alpha^{(3)} m^3 )_{\alpha\, \in\, \{2,3 \}}  \\
(( \zeta_\alpha^{(3)} -\hat \zeta_\alpha^{(3)}) n^3 - \hat \zeta_\alpha^{(3)} m^3, \hat \zeta_\alpha^{(3)} n^3 +  \zeta_\alpha^{(3)} m^3 )_{\alpha\, \in\, \{3,4 \}}
\end{array} \\
\hline
& \zeta_\alpha^{(3)} = \left\{ \begin{array}{lc} (-)^{\tau^{\Z_2^{(3)}}+ \tau^2 } & \alpha=1 \rightarrow 1 \rightarrow 1 \\  (-)^{\tau^{\Z_2^{(3)}}+ \tau^1+ \tau^2 } & \alpha=3 \rightarrow 2 \rightarrow 4 \end{array} \right. & 
 \zeta_\alpha^{(3)} = \left\{ \begin{array}{lc} (-)^{\tau^{\Z_2^{(3)}}+ \tau^2 } & \alpha=2 \rightarrow 3 \rightarrow 2 \\  (-)^{\tau^{\Z_2^{(3)}}+ \tau^1 + \tau^2 } & \alpha=4 \rightarrow 4 \rightarrow 3 \end{array} \right.
& \zeta_\alpha^{(3)} = \hat \zeta_\alpha^{(3)} (-)^{\tau^2}, \quad  \hat \zeta_\alpha^{(3)} = \left\{ \begin{array}{lc} (-)^{\tau^{\Z_2^{(3)}} } & \alpha=1 \rightarrow 1 \rightarrow 1 \\  (-)^{\tau^{\Z_2^{(3)}}+ \tau^1} & \alpha=3 \rightarrow 2 \rightarrow 4 \end{array} \right.
 &  \zeta_\alpha^{(3)} = \hat \zeta_\alpha^{(3)} (-)^{\tau^2}, \quad  \hat \zeta_\alpha^{(3)} = \left\{ \begin{array}{lc} (-)^{\tau^{\Z_2^{(3)}} } & \alpha=2 \rightarrow 3 \rightarrow 2 \\  (-)^{\tau^{\Z_2^{(3)}}+ \tau^1} & \alpha=4 \rightarrow 4 \rightarrow 3 \end{array} \right. \\
\hline
\end{array}
}{exceptional-wrappings-Z2Z6}{Complete list of exceptional wrapping numbers $(x^{(i)}_{\alpha}, y^{(i)}_{\alpha})_{i=1,2,3}$ as a function 
of the discrete parameters per D6-brane, extending the schematic form in table~\protect\ref{tab:Z2Z6ExceptionalWrappingNumbers} of 
section~\protect\ref{Sss:FractionalCycles}. Following section~\ref{Sss:BulkCycles}, $(n^3,m^3)$ is assumed to be (odd,odd) in order to select one specific orbifold representant. The subscript $\alpha$ indicates which exceptional wrapping numbers are non-vanishing, and a chain of the form $\alpha = 3 \rightarrow 2 \rightarrow 4$ should be interpreted as follows: for $(n^1, m^1)=$(odd,odd) the prefactor $\zeta_{\alpha}^{(l=2,3)}$ (or  $\hat \zeta_{\alpha}^{(l=2,3)}$) with subscript $\alpha = 3$ takes the given expression, for $(n^1, m^1)=$(odd, even) it is the prefactor with $\alpha = 2$ and for $(n^1, m^1)=$(even,odd) the prefactor with $\alpha = 4$ has the prescribed form. A similar scheme is understood for the non-vanishing prefactors $z_\alpha^{(1)}$ and $\hat z_\alpha^{(1)}$  in the fourth row of the table, for $(n^2, m^2)=$(odd,odd), (odd,even), or (even,odd) respectively. 
}

\clearpage


\addcontentsline{toc}{section}{References}
\bibliographystyle{ieeetr}
\bibliography{refs_Z2Z6-Models}

\end{document}